\documentclass[aps,prd,superscriptaddress,nofootinbib,amsmath,amsfonts,preprintnumbers,groupedaddress,showpacs,10pt,english]{revtex4-1}
\usepackage{amsmath}
\usepackage{amssymb}
\usepackage{babel}
\usepackage{wrapfig}
\usepackage{cancel}

\usepackage{relsize,exscale}
\makeatletter

\usepackage{array,multirow,graphicx}
\usepackage{dcolumn}
\usepackage{newlfont}
\usepackage{bm}
\usepackage[colorlinks,citecolor=blue,urlcolor=blue,linkcolor=blue]{hyperref}
\usepackage[figtopcap]{subfigure}
\usepackage{color}
\usepackage{verbatim}

\usepackage{scalerel}
\usepackage{tikz}
\usetikzlibrary{svg.path}
\definecolor{orcidlogocol}{HTML}{A6CE39}
\tikzset{
  orcidlogo/.pic={
    \fill[orcidlogocol] svg{M256,128c0,70.7-57.3,128-128,128C57.3,256,0,198.7,0,128C0,57.3,57.3,0,128,0C198.7,0,256,57.3,256,128z};
    \fill[white] svg{M86.3,186.2H70.9V79.1h15.4v48.4V186.2z}
                 svg{M108.9,79.1h41.6c39.6,0,57,28.3,57,53.6c0,27.5-21.5,53.6-56.8,53.6h-41.8V79.1z M124.3,172.4h24.5c34.9,0,42.9-26.5,42.9-39.7c0-21.5-13.7-39.7-43.7-39.7h-23.7V172.4z}
                 svg{M88.7,56.8c0,5.5-4.5,10.1-10.1,10.1c-5.6,0-10.1-4.6-10.1-10.1c0-5.6,4.5-10.1,10.1-10.1C84.2,46.7,88.7,51.3,88.7,56.8z};}}
\newcommand\orcid[1]{\href{https://orcid.org/#1}{\mbox{\scalerel*{
\begin{tikzpicture}[yscale=-1,transform shape]
\pic{orcidlogo};
\end{tikzpicture}
}{|}}}}
\graphicspath{{./}{Figs/}}
\begin{document}
\date{\today}

\title{Constraining $f({\cal R})$ gravity by  Pulsar {\textit SAX J1748.9-2021} observations }

\author{Gamal~G.~L.~Nashed~\orcid{0000-0001-5544-1119}}
\email{nashed@bue.edu.eg}
\affiliation {Centre for Theoretical Physics, The British University, P.O. Box
43, El Sherouk City, Cairo 11837, Egypt.}
\author{Salvatore Capozziello~\orcid{0000-0003-4886-2024}}
\email{capozziello@na.infn.it}
\affiliation {Dipartimento di Fisica E. Pancini, Universit`a di Napoli Federico II\\Complesso Universitario di Monte Sant Angelo, Edificio G, Via Cinthia, I-80126, Napoli, Italy,}
\affiliation {Scuola Superiore Meridionale, Largo S. Marcellino 10, I-80138, Napoli, Italy,}
\affiliation {Istituto Nazionale di Fisica Nucleare (INFN), Sez. di Napoli\\Complesso Universitario di Monte Sant Angelo, Edificio G, Via Cinthia, I-80126, Napoli, Italy.}



\begin{abstract}
We discuss spherically symmetric dynamical systems in the framework of a general model of $f({\cal R})$ gravity,   i.e.  $f({\cal R})={\cal R}e^{\zeta {\cal R}}$, where $\zeta$ is a dimensional quantity in squared length units [L$^2$]. We initially assume that the internal structure of such systems is governed by the Krori-Barua ansatz, alongside the presence of fluid anisotropy. By employing astrophysical observations obtained from the pulsar {\textit SAX J1748.9-2021}, derived from bursting X-ray binaries located within globular clusters, we determine that $\zeta$ is approximately equal to $\pm 5$ km$^2$. In particular, the model can create a stable configuration for  {\textit SAX J1748.9-2021}, encompassing its geometric and physical characteristics. In $f({\cal R})$ gravity, the Krori-Barua approach links $p_r$ and $p_t$, which represent the components of the pressures, to ($\rho$), representing the density, semi-analytically.  These relations are described as $p_r\approx v_r^2 (\rho-\rho_{I})$ and $p_t\approx v_t^2 (\rho-\rho_{II})$.  Here, the expression $v_r$ and $v_t$ represent the radial and tangential sound speeds, respectively. Meanwhile, $\rho_I$ pertains to the surface density and $\rho_{II}$ is derived using the parameters of the model.    Notably, within the frame of $f({\cal R})$ gravity where $\zeta$ is negative, the maximum compactness, denoted as $C$, is inherently limited to values that do not exceed the Buchdahl limit. This contrasts with general relativity or with $f({\cal R})$ with positive  $\zeta$, where $C$ has the potential to reach the limit of the black hole  asymptotically. The predictions of such model suggest a central energy density which  largely exceeds the saturation of nuclear density, which has the value $\rho_{\text{nuc}} = 3\times 10^{14}$ g/cm$^3$. Also, the density at the surface  $\rho_I$  surpasses $\rho_{\text{nuc}}$.  We obtain the relation between mass and radius represent it graphically,  and show that it is consistent with other observational data.
\end{abstract}

\maketitle

\section{Introduction}\label{Sec:Introduction}

Experiments measuring the perihelion advance of Mercury, the gravitational Doppler effect, and light deflection, all of which are used to evaluate the accuracy of Einstein's general relativity (GR) at solar-system scales, have demonstrated exceptionally high precision, as reported in \cite{Will:2014kxa}. Additionally, GR has undergone tests in extreme gravitational conditions: For example, detecting gravitational waves produced by the collision of compact celestial objects \cite{LIGOScientific:2016lio, LIGOScientific:2018dkp}. As a consequence, GR has risen as the most successful and widely accepted theory for explaining gravitational phenomena. However, Einstein's theory presents shortcomings at UV scales for the lack of a viable quantum gravity, and at IR scales because significant issues related  to  observed Universe  remain unanswered. In fact, to account for cosmic acceleration within the framework of GR, we need to introduce dark energy, which is an unusual form of matter-energy characterized by a substantial negative pressure. Furthermore, a huge amount of dark matter is required to address dynamics of cosmic structures like galaxies and clusters of galaxies. In both cases, at the moment, there is no answer at fundamental level explaining such a dark side.

In recent decades, numerous alternative  gravitational theories have arisen, aiming to explain the expansion that accelerates Universe and cosmic structure dynamics. Specifically,  GR can be modified or extended at  UV and IR  scales   in order to achieve  a comprehensive explanation of the early inflationary era involving dark energy, at late epochs. cite{Starobinsky:1980te, Capozziello:2002rd, Carroll:2003wy, Hu:2007nk, Nojiri:2006ri, Amendola:2006we, Appleby:2007vb,  Odintsov:2020nwm, Koyama:2015vza,Nojiri:2003ft, Nojiri:2007cq, Cognola:2007zu, Oikonomou:2020qah, Oikonomou:2020oex}.

A straightforward method for improving GR is  substituting the scalar curvature ${\cal R}$ with  $f{\cal R}$, giving rise to what is known as $f({\cal R})$  gravity. Related models have been explored as potential explanations for dynamics of both  early and late Universe. Comprehensive reviews on this topic are \cite{Sotiriou:2008rp,DeFelice:2010aj,Capozziello:2011et, Nojiri:2010wj, Clifton:2011jh, Nojiri:2017ncd}. Conversely, at astrophysical scales, modifications of GR  impact also the  characteristics of compact stars. In fact, in the context of ${\cal R}^2$ gravity, commonly known as the Starobinsky model \cite{Starobinsky:1980te}, and using a non-perturbative methodology, there is a significant increase in the maximum mass of compact stars on the diagram of mass and radius, primarily attributable to the quadratic term \cite{Yazadjiev:2014cza, Astashenok:2014dja, Yazadjiev:2015xsj, Sbisa:2019mae, Astashenok:2020qds, Astashenok:2021peo, Astashenok:2021xpm, Nobleson:2022giu,Jimenez:2021wik}. Researches  indicate that in the context of ${\cal R}$-squared gravity, the secondary component of the GW190814 event, as reported by the LIGO-Virgo collaboration in \cite{LIGOScientific:2020zkf}, can be effectively characterized as a NS \cite{Astashenok:2020qds, Astashenok:2021peo,Nashed:2020kdb, Astashenok:2021xpm}. In particular,  the physical properties of neutron stars (NSs) have been examined  both for non-rotating \cite{Astashenok:2020cfv} and rotating \cite{Astashenok:2020cqq} systems, within the context of ${\cal R}^2$ gravity coupled with an axion scalar field. In ~\cite{Olmo:2019flu} thoroughly review stellar structure models within modified gravity theories, and develop in metric and metric-affine approaches.

It has been shown in Ref. \cite{Boehmer:2007kx} that the flat rotation curves of galaxies can be explained by only small departures from GR. Specifically, they found that a gravitational Lagrangian taking the form $f({\cal R}) \propto {\cal R}^{1+ \zeta_1}$ represents a relatively natural modification of Einstein gravity, with the dimensionless parameter $\zeta_1$ being communicated in terms of the tangential speed. In addition, there has been interest in this power-law $f({\cal R})$ model lately as a potential solution to the clustered galactic dark matter issue \cite{Sharma:2020vex}, where the parameter $\zeta_1$ is constrained to be of the order of $10^{-6}$. In this context, Ref.~\cite{Sharma:2020vex} was used to calculate the light deflection angle using the rotational velocity profile of typical nearby galaxies in the gravitational background of ${\cal R}^{1+ \zeta_1}$. Additional investigations of the consequences of $f({\cal R})$ power law can be found in.~\cite{Capozziello:2006dp,Nashed:2018piz,Capozziello:2006ph, Martins:2007uf,Nashed:2020mnp,Jaryal:2021lsu,Sharma:2022tce}.

Conversely, it has been demonstrated that the parameter $\zeta_1$, in the context of ${\cal R}^{1+\zeta_1}$ theory, possesses a substantial influence on the mass and radius plots of isotropic NSs \cite{Astashenok:2020qds,Capozziello:2015yza}.Although isotropic perfect fluids are commonly used to describe the matter inside compact stars, there are good reasons to take anisotropies in extremely dense matter into consideration  \cite{Chaichian:1999gd, Ferrer:2010wz, Horvat:2010xf, Doneva:2012rd, Silva:2014fca, Yagi:2015hda, Ivanov:2017kyr, Isayev:2017rci, Biswas:2019gkw, Maurya:2018kxg, Pretel:2020xuo, Rahmansyah:2020gar, Das:2021qaq, Das:2021giz, Deb:2021ftm, Rahmansyah:2021gzt, Bordbar:2022qhl}. These studies  demonstrate that the existence of anisotropy can either raise or decrease the maximal-mass threshold, thereby offering the potential to attain more massive compact stars that align with astronomical observations.

Folomeev conducted a comprehensive investigation of the impact of anisotropies on the internal structure of NSs,  employing a completely autonomous non-perturbative method inside the Starobinsky mode. This study considered three distinct types of realistic equations of state (EoSs) \cite{Folomeev:2018ioy}. The incorporation of anisotropic pressure has been shown to facilitate the utilization of stiffer equations of state (EoSs) to represent configurations that meet observational criteria effectively. Furthermore, Ref.\cite{Panotopoulos:2021sbf} delved into the exploration of anisotropic quark stars within ${\cal R}$-squared gravity, and more recently, in \cite{Nashed:2021gkp} examined anisotropic compact stars in relation to theories of higher-order curvature such as $f({\cal R})$.According to \cite{FarasatShamir:2019ojb,Nashed:2023lqj, Malik:2022lih, Malik:2021bqi}, a study of compact anisotropic structures in the Starobinsky form and within generalised modified gravity has been conducted. The gravitational collapse of an anisotropic system with heat flow was investigated in Ref.~\cite{Usman:2021qjz} using the Karmarkar condition and taking a logarithmic modification of the standard Starobinsky model into account. For a comparative analysis of self-consistent charged anisotropic spheres in an embedded spacetime framework with the Karmarkar condition applied, see Ref.~\cite{Ahmad:2021jkv}.

To  the best of our information, there is no  investigation into anisotropic compact stars within the framework of the ${\cal R}e^{\zeta{\cal R}}$ gravity model. As we will see below, this $f(\cal{R})$ model is particularly suitable to investigate small deviations with respect to GR.   With this objective at the forefront, we aim to formulate the equations that govern the stellar structure  in this gravitational model and explore the impact of compact star anisotropy. For this goal,We'll make use of the so-called Krori-Barua (KB) ansatz to close the system of differential equations.

The structure of the paper is the following: In Section \ref{Sec2}, we offer a concise overview of  $f({\cal R})$ theory. In Section \ref{Sec:Model}, we introduce the specific $f({\cal R})$ model, namely $f({\cal R})={\cal R}e^{\zeta {\cal R}}$, and derive the form of the system of differential equations, which is presented in detail in  the Supplementary Material.
For the sake of computational convenience, we employ the KB ansatz and deduce the expressions for density, radial pressure, and tangential pressure,
 as detailed in the Supplementary Material.
Utilizing matching conditions, we determine the model parameters.

In Section \ref{Sec:Stability},   pulsar's mass and radius from astrophysical observations of {\textit SAX J1748.9-2021} are used  to derive constraints on the model parameter $\zeta$. In Section \ref{Sec:EoS_MR},e examine the physical quantity deviations from GR and provide numerical results. Finally, in Section \ref{Sec:Conclusion}, we summarize our results. It is worth  noticing that we use geometrical units $G=1=c$ and the spacetime signature $(-,+,+,+)$ for our computations. However, final results are reported in physical units.

\section{The Theoretical framework}\label{Sec2}

\subsection{ $f(R)$ gravity in metric formalism}

In this section, we provide a concise overview of $f({\cal R})$ gravity adopting the metric formalism. The action is
\begin{equation}\label{1}
    S = \frac{1}{2\kappa^2}\int d^4x\sqrt{-g}f({\cal R}) + S_m\,.
\end{equation}
In this study, the Ricci scalar in this case is denoted by ${\cal R}$, the matter action is denoted by $S_m$, and the determinant of $g_{\mu\nu}$ is $g$. The equation of motion is obtained as follows when we change the action with respect to the metric:
\begin{equation}\label{2}
f_{\cal R} {\cal R}_{\mu\nu} - \dfrac{1}{2}g_{\mu\nu}f - \nabla_\mu\nabla_\nu f_{\cal R} + g_{\mu\nu}\square f_{\cal R} = \kappa^2 T_{\mu\nu} \,.
\end{equation}
The gravitational coupling is $\kappa^2 = \frac{8\pi G}{c^4}$, with $G$  the Newton  constant and $c$  the speed of light. Additionally, we have the matter energy-momentum tensor denoted as $T_{\mu\nu}$.

Furthermore, we define $f_{\cal R}$ as the derivative of $f({\cal R})$ with respect to ${\cal R}$, and the covariant derivative $\nabla_\mu$ connected to the metric's Levi-Civita connection. Here $\square$ is the d'Alembert operator. Finally, ${\cal R}_{\mu\nu}$ is the Ricci tensor.


In GR, ${\cal R}$ for a stellar fluid is defined by its energy density and pressure, which are ${\cal R}= -8\pi T$, where $T = g_{\mu\nu}T^{\mu\nu}$. However, in $f({\cal R})$ gravity, both the metric tensor $g_{\mu\nu}$ and the Ricci scalar ${\cal R}$ are non-trivial. Consequently, ${\cal R}$ is now prescribed by a   second order differential equation which can be derived from Eq.~(\ref{2}) as:
\begin{equation}\label{3}
   {\mathit  3\square f_{\cal R}({\cal R}) + {\cal R}f_{\cal R}({\cal R}) - 2f({\cal R}) = \kappa^2 T\,,}
\end{equation}
which means that in comparison to GR, where $T=0$ leads to ${\cal R}=0$, if nonlinear functions of ${\cal R}$ are taken into account, additional curvature components must also be taken into consideration even when standard matter is absent. In other words, a curvature fluid has to be taken into account \cite{Capozziello:2002rd,Capozziello:2011et}.

\subsection{Spherical symmetry}

To investigate the internal composition of a stellar in  an equilibrium, we analyze a spherically symmetric system where the spacetime is characterized by the  metric
\begin{equation}\label{4}
    ds^2 = -e^{\mu}dt^2 + e^{\nu}dr^2 + r^2(d\theta^2 + \sin^2\theta d\phi^2)\,,
\end{equation}
where  the spacetime is given by  Schwarzschild-like coordinates $x^\alpha = (t,r,\theta,\phi)$. Here,  $\mu$ and $\nu$ are function of $r$ only.
We assume that the distribution of stellar matter is an anisotropic perfect fluid, i.e., it can be prescribed by:
\begin{equation}\label{5}
   T_{\alpha\beta} = (\rho + p_t) u_\alpha u_\beta + p_t g_{\alpha\beta} - \sigma k_\alpha k_\beta\,,
\end{equation}
  where $u^\alpha$ is the four-velocity of the fluid with the normalization condition $u_\alpha u^\alpha = -1$; $k^\alpha$ is a unit radial four-vector satisfying the condition $k_\alpha k^\alpha = 1$. Here, the energy density is represented by $\rho$, the tangential pressure by $p_t$, the radial pressure by $p_r$, and the anisotropy factor by $\sigma \equiv p_t - p_r$.

According to the above metric, we can express $u^\alpha$ as $e^{-\mu}\delta_t^\alpha$, and  $k^\alpha$ as $e^{-\nu}\delta_r^\alpha$.  As a result, the energy-momentum tensor (\ref{5}) trace has the following form:
\begin{equation}
T= -\rho + 3p_r+ 2\sigma\,.
\end{equation}
The conservation law of momentum and energy can be obtained from the divergence of Equation (\ref{5}) which yields:
\begin{equation}\label{6}
    \nabla_\nu T_r^{\ \nu} = p_r' + (\rho + p_r)\mu' - \frac{1}{r}\sigma = 0 \,.
\end{equation}
Here  $'$ denotes the derivative W.r.t.$r$. Furthermore, it is
\begin{align}
    \square f_{\cal R} &= \frac{1}{\sqrt{-g}}\partial_\mu\left[ \sqrt{-g}\partial^\mu f_{\cal R} \right] = e^{-\nu}\left[ \left( \frac{1}{r} + \mu' - \nu' \right)f_{\cal R}' + f_{\cal R}^{''} \right] .
\end{align}
Hence, the energy-momentum tensor (\ref{5}) and the line element (\ref{4}) provide, from the field equations (\ref{2}), the  following non-zero components:
\begin{widetext}
\begin{equation}\label{8}
    -\frac{f_{\cal R}}{r^2} + \frac{f_{\cal R}}{r^2}\frac{d}{dr}\left( re^{-\nu} \right) + \frac{1}{2}({\cal R}f_{\cal R} - f) + e^{-\nu}\left[ \left( \frac{2}{r} - \nu' \right)f_{\cal R}' + f_{\cal {\cal R}}^{''} \right] = -8\pi\rho ,
\end{equation}
\vspace{-0.4cm}
\begin{equation}\label{9}
    -\frac{f_{\cal R}}{r^2} + \frac{f_{\cal R}}{e^{\nu}}\left( \frac{\mu'}{r} + \frac{1}{r^2} \right) + \frac{1}{2}({\cal R}f_{\cal R}- f) + e^{-\nu}\left( \frac{2}{r} + \mu' \right)f_{\cal R}' = 8\pi p_r ,
\end{equation}
\vspace{-0.4cm}
\begin{equation}\label{10}
    \frac{f_{\cal R}}{r^2}\left[ 1+ e^{-\nu}(r\nu' - r\mu' -1) \right] - \frac{1}{2}f + e^{-\nu}\left[ \left( \frac{1}{r} + \mu' - \nu' \right)f_{\cal R}' + f_{\cal R}^{''} \right] = 8\pi p_t\, .
\end{equation}
\end{widetext}
The equation governing the evolution of  scalar curvature (\ref{3}) now becomes
\begin{align}\label{11}
    3e^{-\nu}\left[ \left( \frac{1}{r} + \mu' - \nu' \right)f_{\cal R}' + f_{\cal R}^{''} \right] =&\ \kappa( p_r + 2p_t-\rho ) + 2f - {\cal R}f_{\cal R}\, .
\end{align}

\section{The $f({\cal R})$ gravity model}\label{Sec:Model}

Let us now adopt the following $f({\cal R})$ model:
 \begin{align}\label{111}
 f({\cal R})= {\cal R}e^{\zeta {\cal R}}\,,\end{align}
 where $\zeta$ represents a parameter with dimensions  [L$^2$], with $L$ a length. When the dimensional parameter $\zeta$ is set to zero,  the standard Einstein GR is recovered. Conversely, for small non-zero  $\zeta$, the expression can be recast as
 \begin{equation}
 f({\cal R})={\cal R}+\zeta {\cal R}^2+\frac{1}2\zeta^2{\cal R}^3 \cdots\,.
 \end{equation}
  For these reasons, Eq. \eqref{111} is a general $f({\cal R})$ model capable of representing both small and large deviations with respect  to GR.

  Upon inserting the expression for $f({\cal R})$ as provided in Eq. (\ref{111}) into Eqs. \eqref{8} through \eqref{10}, we derive a lengthy system, which is reported in the Supplementary Material.
One can readily observe that when considering $\zeta=0$, the matter density and pressure components undergo modifications and ultimately converge to the GR solution, as demonstrated in previous works \cite{Nashed:2020kjh, Roupas:2020mvs}. Nevertheless, the system  in  Eqs.  ({\color{blue} 1}), ({\color{blue} 2}),  and ({\color{blue} 3}), presented in the Supplementary Material, encompasses five unknown functions necessitating the imposition of two extra constraints to  define it. A possible approach  is  assuming  suitable EoS  to establish connections between radial and tangential pressures and density. However, this approach may not be practical due to the presence of a challenging fourth-order differential equation within the system.

An alternative approach is considering reasonable assumptions for the metric potentials $\mu(r)$ and $\nu(r)$. We will adopt this method and consider the so called Krori-Barua (KB) spacetime  to investigate  stellar models in the frame of $f({\cal R})$ theory of gravity.
\subsection{The Krori-Barua framework}\label{Sec:KB}
Let us introduce the metric potentials of the KB form, as described in \cite{Krori1975ASS}:
\begin{equation}\label{eq:KB}
   {  \mu(r)=s_0 (r/R)^2+s_1,\,  \qquad \nu(r)=s_2 (r/R)^2}\,,
\end{equation}
here, $R$ represents the radius of the star, and the parameters $[s_0, s_1, s_2]$ are dimensionless and can be determined by satisfying junction conditions. Equation (\ref{eq:KB}) ensures a singularity-free solution across the whole interior of the star. The ansatz of  KB has been applied in diverse gravitational theories, including GR. However, in the present investigation, we employ observational constraints derived from the  pulsar \textit{SAX J1748.9-2021} to estimate the model parameter $\zeta$. By employing Eq. \eqref{eq:KB} and Eqs.  ({\color{blue} 1}), ({\color{blue} 2}),  and ({\color{blue} 3}) presented in the Supplementary Material, we can expressed the components of the energy-momentum tensor.

Moreover, we present the notion of the anisotropic which is defined as ${ F_a=\frac{2\sigma}{r}}$, which arises due to the pressure difference, equivalently represented by  anisotropy  $\sigma= { p_t}- { p_{r}}$. Significantly, the anisotropy  becomes zero at the central point. In the strong anisotropy situation, $\sigma>0$, for $0<r \leq R$, it necessitates that $p_t> p_r$ across the  star's interior. Conversely, in mild anisotropy case, i.e.,  $\sigma<0$, it requires that $p_r> p_t$ throughout the  interior of the star.

\subsection{Matching Conditions}\label{Sec:Match}
For a stellar system,  we can suppose that the vacuum solutions of  GR  correspond to those of  $f({\cal R})$  given in Eq. \eqref{111}. This means that the external solution simply aligns with  Schwarzschild spacetime. Thus, we choose to represent vacuum  space as:
\begin{equation}
{  ds^2=-\left(1-\frac{2GM}{c^2r}\right) c^2 dt^2+\left(1-\frac{2GM}{c^2 r}\right)^{-1}dr^2+d\Omega^2, \qquad \mbox{where} \quad d\Omega^2={\mathit r^2 (d\theta^2+\sin^2 \theta d\phi^2)}}\,.
\end{equation}
In this context, $M$ describes  star's mass. Applying the junction conditions we get:
\begin{equation}\label{eq:bo}
  \mu_{_\mathrm {(r={R})}}=\ln{{\mathit (1-C)}},  \qquad \nu_{_\mathit {(r={R})}}=-{\mathrm \ln(1-C)}\,,  \qquad \mbox{\textit  and} \, \qquad {  {p}_{_r}{_{_\mathit{(r={R})=0}}}}\,,
\end{equation}
with ${\mathit C}$ represents  compactness described as:
\begin{align}\label{comp11}
 {\mathit  C=\frac{2GM}{c^2 {R}}}\,.
\end{align}
By adopting Eq.~\eqref{eq:KB} and using $p_t$ from Eqs. (4) in the Supplementary Material, we can establish the specified surface limits. This allows us to express the model parameters $[s_0, s_1, s_2]$ in terms of $\zeta_1=\frac{\zeta}{R^2}$ and $C$.The pulsar mass and radius  can be constrained by astrophysical observations, which, in turn, determine the compactness; consequently, the next step is observationally  constraining  the  parameter $\zeta_1$.
\section{Constraints and    stability considerations from  {\textit SAX J1748.9-2021} observations}\label{Sec:Stability}

In this section, we utilize observational limitations, with a specific focus on $M$ and $R$ of stellar  {\textit SAX J1748.9-2021}, to determine the value of the  parameter $\zeta_1$ in the frame of $f({\cal R})$ theory of gravity. Furthermore, we subject the obtained solution to a thorough evaluation of its stability based on different physical limits. As mentioned earlier in the Introduction, accurate observational data plays a vital role in limiting the parameter space. To begin, our selection criteria are presented for the pulsar {\textit SAX J1748.9-2021} aiming to limit $f({\cal R})$ theory of gravity. This choice is based on the availability of time-resolved spectroscopic data from EXO 1745-248 during thermonuclear bursts, which has yielded precise measurements for the pulsar   ($M= 1.81 \pm 0.3 M_\odot$) and   ($R=11.7\pm 1.7$ km) \citep{Ozel:2008kb,}. By leveraging $M$ and $R$ of the pulsar {\textit SAX J1748.9-2021}, we narrow  $\zeta_1$  of  $f({\cal R})$ model. This parameter space comprises the set of parameters [$\zeta$, $s_0$, $s_1$, $s_2$].

\subsection{The material component}\label{Sec:matt}

\begin{figure*}
\centering
\subfigure[~$\rho$]{\label{fig:density}\includegraphics[scale=0.3]{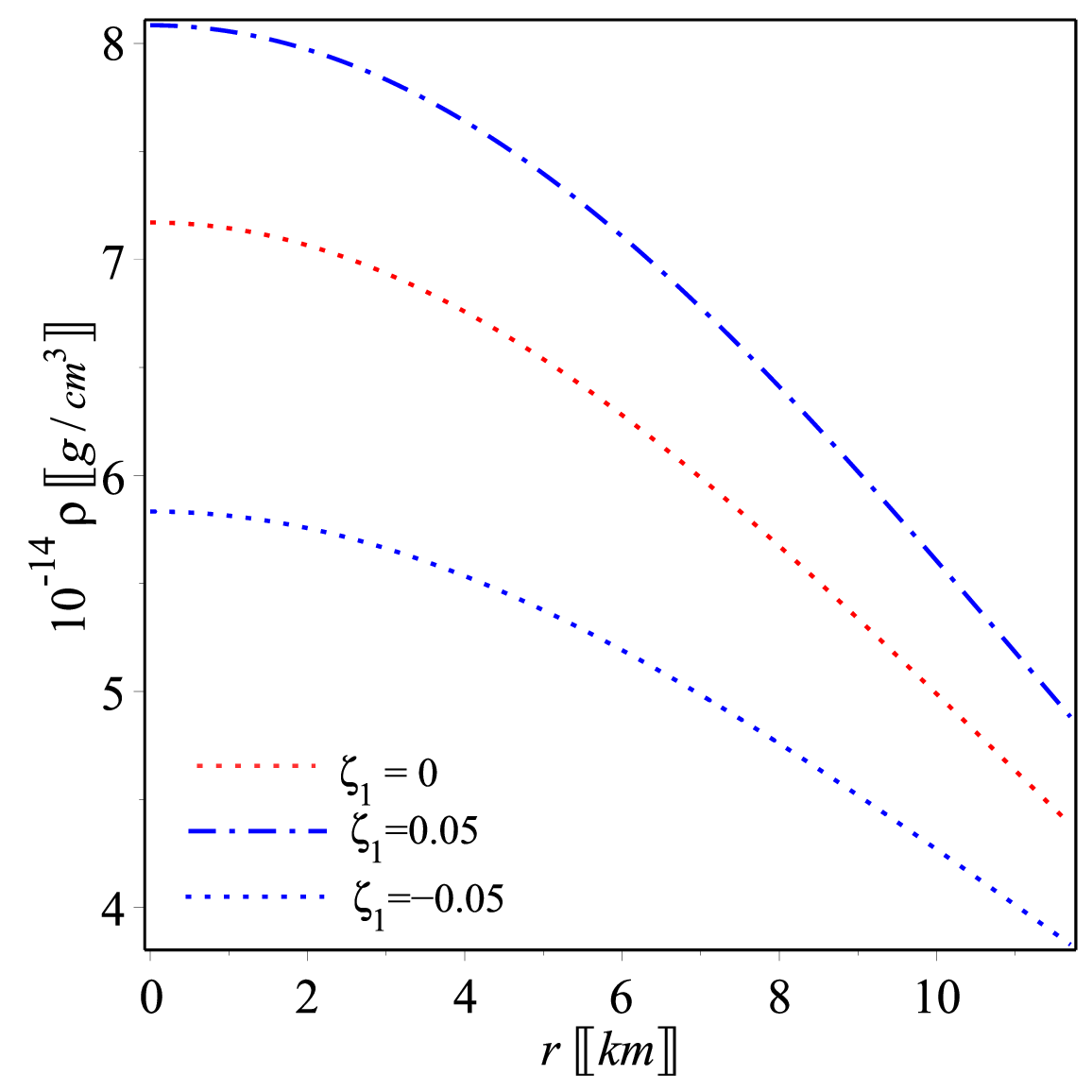}}\hspace{0.5cm}
\subfigure[~$p_r$]{\label{fig:radpressure}\includegraphics[scale=0.3]{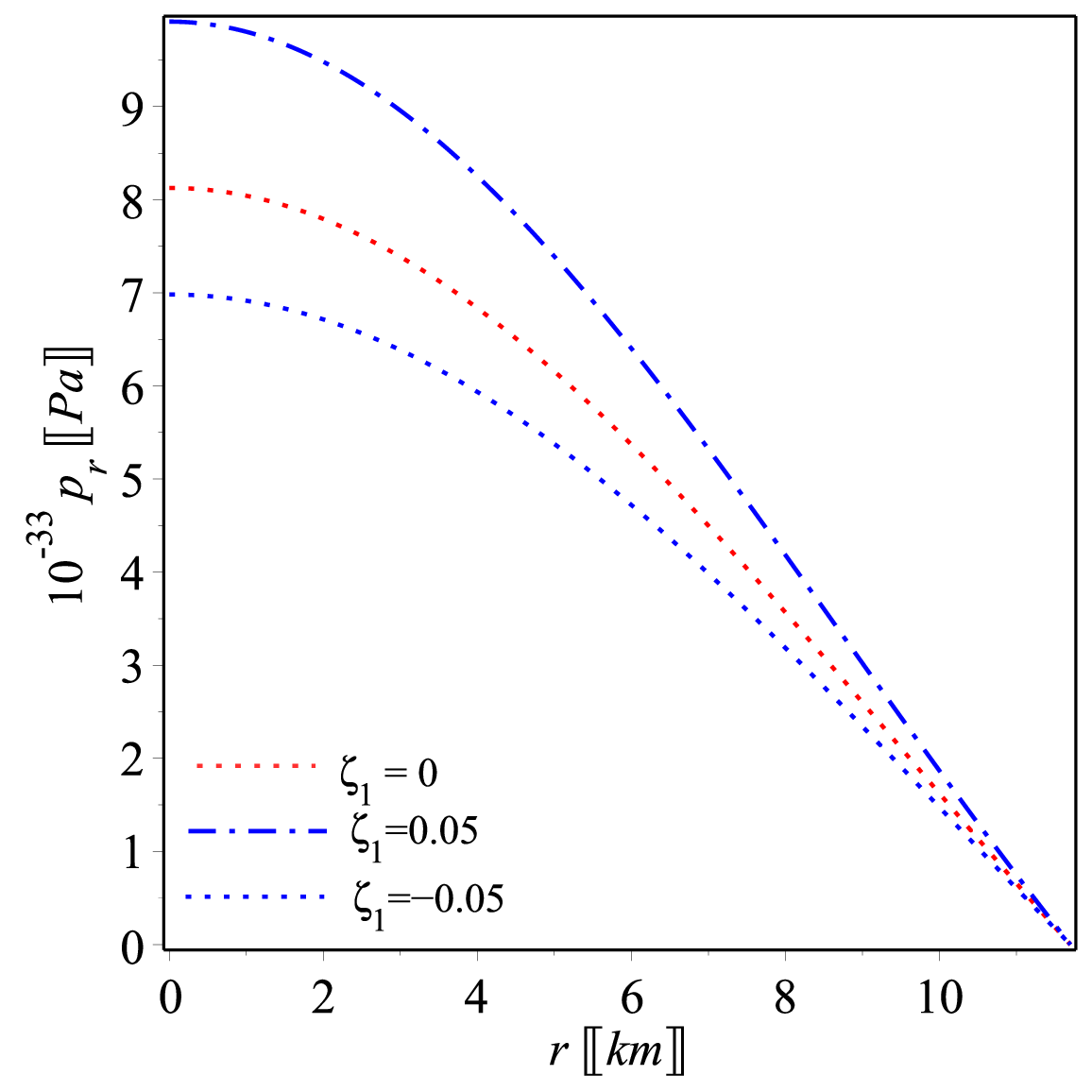}}\\
\subfigure[~$p_t$]{\label{fig:tangpressure}\includegraphics[scale=0.3]{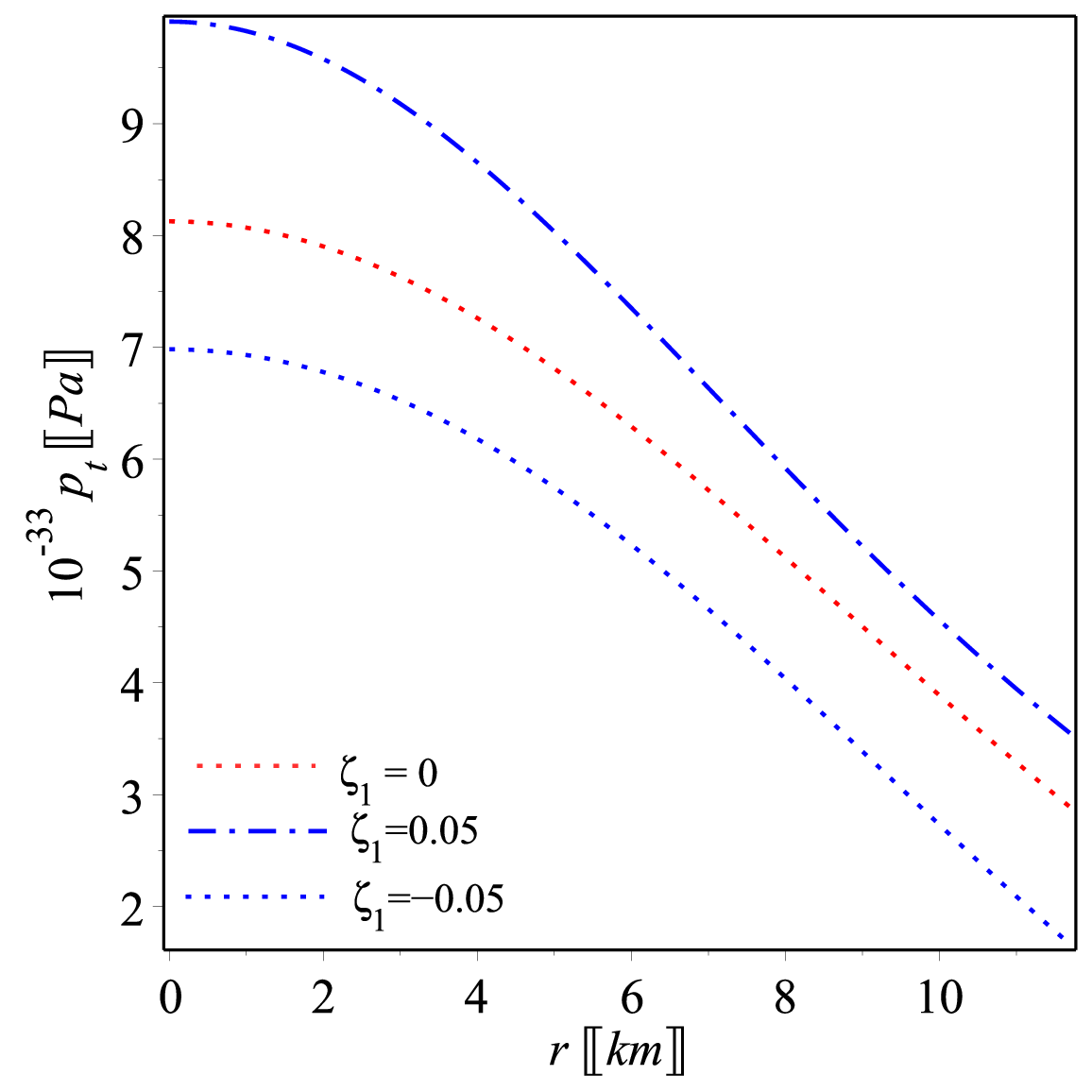}}\hspace{0.5cm}
\subfigure[~$\Delta$ ]{\label{fig:anisotf}\includegraphics[scale=0.3]{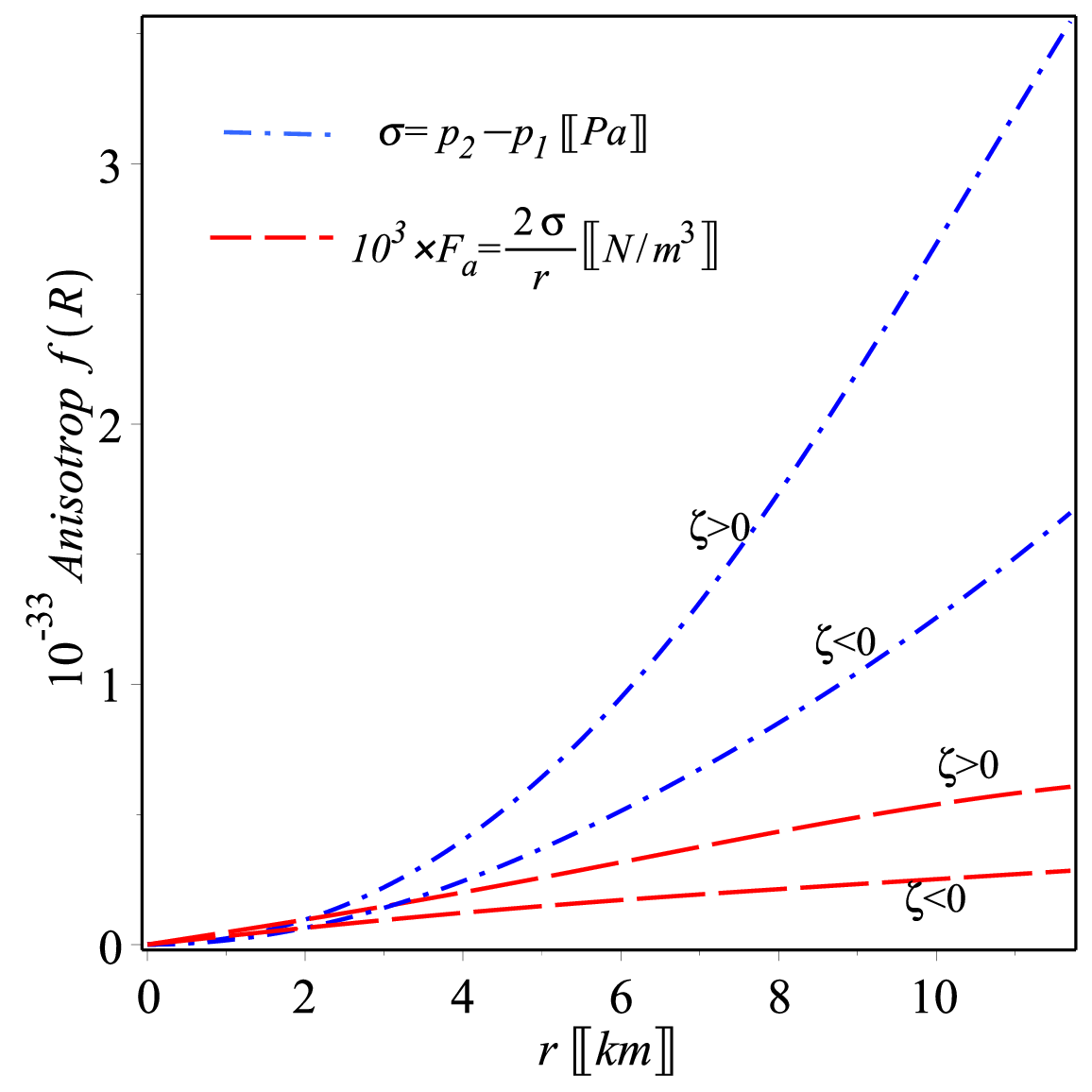}}
\caption{With regard to the pulsar's matter sector  SAX J1748.9-2021, Figs. \subref{fig:density} through \subref{fig:tangpressure} represent the profiles of density as well as $p_r$ and $p_t$  pressure components,   presented in the Supplementary Material, for different values of $\zeta_1$, specifically $\zeta_1=0,~\pm 0.05$. The  figures confirm that $\rho$, $p_r$ and $p_t$ remain limited throughout the interior of the stellar and exhibit a monotonic decrease towards its surface. In Fig. \subref{fig:anisotf}, we present the distribution of anisotropy, denoted as $\sigma(r)$, within the stellar for varying values of $\zeta_1$, specifically $\zeta_1=0,~\pm 0.05$. Notably, the anisotropy vanishes at the central point where $p_t=p_r$, while it assumes positive values elsewhere. }
\label{Fig:dens_press}
\end{figure*}

Referring  to equations   presented in the Supplementary Material, and taking into account  numerical estimation for the model parameters, we generate plots illustrating  $\rho$, $p_r$ and $p_t$ as functions dependent on $r$. These plots are reported in Figures \ref{Fig:dens_press}\subref{fig:density}--\subref{fig:tangpressure}.  The density and pressure profiles clearly adhere to the stability criteria within the material sector, which we are going to discuss below. They reach their maximum values at the core, maintain positivity, and are free of singularities throughout the star's interior. They exhibit a monotonic decrease towards the star surface.
 Furthermore, we generate a figure  representing the difference between pressures (i.e.,  anisotropy$\equiv \Delta$), as illustrated in plot \ref{Fig:dens_press}\subref{fig:anisotf}. Such graph demonstrates that $\Delta$ adheres  the requirement of stability, as it reaches zero at the core and steadily grows in a monotonic way towards the star surface.  It's important to note that the strong anisotropy criteria, as analyzed in this study, introduces an extra positive force, proportional to $\sigma/r$, which impacts the material's overall performance and influences hydrodynamic equilibrium. Such force acts against the force of gravity, playing a pivotal role in adjusting the size of the star, which enables it to support a larger mass compared to scenarios with isotropy or mild anisotropy. A more thorough explanation of this impact can be found in Subsection \ref{Sec:TOV}.

{ It is worth  providing numerical values for  physical quantities associated with the pulsar {\textit SAX J1748.9-2021}, as predicted by the current model.\\ For instance, with $\zeta_1=0.05$, the core density is approximately ${\rho_\text{core}\approx 8.08\times 10^{14}}$ g/cm$^{3}$, that roughly 3 times  $\rho_\text{nuc}$, and both  $p_{r\text{(core)}}$ and $p_{t\text{(core)}}$, are approximately $9.91\times 10^{33}$ dyn/cm$^2$. At the star boundary, we show ${\rho_{I}= 4.88\times 10^{14}}$ g/cm$^{3}$, that $\approx 1.8$ times ${\rho_{I}= 4.88\times 10^{14}}$ g/cm$^{3}$, $p_{r(r=R)}=0$ dyn/cm$^2$, and $p_{t(r=R)}\approx 3.55\times 10^{33}$ dyn/cm$^2$.}\\

{ When $\zeta_1=-0.05$,  ${\rho_\text{core}\approx 5.83\times 10^{14}}$ g/cm$^{3}$, that 2.16 times ${\rho_{I}= 4.88\times 10^{14}}$ g/cm$^{3}$, and  pressure components at the center are approximately $7\times 10^{33}$ dyn/cm$^2$. At boundary, we get  ${\rho_I=3.83\times 10^{14}}$ g/cm$^{3}$, that $\approx 1.4$ times ${\rho_{I}= 4.88\times 10^{14}}$ g/cm$^{3}$, $p_{r(r=R)}= 0$ dyn/cm$^2$, $p_{t(r=R)}\approx 1.66\times 10^{33}$ dyn/cm$^2$.}

{In this context, it is noteworthy that the stellar model under consideration doesn't eliminate the chance that the pulsar {\it SAX J1748.9-2021} has a core composed by neutrons. }
As indicated in Section \ref{Sec:Model}, the KB ansatz \eqref{eq:KB} has been employed as an alternative to  EoSs in order to close the system  presented in the Supplementary Material. Furthermore, we demonstrate that the KB  efficiently establishes relations between the components of the energy-momentum tensor. In order to realize this,   we present   $\xi:=r/R$ and subsequently expand the power series of Eqs. ({\color{blue} 4}) presented in the Supplementary Material up to $O(\xi^4)$. It is evident that the procedure leads to the following relations:
\begin{equation}\label{eq:KB_EoS}
    p_r(\rho)\approx a_1 \rho+a_2\,, \qquad  p_t(\rho) \approx a_3 \rho+a_4\,.
\end{equation}
The constants denoted as $a_1, ..., a_4$ are entirely determined in  the model parameter space, which is  $[\zeta_1,s_0,s_1,s_2]$, as detailed in  the Supplementary Material. Remarkably, we can express the equations above in a more physically meaningful way as follows:
\begin{equation}\label{eq:KB_EoS2}
   {\mathit  p_r(\rho)\approx v_r^2(\rho-\rho_{I})}\,, \qquad  {\mathit p_t(\rho) \approx v_t^2 (\rho-\rho_{II})}\,.
\end{equation}
In this context, we can interpret the following physical parameters as follows: $v_r^2$ represents the sound of speed in the $r$ direction, $\rho_I$ is the density (specifically, the surface density $\rho_I$ satisfying the boundary condition $p_r(\rho_I)=0$), $v_t^2$ denotes the speed of sound in the tangential direction, and $\rho_{II}$ represents another density. It is worth noticing that the condition $p_r(\rho_I)=0$ applies to $\rho_I$, but not necessarily to $\rho_{II}$, since $p_t$  may not vanish on the surface. These parameters encompass two special cases: For hadron matter, the equation of state (EoS), which is maximally compact, where $v_r^2$=$c^2$, and the  EoS based on the MIT which represents the bag model of quark matter, where $v_r^2$=$c^2/3$.    For instance, when $\zeta_1=0.05$, using Eqs.  ({\color{blue} 4})  as detailed in  the Supplementary { Material,  we obtain the following values: $v_r^2=a_1\approx 0.43c^2$, $v_t^2=a_3\approx 0.337c^2$, $\rho_I=-a_2/a_1= 5.9\times 10^{14}$ g/cm$^3$, and $\rho_{II}=-a_4/a_3\approx 4.3\times 10^{14}$ g/cm$^3$. Likewise, for $\zeta_1=-0.05$, we obtain the following values: $v_r^2\approx 0.385c^2$, $v_t^2\approx 0.29c^2$, $\rho_I= 3.9\times 10^{14}$ g/cm$^3$, and $\rho_{II}\approx 4.1 \times 10^{14}$ g/cm$^3$.}

\subsection{The geometric component}\label{Sec:geom}

The function of the  gravitational redshift,  corresponding to our ansatz is  as follows:
\begin{equation}\label{eq:redshift}
   {\mathit  Z(r)=\frac{1}{\sqrt{-g_{tt}}}-1=\frac{1}{\sqrt{e^{s_0 (r/R)^2+s_1}}}-1}\,.
\end{equation}

We creat a figure of the pulsar's {\textit SAX J1748.9-2021}  redshift function  as reported in Fig. \ref{Fig:Mass1}\subref{fig:redshift}. In the GR case, i.e., $\zeta_1=0$, $Z$ at the core is approximately $Z(0)\approx 0.675$, and at the boundary $Z_{R}\approx 0.35$. If  $\zeta_1=- 0.05$, $Z$ at $r=0$ yields  $Z(0)\approx 0.646$ less than GR.  Importantly, this value remains below the maximum  limit $Z_s=2$, as discussed in \citep{Buchdahl:1959zz,Ivanov:2002xf,Barraco:2003jq,Boehmer:2006ye}.
 Likewise, for $\zeta_1= 0.05$, we observe that the upper limit of $Z$ at the center is $Z(0)\approx 0.69$ which remains below the maximum limit.

 It effectively conveys that the redshift distributions in the frame of $f({\cal R})$ satisfy stability limits in both cases discussed. It effectively communicates that the value of
$Z>0$ and has a finite value within the   star's interior, decreasing monotonically towards the boundary, and has a positive redshift and its first derivative is negative    as demonstrated in Fig.~\ref{Fig:Mass1}\subref{fig:redshift}.

\subsection{Limits on the mass radius relation  from  {\textit SAX J1748.9-2021} observations}\label{Sec:obs_const}

In the following analysis, we utilize the value of mass and the value of radius evaluated from observation of the stellar  {\it SAX J1748.9-2021}, which are reported as $M= 1.81 \pm 0.3 M_\odot$ and $R=11.7\pm 1.7$ km according to \citep{Ozel:2008kb}, to impose constraints on the parameter $\zeta_1$ within the framework of   $f({\textit R})$ theory.
\begin{figure*}
\centering
\subfigure[~Redshift]{\label{fig:redshift}\includegraphics[scale=0.38]{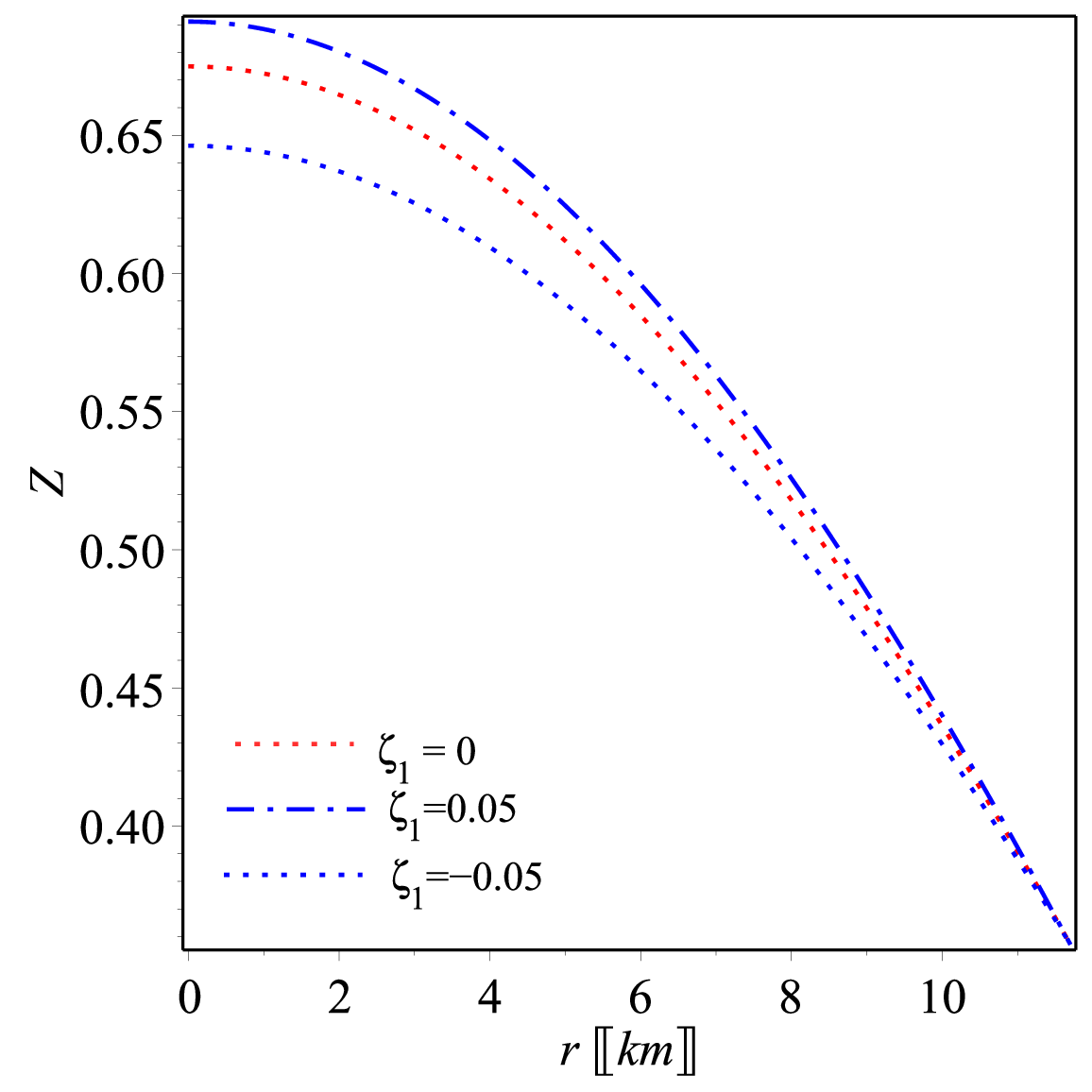}}\hspace{1cm}
\subfigure[~Mass function]{\label{Fig:Mass}\includegraphics[scale=0.38]{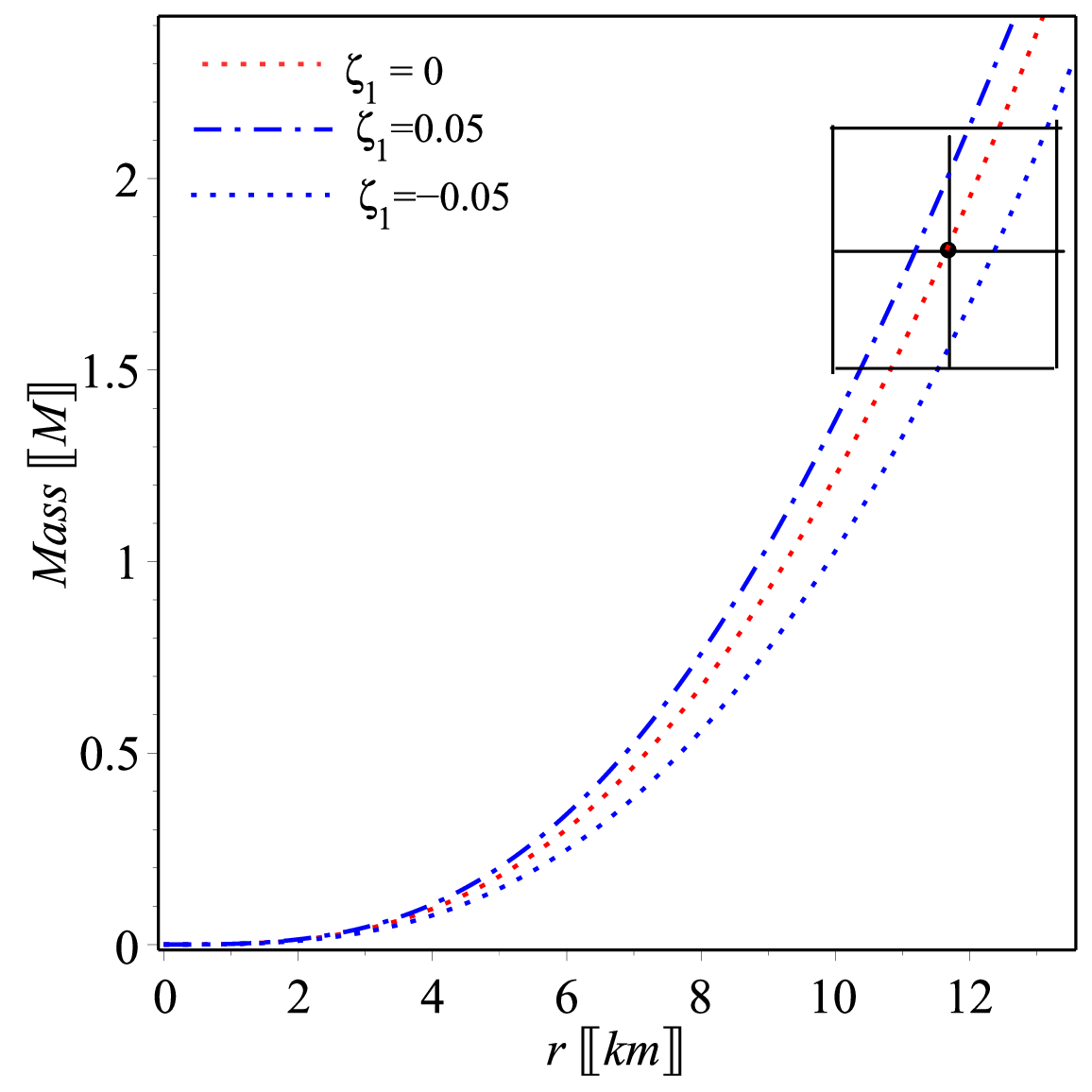}}
\caption{  In all the figures presented in this study, we have consistently employed identical numerical values assigned to the model parameters. \subref{fig:redshift}: This graph displays the redshift function \eqref{eq:redshift} for various cases, including $\zeta_1=0~ \pm 0.05$. In all scenarios, the high redshift is observed at the core with $Z(r=0)\approx 0.69$, gradually decreasing to approximately $Z_s\approx 0.35$ at  the boundary of the star.  \subref{Fig:Mass}: This figure illustrates the mass \eqref{Mf3} for the stellar SAX J1748.9-2021.  The  region denotes the observational limits on  ($M=1.85\pm 0.25 M_\odot$ and ${R}=12.35\pm0.11$ km) \citep{Legred:2021hdx}. For $\zeta_1=-0.05$, we employ the parameters  of the model, i.e., [$C=0.401$, $s_0 =0.39$, $s_1 =-0.997$, $s_2 =0.61$]. For $\zeta_1=0.05$, we use [$\zeta_1=0.05$, $C=0.514$, $s_0 =0.44$, $s_1 =-1.05$, $s_2 =0.61$]. For $\zeta_1=0$ GR, we use [$s_0 =0.42$, $s_1 =-1.03$, $s_2 =0.61$].  }
\label{Fig:Mass1}
\end{figure*}
Mass distribution, within the  radius of the stellar is given as:
 \begin{align}\label{Mf3}
     { m(r)} =  4\pi\int_{0}^{r} \rho(\eta) \eta^2 d\eta \,.
 \end{align}
Recalling the density profile presented in the Supplementary Material, for the   $f({\cal R})={\cal R}e^{\zeta {\cal R}}$, we show in the  Fig. \ref{Fig:Mass1}\subref{Fig:Mass} that:
\begin{itemize}
        \item When $\zeta_1=-0.05$,  $M=1.58 M_\odot$ for ${R} =12.45$ km and  $C=0.401$. This determines the constant parameters to be [$\zeta_1=-0.05$, $C=0.401$, $s_0 =0.39$, $s_1 =-0.997$, $s_2 =0.61$].
            \item For $\zeta_1=0.05$,   $M\approx 2.07 M_\odot$ for  ${R} =11.21$ km and  $C=0.514$. This determines the  constant parameters ro be [$\zeta_1=0.05$, $C=0.514$, $s_0 =0.44$, $s_1 =-1.05$, $s_2 =0.0.61$].
         \item For $\zeta_1=0$,  the constant parameters are [$s_0 =0.42$, $s_1 =-1.03$, $s_2 =0.61$].
\end{itemize}
This imposes limits on   $0\leq |\zeta| \leq 5$ km$^2$. Generally, as represented in Fig. \ref{Fig:Mass1}\subref{Fig:Mass}, the inclusion of $f({\cal R})$ gravity results in variations in the stellar mass.
In the analysis that follows, we'll utilize the specified numerical values to evaluate how robust current stellar is under different stability conditions.
\subsection{The Zeldovich condition}
An important criterion to confirm  the pulsar's stability  is provided by the study of Zeldovich \citep{1971reas.book.....Z}. According to this criterion, $p_r(r\to 0)$ will never be higher than the core energy density, which means that,
\begin{equation}\label{eq:Zel}
    {\frac{{p}_r(0)}{c^2{\rho}(0)}\leq 1\,.}
\end{equation}
Referring back to Eqs. ({\color{blue} 4}) as detailed in  the Supplementary Material, we can derive ${\rho}(r\to 0)$ and ${p_r}(r\to0)$  as:
\begin{align}
 & {c^2 {\rho}(r\to 0)} =-{\frac {3{e^{-6\zeta_1 \left(s_0 -s_2 \right) }} \left(156{\zeta_1}^{2}s_0{s_2}^{2}-s_2-2\zeta_1{s_0}^{2}-120{\zeta_1}^{2}{s_0}^{2}s_2-60{\zeta_1}^{2}{s_2}^{3}+24{\zeta_1}^{2}{s_0}^{3}+26\zeta_1 s_2 s_0-20\zeta_1{s_2}^{2} \right) }{{R}^{2}{c}^{2}{\kappa}^{2}}}\,, \nonumber \\
 &  {p_r(r\to0) = {p}_{t}(r\to 0)}\nonumber\\
 &={\frac { \left( {e^{\zeta_1s_2}} \right) ^{6} \left(2s_0 -s_2-240{\zeta_1}^{2}{s_0}^{2}s_2+312{\zeta_1}^{2}s_0{s_2}^{2}+46\zeta_1s_2s_0-28\zeta_1{s_2}^{2}-10\zeta_1{s_0}^{2}+48{\zeta_1}^{2}{s_0}^{3}-120{\zeta_1}^{2}{s_2}^{3} \right) }{{R}^{2}{\kappa}^{2} \left( { e^{\zeta_1 s_0}} \right) ^{6}}}.\quad
&\end{align}
Utilizing the previous numerical values for the pulsar {\textit SAX J1748.9-2021}, as discussed in Subsection \ref{Sec:obs_const}, we can assess the Zeldovich inequality \eqref{eq:Zel}. For $\zeta_1=0.05$, the inequality becomes $\frac{{p}_r(0)}{c^2{\rho}(0)}=0.136$, which is less than 1. Similarly, for $\zeta_1=-0.05$, the inequality reads $\frac{{p}_r(0)}{c^2{\rho}(0)}=0.133$, also less than 1. This ensures the validation of the condition of  Zeldovich.
\subsection{The conditions of energy}\label{Sec:Energy-conditions}

It is advantageous to express Eqs. \eqref{2} as:
\begin{equation}\label{eq:fR_MG}
    G_{\mu\nu}=\kappa\left(\mathbb{T}_{\mu\nu}+\mathbb{T}_{\mu\nu}^{geom}\right)=\kappa \bar{\mathbb{T}}_{\mu\nu}\,.
\end{equation}
In this context, we use the notation $G_{\mu\nu}:={\cal R}_{\mu\nu}-g_{\mu\nu}{\cal R}/2$ to represent the Einstein tensor, which accounts for the correction arising from  $f({\cal R})$ gravity \cite{DeFelice:2010aj, Capozziello:2011et} as:
\begin{equation}
  {\mathit  \mathbb{T}_{\mu\nu}^{geom}=\frac{1}{\kappa}\left(g_{\mu\nu}(f-{\cal R})/2+\nabla_\mu \nabla_\nu f_{\cal R}- g_{\mu\nu} \square f_{\cal R}+(1-f_{\cal R}){\cal R}_{\mu\nu}\right)\,.}
\end{equation}
The energy-momentum tensor can be written as $\mathit{ \bar{\mathbb{T}}_{\mu}{^\nu}=diag(-\bar{\rho} c^2, \bar{p}_r, \bar{p}_t, \bar{p}_t)}$.
\begin{figure*}
\centering
\subfigure[~NEC \& WEC (in the radial direction)]{\label{fig:Cond1}\includegraphics[scale=0.27]{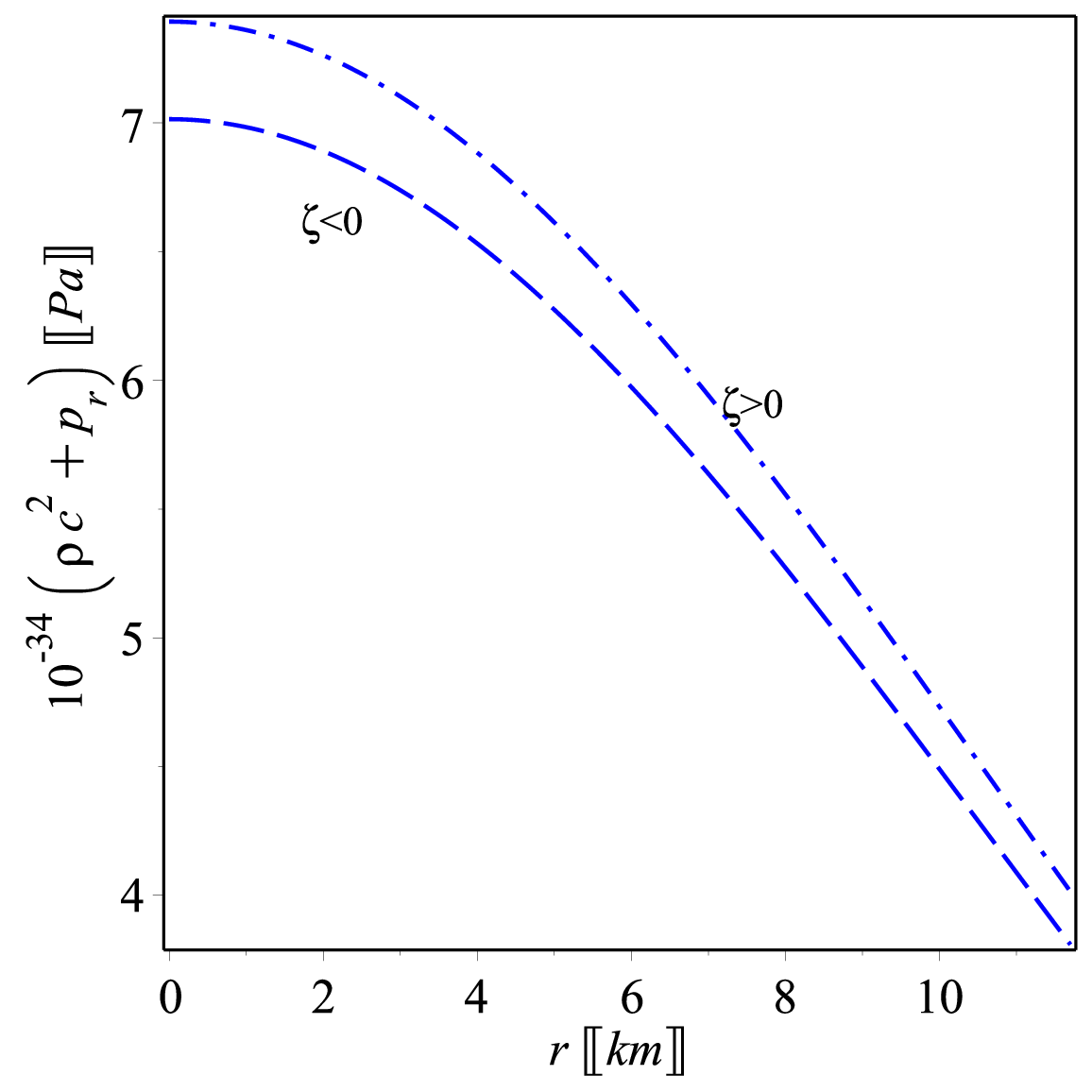}}\hspace{0.2cm}
\subfigure[~NEC \& WEC (in the tangential  direction)]{\label{fig:Cond2}\includegraphics[scale=0.27]{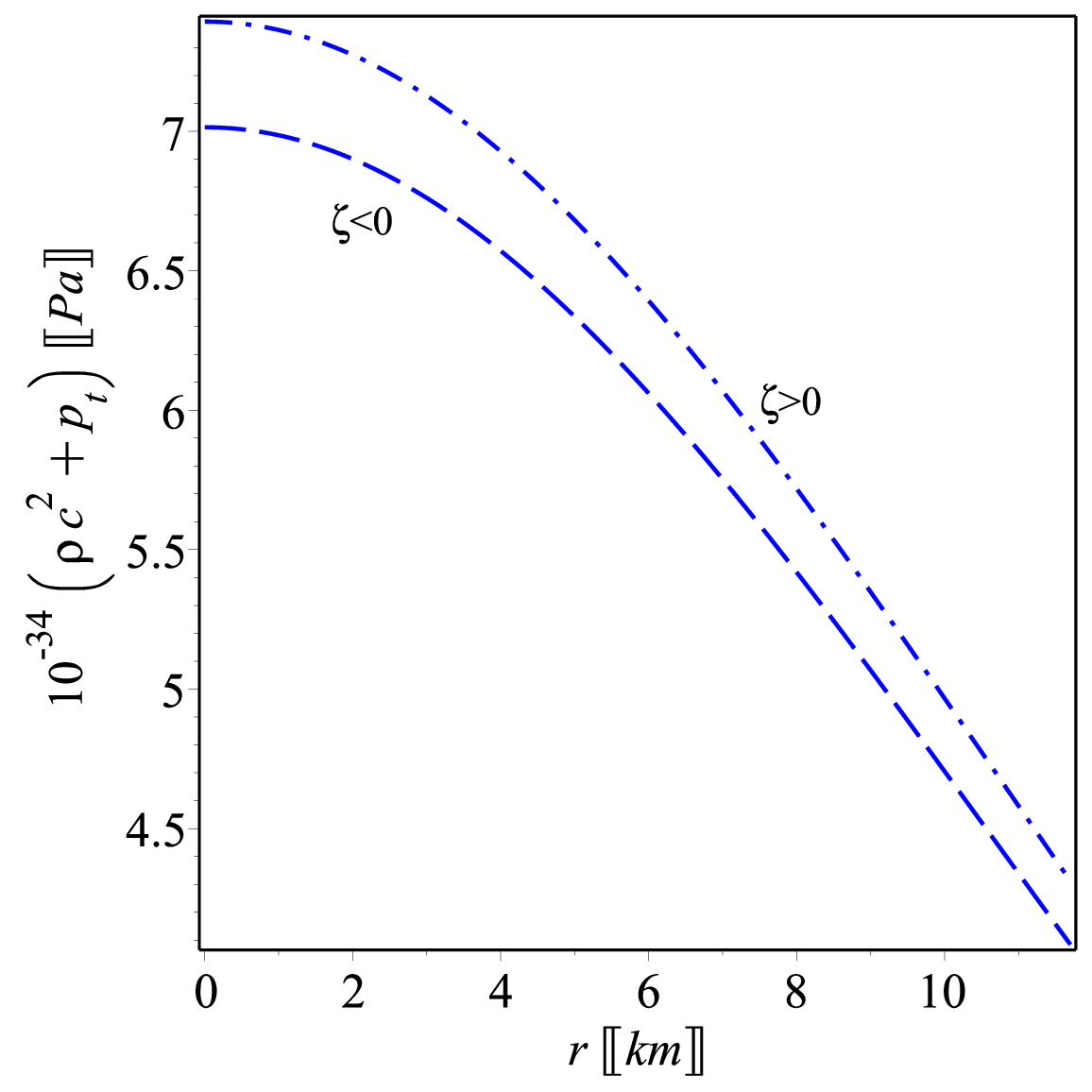}}\hspace{0.2cm}
\subfigure[~The SEC]{\label{fig:Cond3}\includegraphics[scale=.27]{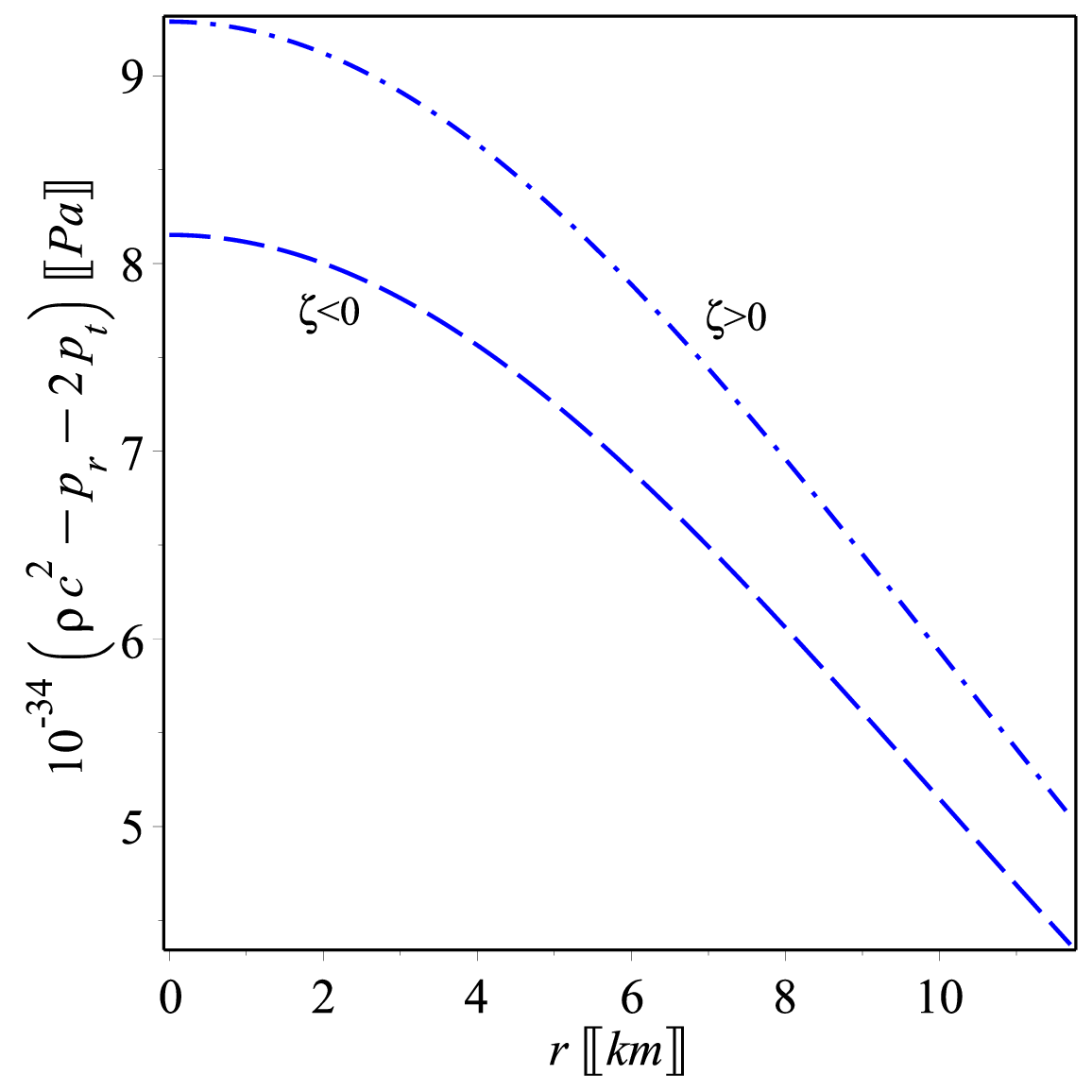}}\\
\subfigure[~The DEC (in the radial direction)]{\label{fig:DEC}\includegraphics[scale=.27]{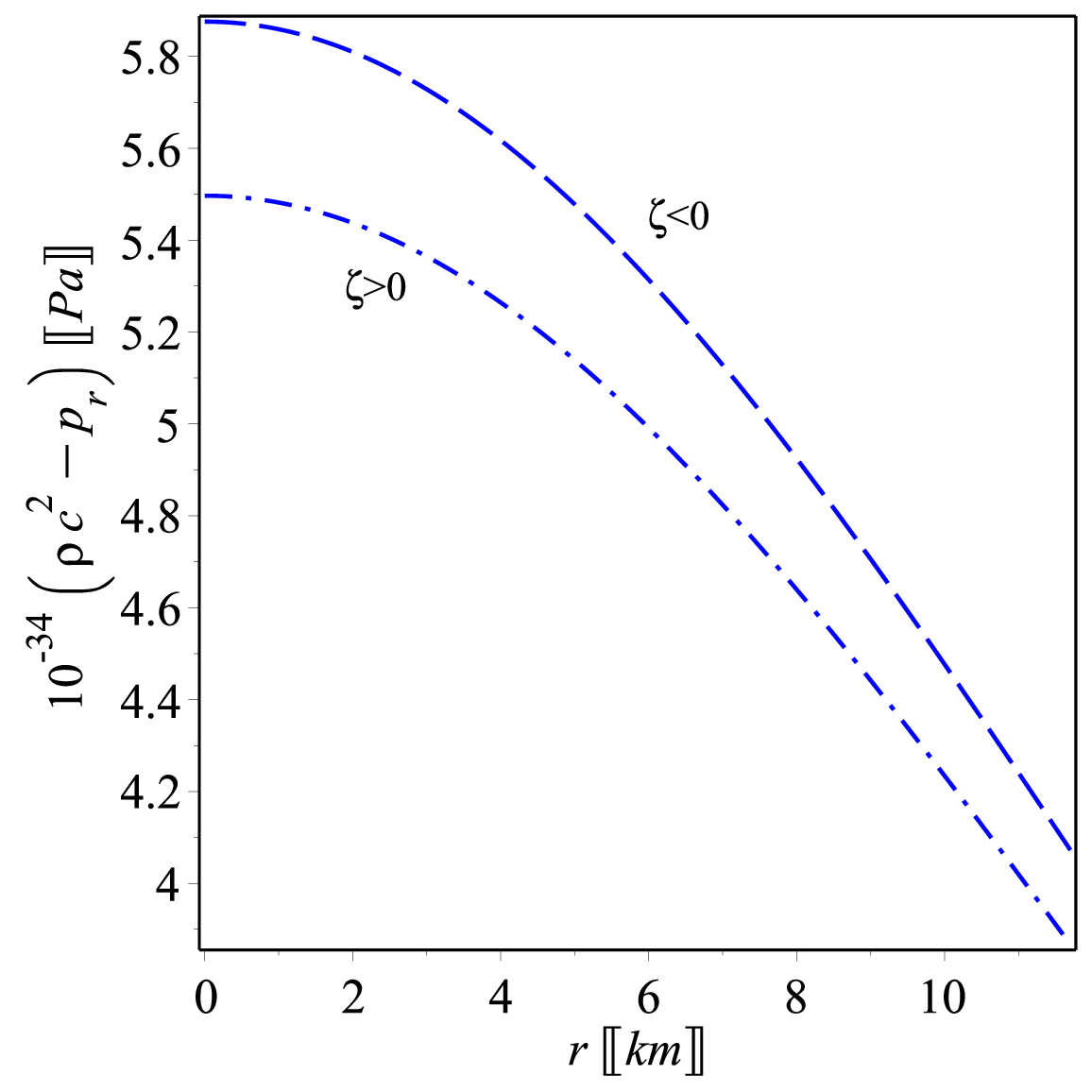}}\hspace{0.2cm}
\subfigure[~The DEC (in the tangential  direction)]{\label{fig:DEC}\includegraphics[scale=.27]{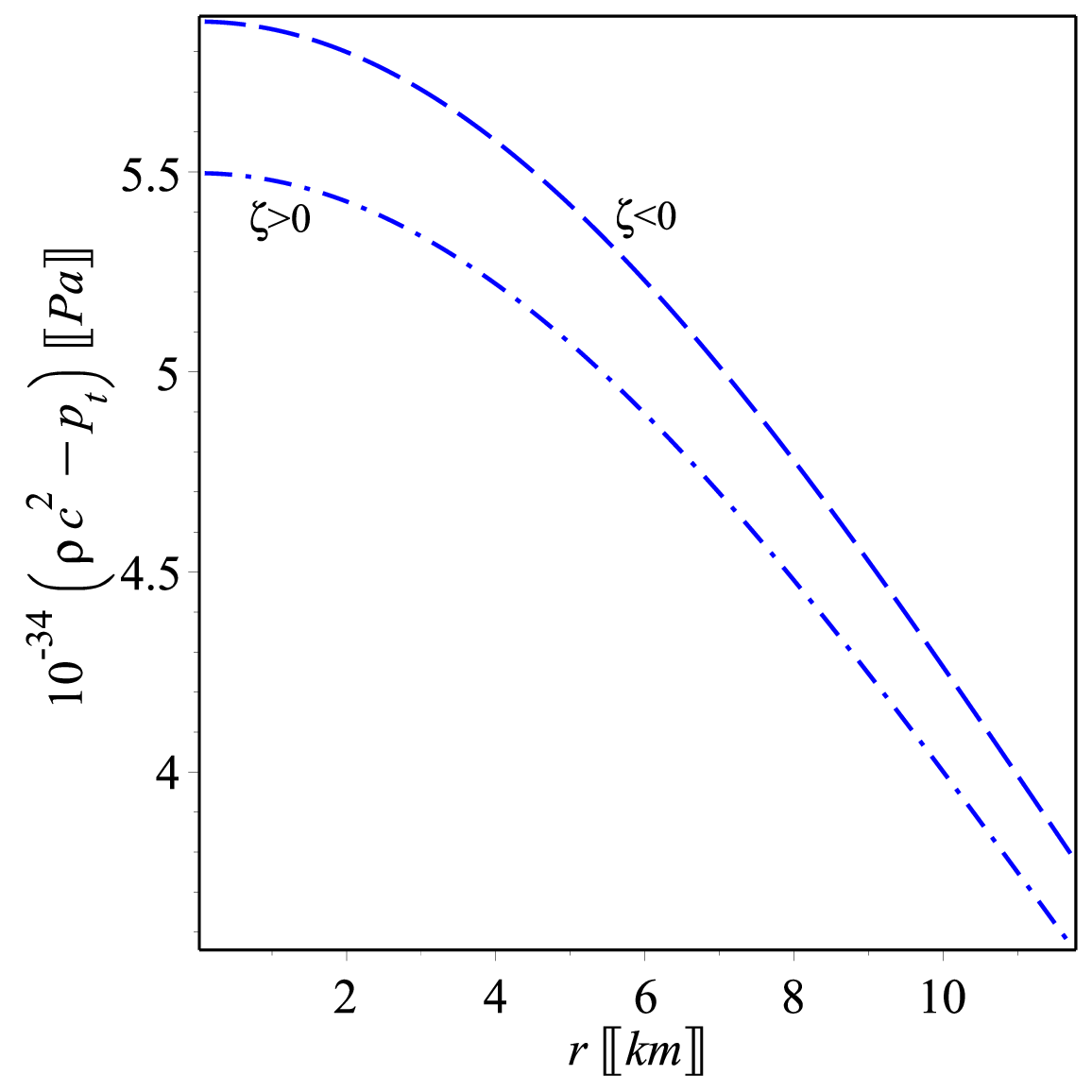}}
\caption{The reported graphs show that the effective matter $\bar{\mathbb{T}}_{\mu\nu}$ satisfies all the energy conditions, as elaborated in Subsection \ref{Sec:Energy-conditions}, for the pulsar model SAX J1748.9-2021. The cases where $\zeta>0$ and $\zeta<0$ correspond to $\zeta_1=0.05$ and $\zeta_1=-0.05$, respectively.}
\label{Fig:EC}
\end{figure*}

The focusing theorem and the Raychaudhuri equation imply the following inequalities for the tidal tensor:   ${\textit R}_{a b} u^{a} u^{b} \geq 0$ and ${\textit R}_{a b} n^{a} n^{b} \geq 0$, where $n^{\mu}$ signifies an arbitrary future-directed null vector and $u^{\mu}$ represents an arbitrary timelike vector.  It is worth noticing that, within $f({\cal R})$  theory, the ${\cal R}$ could be expressed as ${\cal R}_{\mu\nu}=\kappa\left(\bar{\mathbb{T}}_{\mu\nu}-\frac{1}{2} g_{\mu\nu} \bar{\mathbb{T}}\right)$, as can be inferred from Eq. \eqref{eq:fR_MG}. With this consideration in mind, the energy conditions (ECs) can be extended to $f({\textit R})$ gravity in the following way \cite{Capozziello:2014bqa}:
\begin{itemize}
  \item[1.]  $ \bar{\rho} c^2+ \bar{p}_r > 0$, $\bar{\rho} c^2+\bar{p}_t > 0$, and $\bar{\rho}\geq 0$, which corresponds to  the Weak Energy Condition (WEC).
  \item[2.] $\bar{\rho} c^2+  \bar{p}_t \geq 0$, and  $\bar{\rho} c^2+ \bar{p}_r \geq 0$, which corresponds to the Null Energy Condition (NEC).
  \item[3.]  $\bar{\rho} c^2+\bar{p}_r \geq 0$,  $\bar{\rho} c^2+\bar{p}_t \geq 0$,  $\bar{\rho} c^2+\bar{p}_r+2\bar{p}_t \geq 0$, which corresponds to the Strong Energy Condition (SEC).
  \item[4.]  $\bar{\rho} c^2-{ |\bar{p}_r| }\geq 0$, $\bar{\rho} c^2- |\bar{p}_t| \geq 0$,  $\bar{\rho}\geq 0$, which corresponds to the Dominant Energy Conditions (DEC).\\
\end{itemize}
Figure \ref{Fig:EC} provides a visual representation of the ECs when $\zeta_1 \gtrless 0$. These figures confirm that the current model of pulsar {\textit SAX J1748.9-2021} verifies all the conditions listed above.
\subsection{Causality conditions}\label{Sec:causality}

\begin{figure*}
\centering
\subfigure[~$v_r{}^2$]{\label{fig:vr}\includegraphics[scale=0.28]{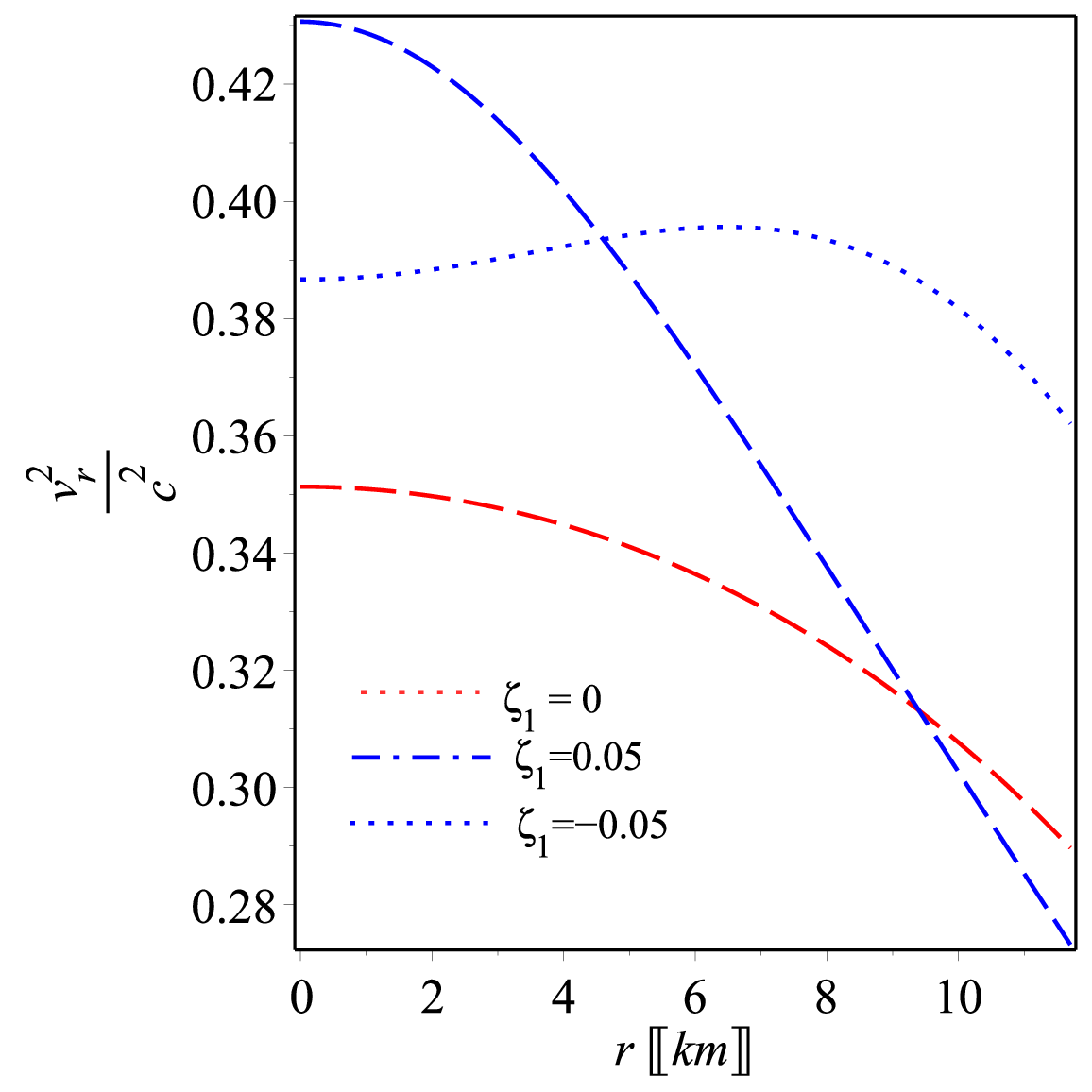}}\hspace{0.2cm}
\subfigure[~$v_t{}^2$]{\label{fig:vt}\includegraphics[scale=.28]{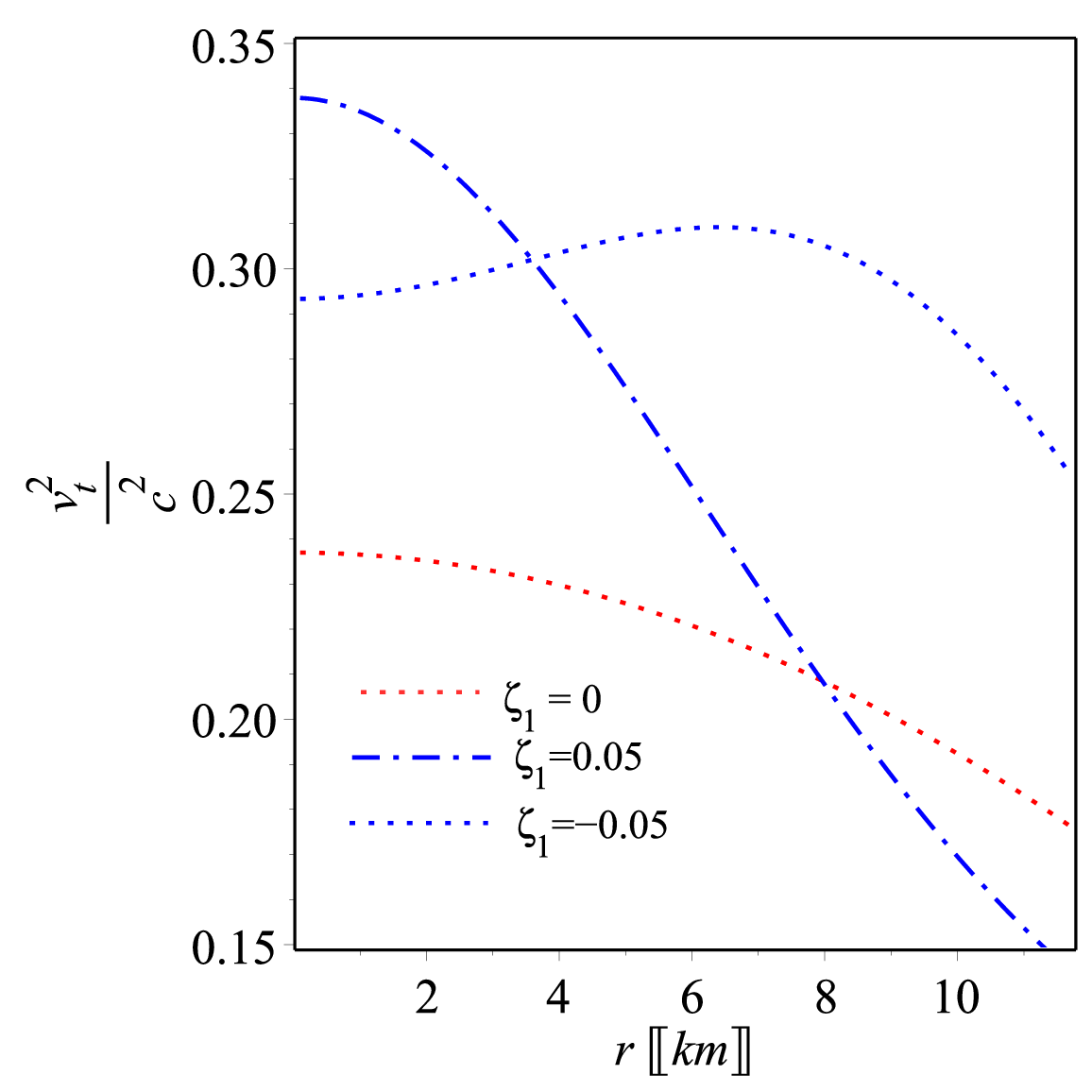}}\hspace{0.2cm}
\subfigure[~Stability in the presence of strong anisotropy]{\label{fig:vt-vr}\includegraphics[scale=.28]{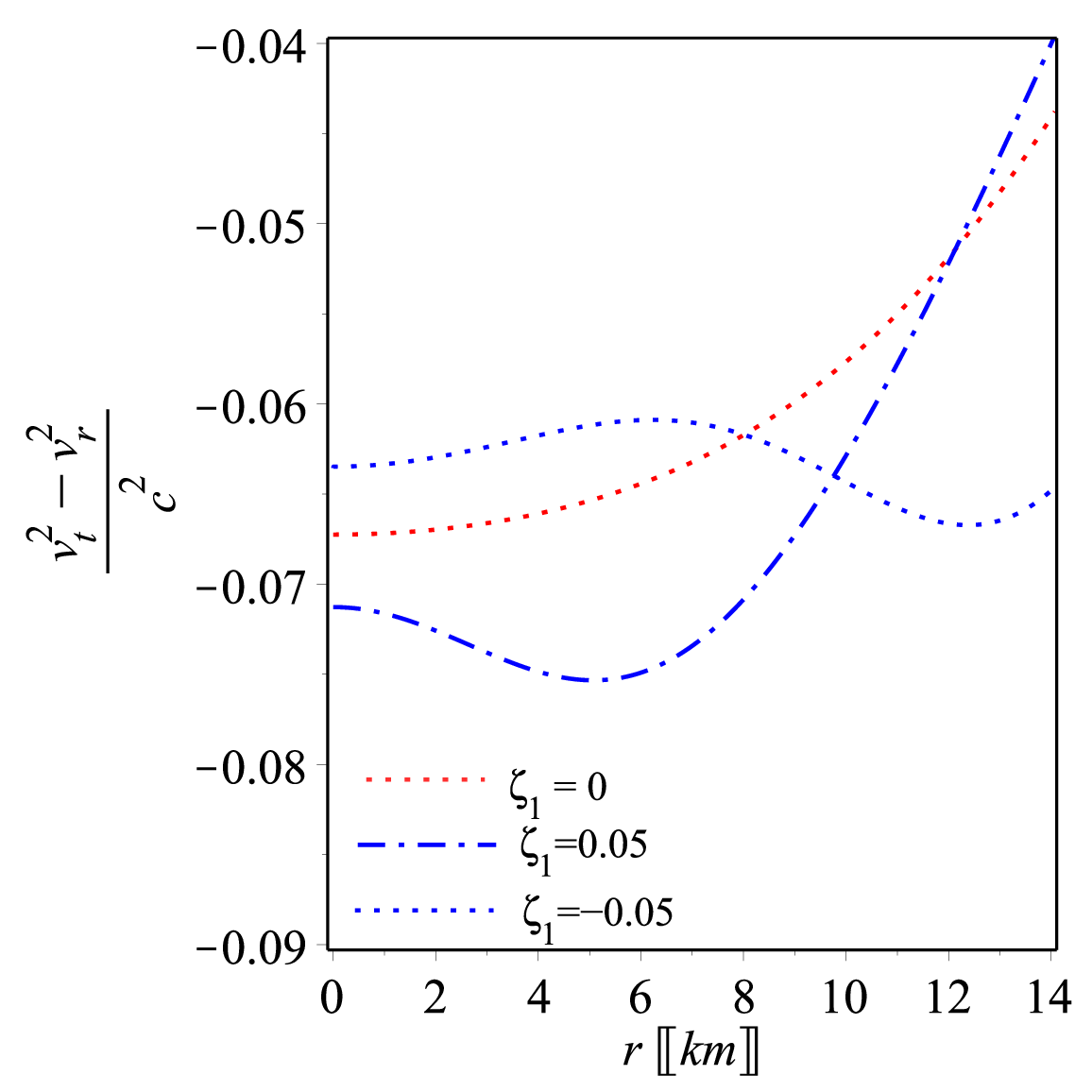}}
\caption{The sound speed   of stellar {\textit SAX J1748.9-2021} for $\zeta_1=0,~ \pm 0.05$ is illustrated in the following way: \subref{fig:vr} and \subref{fig:vt} describe $v_r{}^2$ and $v_t{}^2$, respectively, as defined by \eqref{eq:sound_speed}. \subref{fig:vt-vr} demonstrates that the model adheres to the stability requirement $(v_t^2-v_r^2)/c^2 < 0$ for the strong anisotropic case.}
\label{Fig:Stability}
\end{figure*}

The condition of causality is consider as one of the fundamental aspects which states that the sound speed should not exceed  the light speed. Referring to Eqs. \eqref{eq:KB_EoS2}, $p_r$ and $p_t$ can be fixed as:
\begin{equation}\label{eq:sound_speed}
  v_r^2 =  \frac{ d{ p}_r}{d { \rho}}=  \frac{p'_r}{{ \rho'}}, \quad
  v_t^2 = \frac{d{  p}_t}{d{   \rho}}= \frac{p'_t}{{ \rho'}}\,.
\end{equation}
Utilizing Eqs. ({\color{blue} 4}) in the Supplementary Material, we can determine the gradients of density and pressure components, as provided by Eqs. ({\color{blue} 7}) in the Supplementary Material.
We represent  $v_r$ and $v_t$ of pulsar SAX J1748.9-2021 when $\zeta_1\gtrless 0$, as in Figs. \ref{Fig:Stability}\subref{fig:vr} and \subref{fig:vt}. These figures show $0\leq {v_r^2}/c^2\leq 1$ and $0\leq {v_t^2/c^2} \leq 1$, satisfying the  causality and stability conditions. Moreover, Fig. \ref{Fig:Stability}\subref{fig:vt-vr},  illustrates that $-1< (v_t^2-v_r^2)/c^2 < 0$ throughout the interior of pulsar SAX J1748.9-2021 \citep{Herrera:1992lwz}.

It is important to notice  that the sound speed exhibits variations with $r$, as reported in Figs. \ref{Fig:Stability}\subref{fig:vr} and \subref{fig:vt}. Specifically, when $\zeta_1=0.05$, we observe $0.27< v_r^2/c^2<0.44$ and $0.15<v_t^2/c^2<0.34$. Conversely, for $\zeta_1=-0.05$, the intervals are $0.36< v_r^2/c^2<0.39$ and $0.25<v_t^2/c^2<0.30$. The maximum  limits  correspond to speed sound at $r=0$,  align perfectly  the values derived in  \ref{Sec:matt} from the induced EoSs \eqref{eq:KB_EoS2}, specifically ($\zeta_1=0.05, v_r^2\approx 0.43c^2, v_t^2\approx 0.337c^2$) and ($\zeta_1=-0.05, v_r^2\approx 0.385c^2, v_t^2\approx 0.29c^2$).
\subsection{The adiabatic and the equilibrium of hydrodynamic forces}\label{Sec:TOV}
In Newtonian gravity, it's widely accepted that there is no maximum limit on the mass of a stable configuration when the adiabatic index  for a specific  EoS increases $4/3$. Conversely, within Newtonian gravity, it is necessary for a stable configuration that $\gamma < 4/3$. Nevertheless, it has been shown that a star can withstand radial perturbations in a fully relativistic anisotropic neutron star model, even when $\gamma > 4/3$. To account for this, we define the adiabatic index \citep{Chandrasekhar:1964zz,chan1993dynamical} as follows:
\begin{equation}\label{eq:adiabatic}
{\gamma}=\frac{4}{3}\left(1+\frac{{ \sigma}}{r|{  p}'_r|}\right)_{max},\qquad \qquad
{\Gamma_r}=\frac{{\rho c^2}+{p_r}}{{p_r}}{v_r^2}, \qquad \qquad
{\Gamma_t}=\frac{{\rho c^2}+{p_t}}{{p_t}}{v_t^2 .}
\end{equation}
Obviously, in the case of isotropy ($\sigma=0$), we obtain $\gamma=4/3$. In the mildly anisotropic case ($\sigma<0$), which is similar to the Newtonian theory, we have $\gamma<4/3$, consistent with the standard stability requirement. On the other hand, when strong anisotropy is involved ($\sigma>0$), similar to what is considered in this study, we find $\gamma>4/3$ \cite{chan1993dynamical,1975A&A....38...51H}.

\begin{figure}
\centering
\subfigure[~$\Gamma$]{\label{fig:gamar1}\includegraphics[scale=0.28]{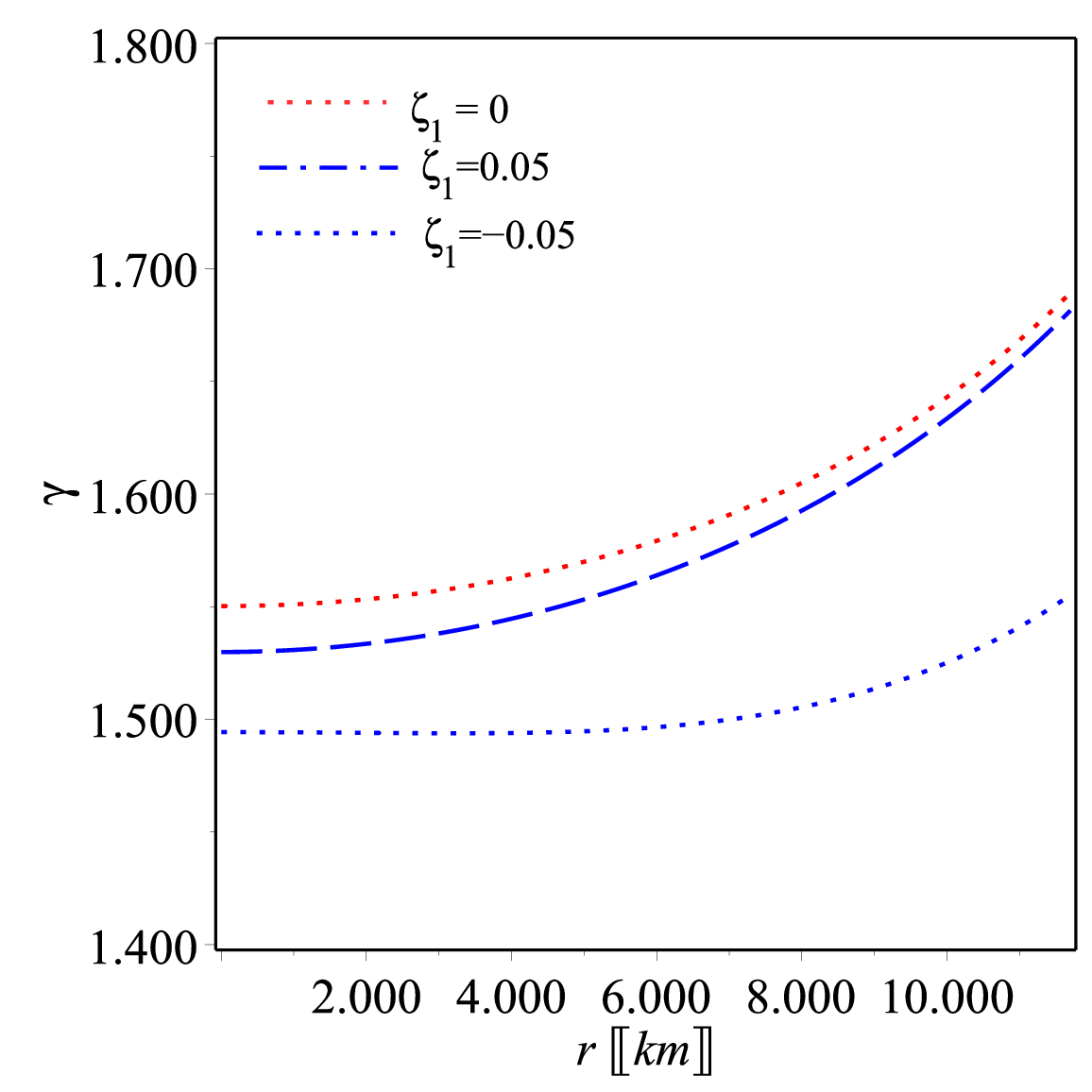}}\hspace{0.2cm}
\subfigure[~$\Gamma_r$]{\label{fig:gamar}\includegraphics[scale=0.28]{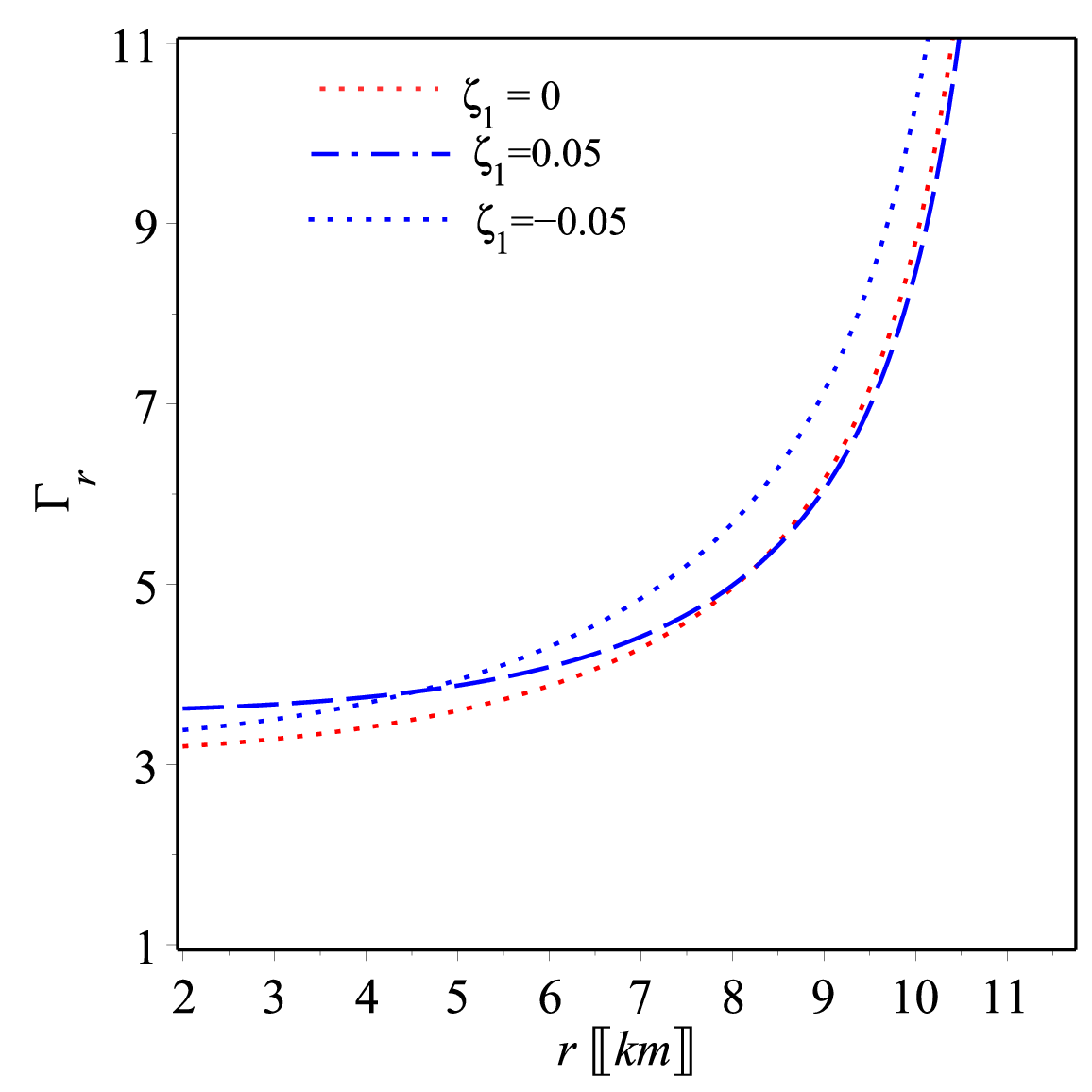}}\hspace{0.2cm}
\subfigure[~$\Gamma_t$]{\label{fig:gamar}\includegraphics[scale=0.28]{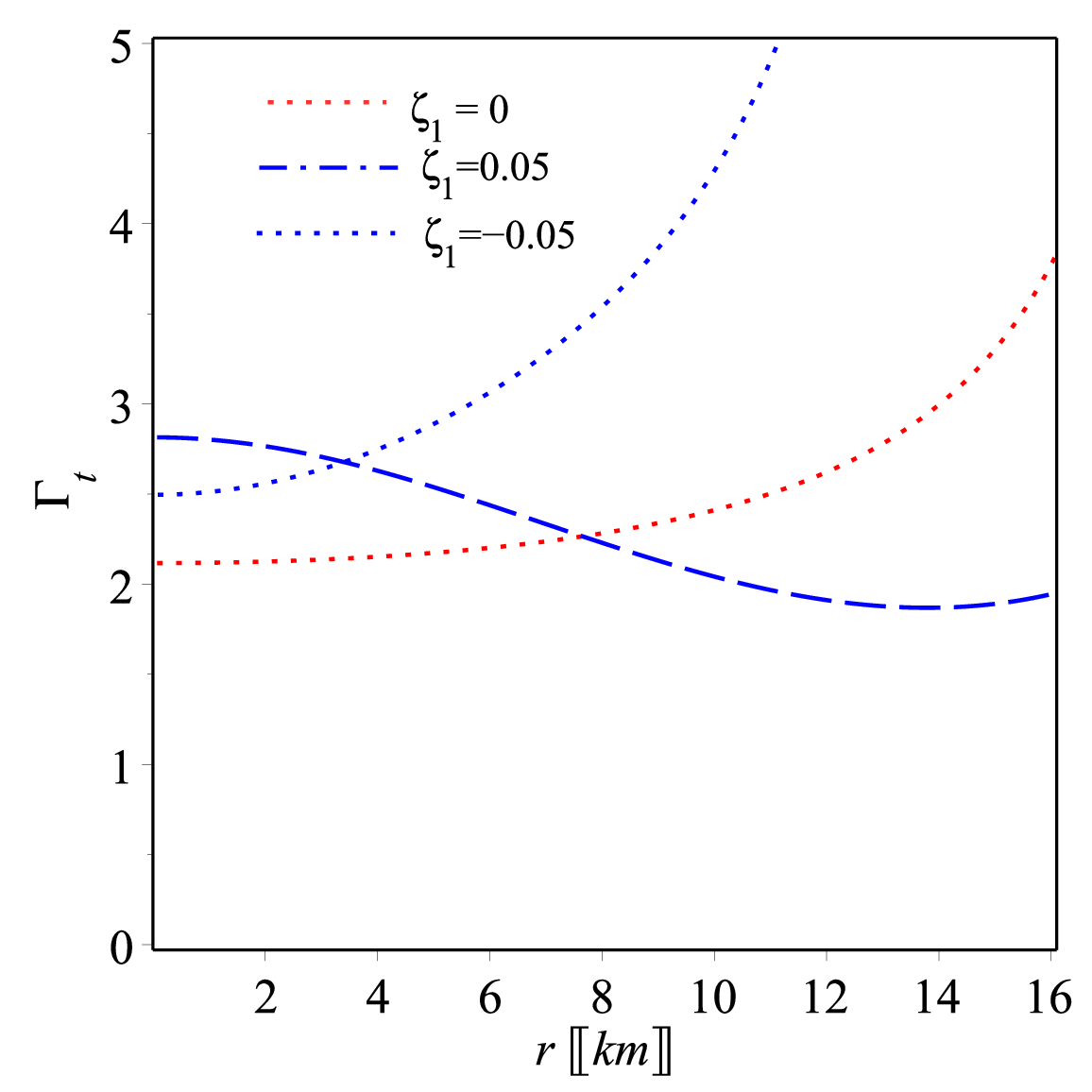}}
\caption{The  indices of  adiabatic,  provided by \eqref{eq:adiabatic}, for  {\textit SAX J1748.9-2021}. These figures ensure that throughout the interior of the pulsar that $\gamma$ exceeds 4/3 and both $\Gamma_r$ and $\Gamma_t$ surpass $\gamma$.}
\label{Fig:Adiab}
\end{figure}

By utilizing the field equations and Eqs.  ({\color{blue} 4}),  and the gradients given by Eq.  ({\color{blue} 7}) presented in the Supplementary Material, we demonstrate that our $f(R)$ gravity offers a stable anisotropic model for the pulsar {\textit SAX J1748.9-2021} when  $\zeta_1\gtrless 0$, as illustrated in Figure \ref{Fig:Adiab}."

\begin{figure}
\centering
\subfigure[~The TOV  of the GR case ($\zeta=0$)]{\label{fig:GRTOV}\includegraphics[scale=0.28]{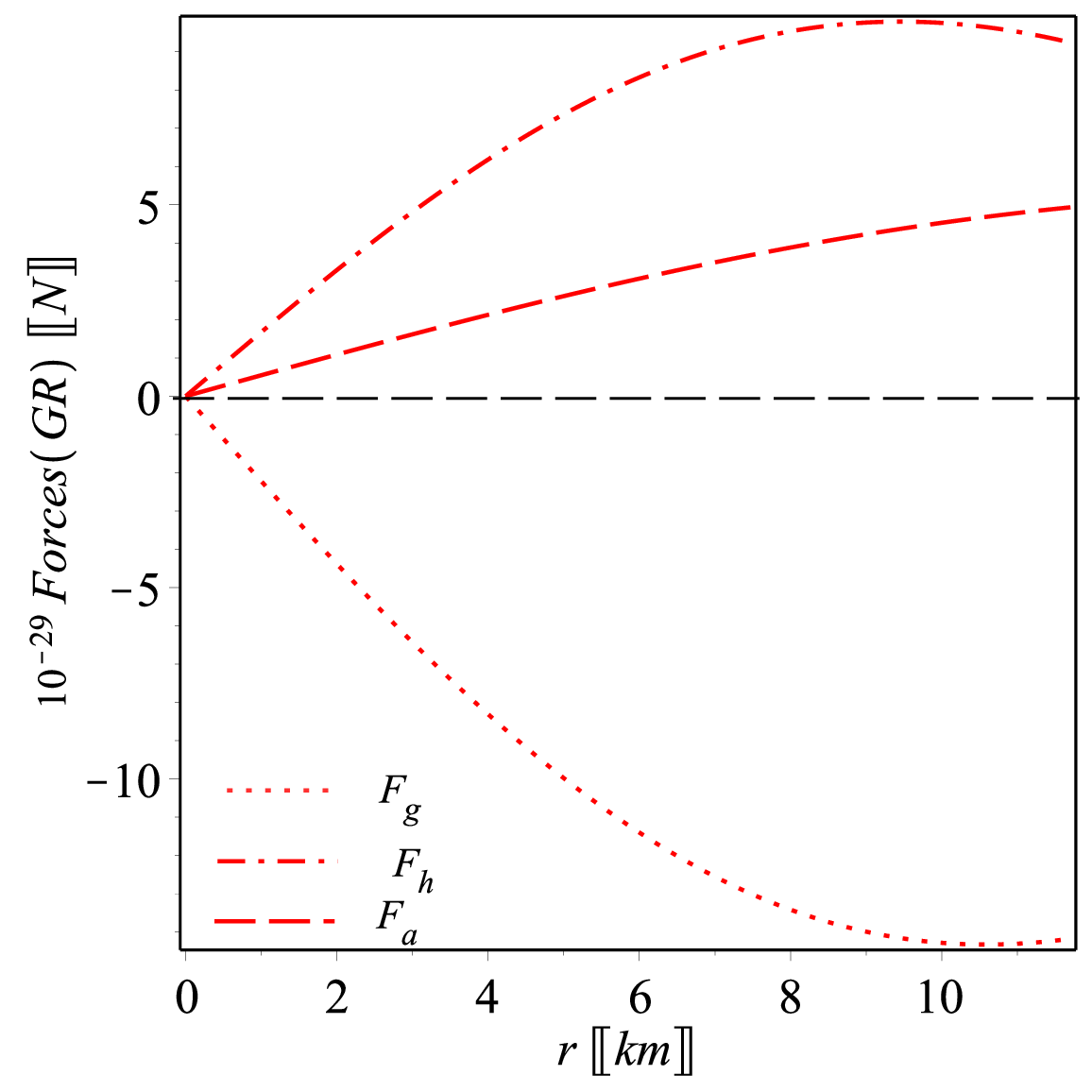}}\hspace{0.2cm}
\subfigure[~The TOV of $f({\cal R})$: $\zeta>0$]{\label{fig:FRp}\includegraphics[scale=0.28]{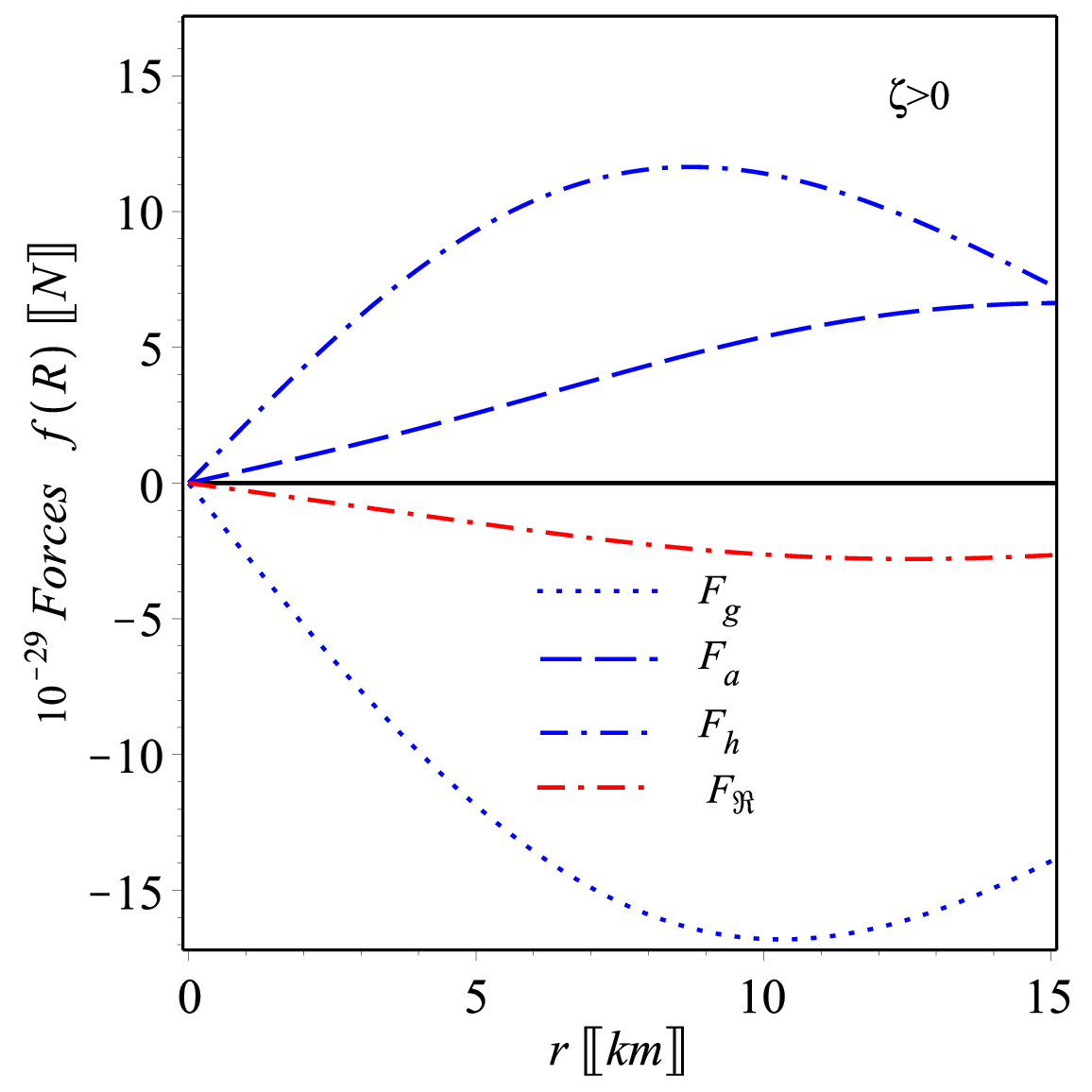}}\hspace{0.2cm}
\subfigure[~The TOV of $f({\cal R})$: $\zeta<0$)]{\label{fig:FRn}\includegraphics[scale=0.28]{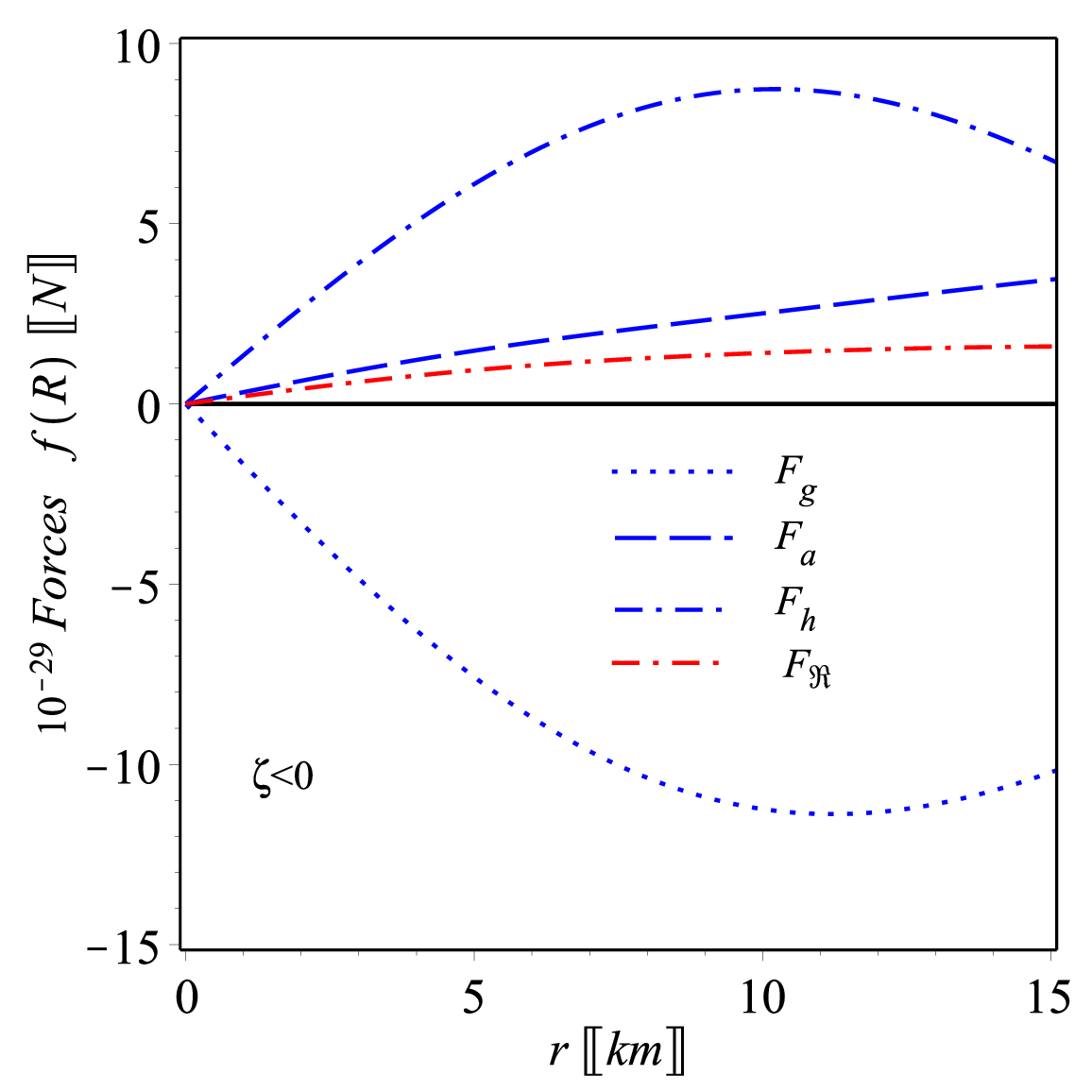}}
\caption{The Tolman-Oppenheimer-Volkoff constraint, as described in Eq.\eqref{eq:RS_TOV}, imposes various forces within the pulsar {\textit SAX J1748.9-2021}. These forces are represented for $\zeta_1=0,\pm 0.05$. In the case of $\zeta_1=0.05$, the $f({\cal R})$ the gravitational collapse force is further enhanced by correction, which adds an additional negative force. Conversely, for $\zeta_1=-0.05$, $f({\cal R})$ theory provides an extra positive force that lessens the force of gravitational collapse.}
\label{Fig:TOV}
\end{figure}

Next, we examine the hydrodynamic equilibrium of the current model using the Tolman-Oppenheimer-Volkoff (TOV) equation, which is defined as follows:
\begin{equation}\label{eq:RS_TOV}
{\mathit F_a}+{\mathit F_g}+{\mathit F_h}+{\mathit F_{\cal R}=0}\,.
\end{equation}
Here ${\mathrm F_g}$, ${\mathrm F_h}$, and ${\mathrm F_a}$ represent the typical  gravitational,  hydrostatic, and anisotropic forces, respectively,  in additional to the force of $F_{\textit R}$ originating from the $f({\textit R})$ component. These forces  defined as:
\begin{eqnarray}\label{eq:Forces}
  {\mathrm F_a} =&\frac{ 2{\mathit  \Delta}}{\mathit r} ,\qquad
  {\mathit F_g} = -\frac{{\mathit  M_g}}{r}({\mathit  \rho c^2}+{\mathit p_r})e^{\varepsilon/2} ,\qquad\nonumber\\
  {  \mathit F_h} =&-{\mathit  p'_t} ,\qquad
  {\mathit F_{\cal R}} = \zeta_1({  c^2 \rho}'-{ p}'_r-2{  p}'_{t})\,.
\end{eqnarray}
Within the formula for the gravitational force, denoted as $F_g$, we have introduced the quantity $\varepsilon := \mu - \nu$, in addition to   $M_g$ which denotes the gravitational mass of a system that is isolated within the 3-space volume ${\mathit V}$ (at constant time $t$). This can be described by the Tolman mass formula within the framework of $f({\cal R})$ gravity. gravity, as described by Tolman in his 1930 work \citep{1930PhRv...35..896T}.
\begin{eqnarray}\label{eq:grav_mass}
{\mathit M_g(r)}&=&{\int_{\mathit V}}\Big(\mathbb{{\bar{\mathfrak T}}}{^r}{_r}+\mathbb{\bar{\mathfrak T}}{^\theta}{_\theta}+\bar{\mathfrak{T}}{^\phi}{_\phi}-\bar{\mathfrak{T}}{^t}{_t}\Big)\sqrt{-g}\,dV\nonumber\\
&=&e^{-\mu}(e^{\mu/2})'  e^{\nu/2} r =\frac{1}{2} r \mu' e^{-\varepsilon/2}\,.
\end{eqnarray}
Consequently, the gravitational force can be expressed as ${\mathbb F_g} = -\frac{s_0 r}{R^2}({\mathit \rho c^2}+{ p_r})$. Utilizing the field equations ({\color{blue} 4}) together with the gradients ({\color{blue} 7}) in the Supplementary Material, we can demonstrate that  $f({\cal R})$ gravity  complies with \eqref{eq:RS_TOV}, giving a stable model for the pulsar {\textit SAX J1748.9-2021} when  $\zeta_1 \gtrless 0$, including  ($\zeta_1=0$), as shown  in Fig. \ref{Fig:TOV}.

Note that ($\sigma>0$) introduces  positive extra force which counters the field of gravity. This is an important to enlarge the  star's size. As evident Figs. \ref{Fig:TOV}\subref{fig:FRp} and \ref{Fig:TOV}\subref{fig:FRn} demonstrate such phenomena with  the additional force creating from $f({\cal R})$ theory when $\zeta \gtrless 0$. This investigation  aligns with the previously obtained results for the star {\textit SAX J1748.9-2021} in Subsection \ref{Sec:obs_const}, which yielded ($\zeta_1=0.05, M=2.070 M_\odot, {R} =11.21~\text{km}, C=0.514$) and ($\zeta_1=-0.05, M=1.58 M_\odot, {R} =12.45~\text{km}, C=0.401$).
\section{ Mass-radius relation and equation of state}\label{Sec:EoS_MR}

\begin{figure}[th!]
\centering
\subfigure[~EoS in the radial direction]{\label{fig:RfEoSp}\includegraphics[scale=0.45]{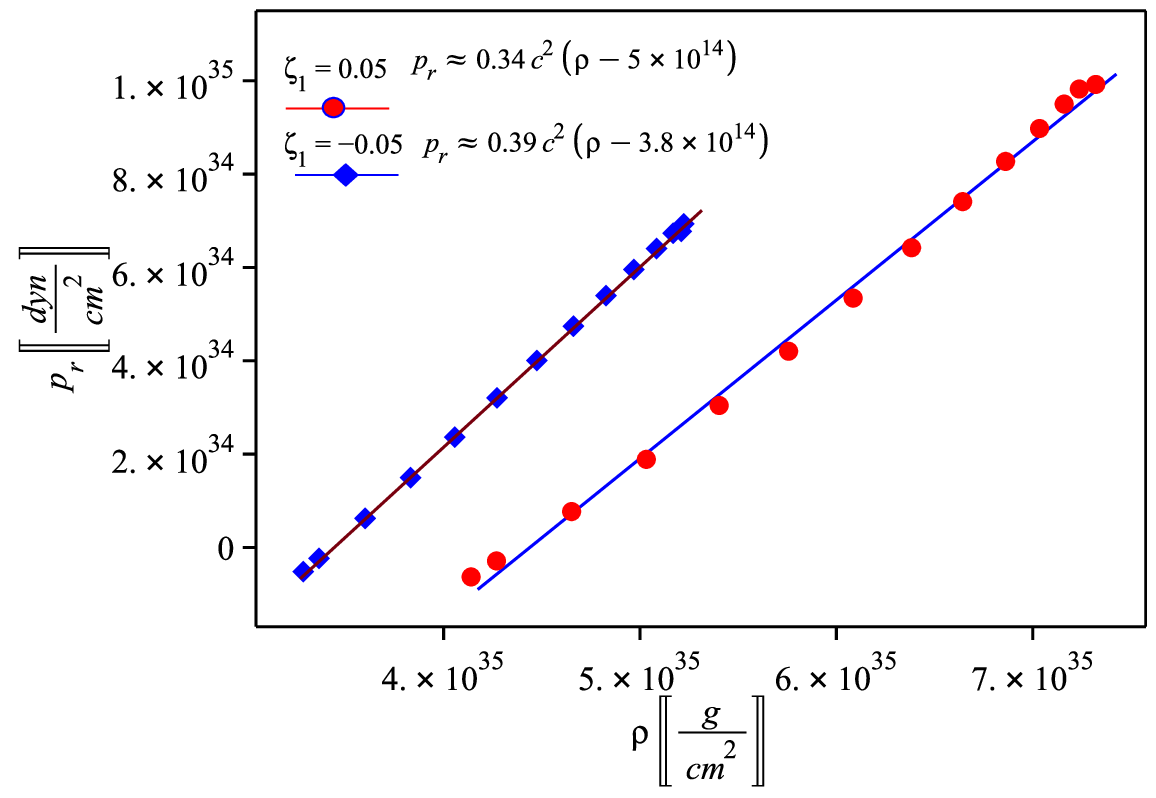}}
\subfigure[~EoSin the tangential direction]{\label{fig:TEoSn}\includegraphics[scale=0.45]{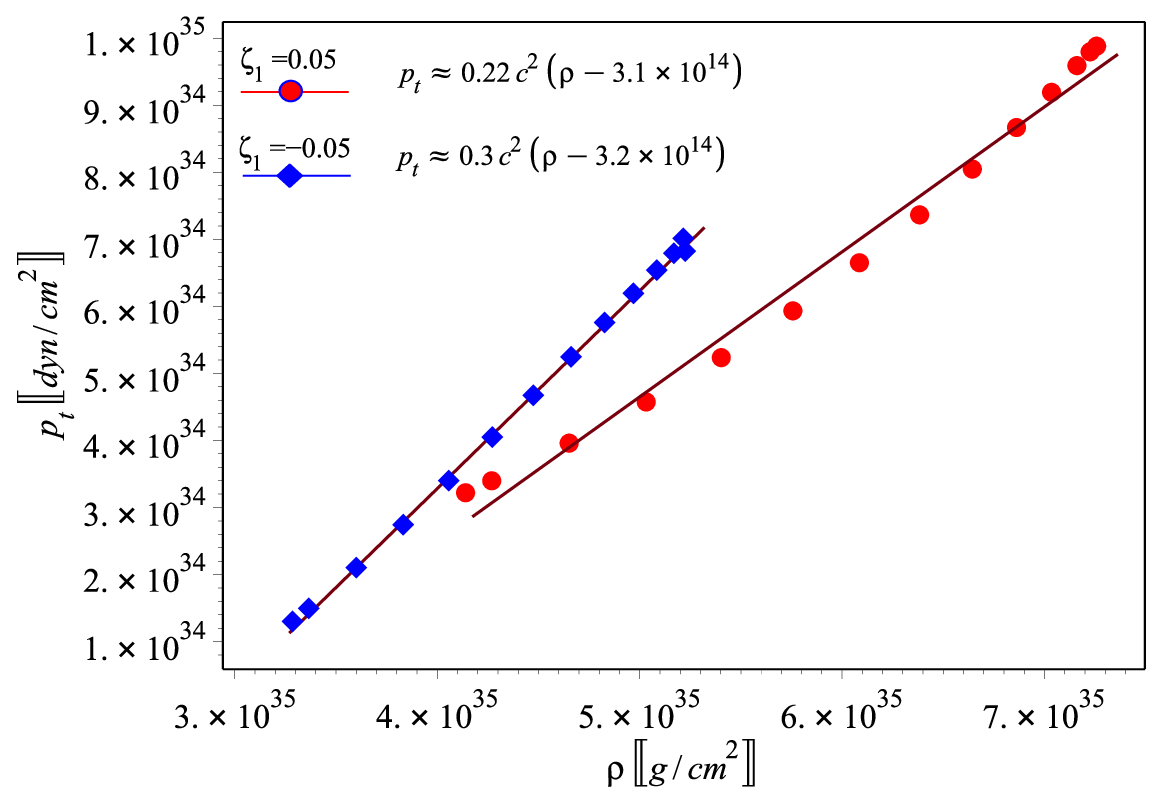}}
\caption{The optimal-fitting  EoSs  for the pulsar \textit{SAX J1748.9-2021} are reported as follows: \subref{fig:RfEoSp}: We  create a set of points for $\rho$ and $p_r$ using Eqs.~({\color{blue} 4}),   presented in the Supplementary Material for $\zeta=\pm 0.05$, where such points exhibit good alignment with a linear equation of state (EoS) pattern \subref{fig:TEoSn}: Similarly, the data points also demonstrate a significant correlation with a linear model when tangential equations of state (EoS) with $\zeta=\pm 0.05$ are used. The correctness  of these relations throughout the star's interior is confirmed by the linear EOSs best-fit, which agree with the equations derived before, specifically \eqref{eq:KB_EoS2}. It is worth noticing that, for the case  $\zeta_1=0.05$, a slight deviation from the linear pattern can be observed, suggesting that  $f({\cal R})$ might present good  fit, i.e., $p_{(rad.,tang.)}(\rho)\approx \tilde{a}_0+\tilde{a}_1\rho+\tilde{a}_2\rho^2$ in this particular instance.}
\label{Fig:EoS}
\end{figure}

 It is worth noticing that our study refrains from imposing specific EoS assumptions; rather, we use the ansatz of KB  as outlined in Eqs. \eqref{eq:KB}.  Nevertheless, the resulting EoSs in Eqs.\eqref{eq:KB_EoS2} show how pressures and density are related in this ansatz, and they are mainly valid at the core because of the power series assumptions that underlie it. By creating a range of density and pressure values that span from the core to the surface and take taking into account the model parameter's $\zeta_1\gtrless 0$, we confirm the accuracy of these equations. $\zeta_1$. We construct these sequences, as shown in Fig.\ref{Fig:EoS}., using the numerical values given in Section  \ref{Sec:obs_const}  for the pulsar \textit{SAX J1748.9-2021}   and the  equation of motions of $f(\cal R)$ theory of gravity, which are specifically described in Eqs. ({\color{blue} 4}), given in the Supplementary Material.
%


It is noteworthy that the linear best fit, in the both cases of positive/negative of $\zeta_1$, induces slight alterations in the surface density and sound speeds. 
 Consequently, we suggest that a higher polynomial, namely ${\mathit p_{r,t}(\rho)\approx \tilde{c}_0+\tilde{c}_1\rho+\tilde{c}_2\rho^2}$, could offer a more accurate fit, yielding non-linear equations of state.
\begin{figure*}[t]
\subfigure[~C]{\label{fig:Comp}\includegraphics[scale=0.4]{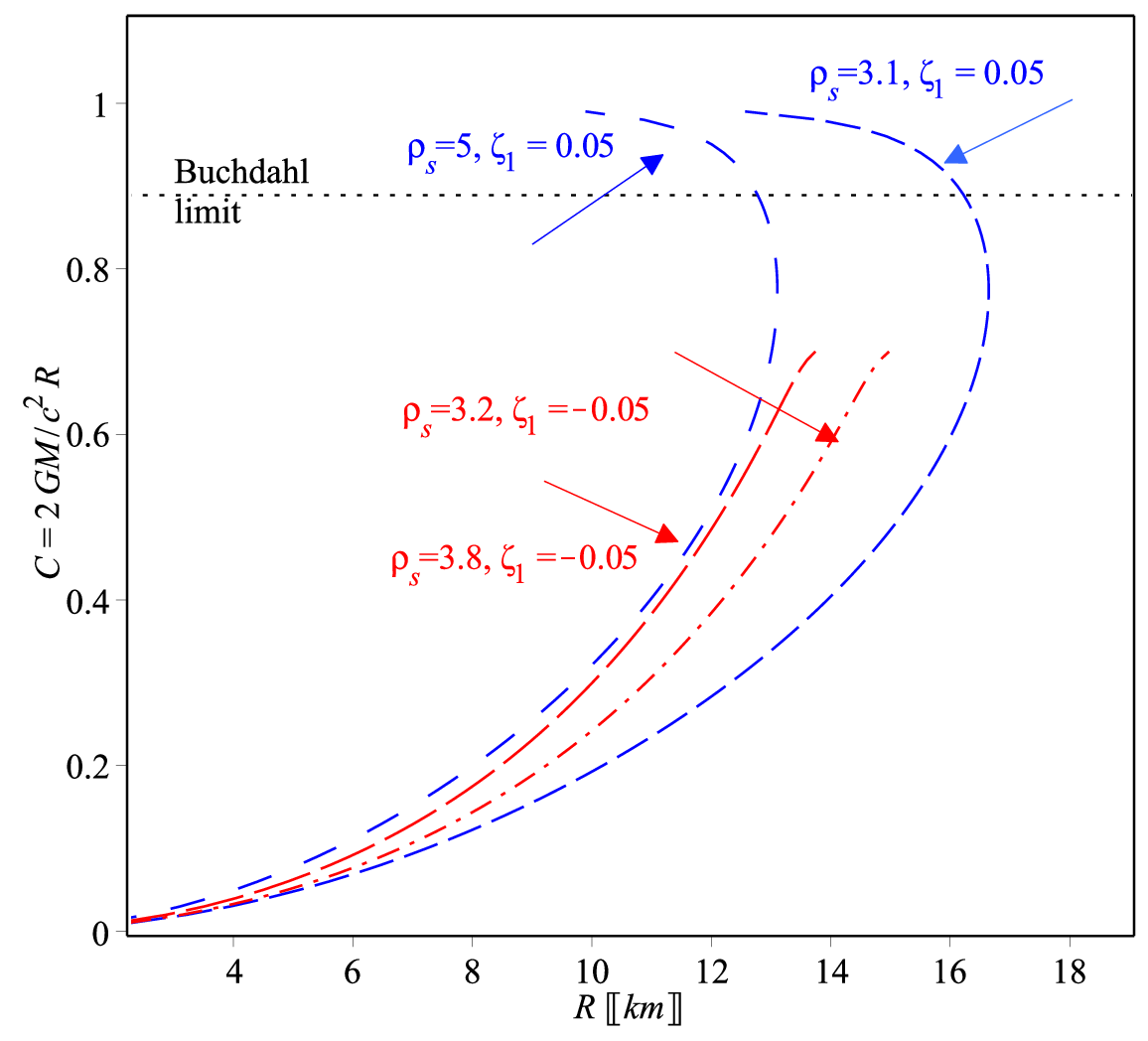}}\hspace{0.5cm}
\subfigure[~MR]{\label{fig:MR}\includegraphics[scale=.4]{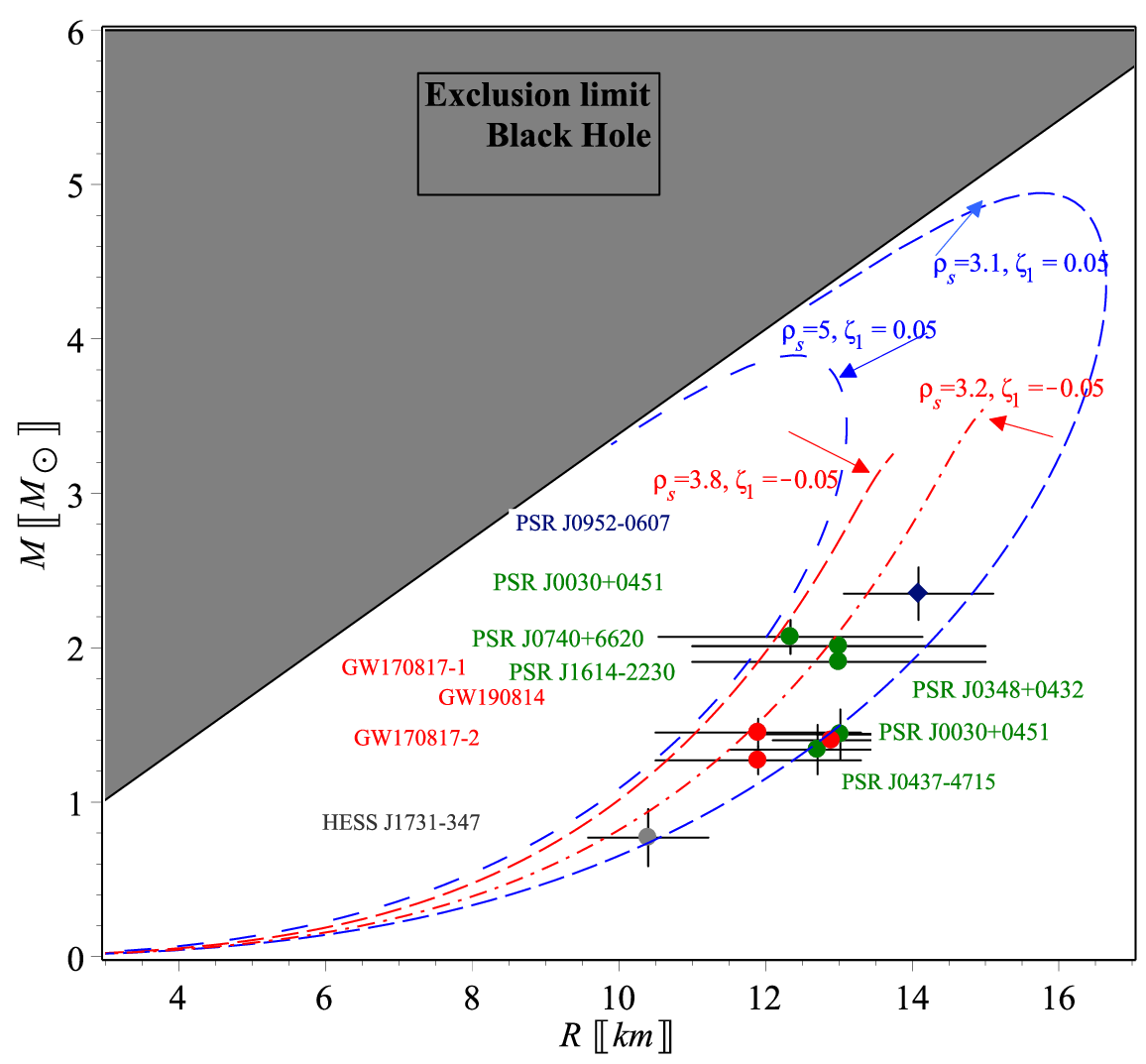}}
\caption{\subref{fig:Comp} Compactness-radius (CR) chart:
 The compactness of the Buchdahl limit is represented by the dotted horizontal line, where ($C=8/9$). We graph the CR  according to (EoSs) that fit best as outlined in Fig. \ref{Fig:EoS}. Additionally, we represent CR relation for distinct surface densities, specifically $\rho_I=3.1\times 10^{14}$ g/cm$^3$, $\rho_I=3.2\times 10^{14}$ g/cm$^3$,  $\rho_I=5\times 10^{14}$ g/cm$^3$, and $\rho_I=3.8\times 10^{14}$ g/cm$^3$. It is worth noticing that, for $\zeta_1=0.05$, $C$ can exceed the Buchdahl upper bound, a feature that is similar to the GR mode \citep{Roupas:2020mvs}. Interestingly, for $\zeta_1=-0.05$, the  higher value of $C$ falls below the Buchdahl maximum limit, setting  the difference between $f({\cal R})$  gravity model and GR in such a study. \subref{fig:MR} MR chart: When $\zeta\gtrless0$, the MR curves cannot extend into non-physical branches.}
\label{Fig:CompMR}
\end{figure*}

As stated in \cite{PhysRev.116.1027}, Buchdahl constructed a critical constraint on stable stellar configurations that places an upper limit on the $C$, i.e.,  $C<8/9$. Notably, such boundary was initially developed for spherically symmetric isotropic (or slightly anisotropic) solutions inside GR.However, subsequent investigations have revealed that such constraints could be violated by an abundance of at least a single or many of these presumptions. With regard to stronger anisotropic models that are more realistic,  the compactness can approach the limit of black hole, i.e., $C\to 1$, also within the framework of GR, as demonstrated in \cite{Alho:2022bki}. intriguingly, other limits impose more strict limits  on $C$ in cases featuring strong anisotropy, as indicated in \cite{Alho:2021sli,Roupas:2020mvs,Raposo:2018rjn,Cardoso:2019rvt}. A similar conclusion has been reached when exploring scenarios involving nonminimal coupling between  matter and geometry \citep{Nashed:2022zyi}.

Within this context, it is worthwhile to look into this constraint in the context of $f({\cal R})$ of gravity, as this study does. For generalized $f({\cal R})$ theory of gravity, the Buchdahl limit is expressed as detailed in \cite{Goswami:2015dma}.
\begin{align}
    C=\frac{2GM}{c^2 R}< \frac{4\frac{f_{\cal R}(r=R)}{f_{\cal R}(r=0)}\left[1+\frac{f_{\cal R}(r=R)}{f_{\cal R}(r=0)}\right]}{\left[1+2\frac{f_{\cal R}(r=R)}{f_{\cal R}(r=0)}\right]^2}.\nonumber
\end{align}
In this study where,   $f({\cal R})={\cal R}e^{\zeta {\cal R}}$,  for the compactness, we express the modified Buchdahl upper limit as
\begin{equation}\label{comp1}
    C < {{4(1+\zeta {\cal R}|_{r=R})e^{\zeta {\cal R}|_{r=R}}\left[(1+\zeta {\cal R}|_{r0})e^{\zeta R|_{r=0}}+(1+\zeta {\cal R}|_{r=R})e^{\zeta {\cal R}|_{r=R}}\right]} \over {\left[(1+\zeta {\cal R}|_{r0})e^{\zeta R|_{r=0}}+2(1+\zeta {\cal R}|_{r=R})e^{\zeta {\cal R}|_{r=R}}\right]^2}}.
\end{equation}
We repeat that $\zeta=\zeta_1 R^2$ where $R$ being the radius of the NS $11.7$ km. Clearly  Eq.(\ref{comp1}) reduces to Buchdahl's  GR if $\zeta=0$ \cite{PhysRev.116.1027}.

The Buchdahl limit for the pulsar {\textit SAX J1748.9-2021} can be computed as follows: We find $C\lesssim 0.891$ for $\zeta_1=0.05$, and $C\lesssim 0.895$ for $\zeta_1=-0.05$, which different from GR.

We provide the corresponding MR relation  in Fig. \ref{Fig:CompMR}\subref{fig:MR} when   $\zeta_1\gtrless 0$,where the matching condition \eqref{eq:bo} determines $M$; that is, $M= \frac{c^2 R}{2 G} (1-e^{-s_2})$. This refers to the optimal  EoS  that was previously derived in this section.  Consequently, we use $\zeta_1=0.05$ and the boundary density $\rho_I=5\times 10^{14}$ g/cm$^{3}$. This results in a maximal mass $M\approx 3.93 M_\odot$  with  $R\approx 12.7$ km. With $\zeta_1=-0.05$ and a boundary density of $\rho_I=3.8\times 10^{14}$ g/cm$^{3}$, the maximum mass $M=3.275 M_\odot$ is obtained at a radius of $R=13.7$ km.  Furthermore, we employ alternative boundary density values that yield a better fit with other pulsars while remaining consistent with the pulsar {\textit SAX J1748.9-2021}, as shown in Fig. \ref{Fig:CompMR}\subref{fig:MR}. Based on $\zeta_1=0.05$ and the boundary density $\rho_I=3.1\times 10^{14}$ g/cm$^{3}$, the maximum mass $M=5.01 M_\odot$ is found at radius $R=16.05$ km. With $\zeta_1=-0.05$ and $\rho_I=3.2\times 10^{14}$ g/cm$^{3}$, the maximum mass at radius $R=15.02$ km is $M=3.565 M_\odot$. The MR curves clearly do not extend to the limit of black hole  for positive $\zeta_1$ cases, as depicted by the grey region in Fig. \ref{Fig:CompMR}\subref{fig:MR}.

\section{Discussion and Conclusions}\label{Sec:Conclusion}
Several captivating and noteworthy discoveries emerge as soon as  we delve into the realm of NSs within the framework of $f({\cal R})$  theory of gravity. In our exploration, we placed our primary emphasis on the particular $f({\cal R})$ model put forth as $f({\cal R})={\cal R}e^{\zeta {\cal R}}$, and  scrutinized its ramifications on the structural integrity and stability characteristics of neutron stars. To summarize our results:
\begin{itemize}

\item  We investigated how  $f(\cal{R})$ gravity (with $f(\cal{R})\neq {\cal R}$) models alters the conventional understanding of neutron stars as described by GR. Examining the $f({\cal R})={\cal R}e^{\zeta {\cal R}}$ function, we discovered that such adjustments can wield a substantial influence on the composition and dynamics of neutron stars. Our examination delved with  the more practical setting of an anisotropic fluid behavior, assuming that the KB ansatz is followed by the inner region of the  spherically symmetric stellar model. Concretely, we employed precise measurements derived from the radius and mass  of star {\textit SAX J1748.9-2021}, which boasts a mass of $2.07 \pm 0.11 M_\odot$ and a radius within the range of $12.34^{+1.89}_{-1.67}$ km. These measurements served as valuable constraints on the parameter space of our current model, specifically influencing the value of $\zeta$.

\item One of our central concerns revolved around assessing the stability of neutron star configurations within the context of $f({\cal R})$ gravity. We conducted an exhaustive analysis of various stability criteria, including the Zeldovich condition, energy conditions, and the speed of sound. Our results showed that  $f({\cal R})$  theory of gravity can verify these stipulations for certain parameter ranges. As we investigated, the allowable  zone of $f({\cal R})$ theory parameter, $\zeta$   is given  by $|\zeta_1| < 0.05$. Different  criteria of stability applied to our model from the viewpoint of geometry and matter  confirmed its validity.  Regarding the central density of pulsar {\textit SAX J1748.9-2021}, our predictions are: When $\zeta_1=0.05$, ${\rho_\text{core}}\approx 7.58\times 10^{14}$ g/cm$^{3}$, that is  2.8 times the nuclear density ($\rho_\text{nuc}$). Conversely, for $\zeta_1=-0.05$, ${\rho_\text{core}}\approx 6.78\times 10^{14}$ g/cm$^{3}$, that is 2.5 times the nuclear density.

\item In the frame of $f({\cal R})$ theory, we derived the MR curve for NSs. Our results confirmed that NS mass and radius might be  differ from the predictions of GR relying on how the form of $f({\cal R})$.  Such differences  have critical consequences on the astrophysical observations and could be employ to calculate  the validity of  ${\cal R}e^{\zeta {\cal R}}$.
More specific, we calculate the maximal mass  $M=4.0 M_\odot$ corresponding to a radius of $R=12.6$ km when $\zeta_1=0.05$ and a surface density of $\rho_I=4.4\times 10^{14}$ g/cm$^{3}$. On the other hand, we derive the maximal  mass of $M=4.1 M_\odot$ with a corresponding radius  $R=14.4$ km when $\zeta_1=-0.05$ and a surface density  $\rho_I=4.0\times 10^{14}$ g/cm$^{3}$.  Such findings are consistent with the properties of pulsar {\textit SAX J1748.9-2021} \cite{Nashed:2023pxd}.

\item We discussed the anisotropic properties  and causality criteria  of  NSs in the frame of $f({\cal R})$. Our discoveries revealed that, even in the presence of $f({\cal R})$ modifications, the speed of sound remains consistent with causality constraints.
      We observed that the combined effects of strong anisotropy and $f({\cal R})$ gravity, particularly when $\zeta$ is negative, act to counteract gravitational collapse and successfully lower the pulsar fluid's sound speed.  Such results of  maximal $v_r$, is  $v_r^2\approx 0.35c^2$, at the center of the neutron star. This finding aligns more closely with the soft EoSs predicted by gravitational wave observations, indicating a better agreement with astrophysical reality.

     \item{ In this study we show that the effect of anisotropy and within KB ansatz the pulsar could have exceed mass than that of GR. This case is similar to the isotropic case presented in \cite{Astashenok:2021btj} where the authors show that the maximum baryonic mass can exceed  gravitational mass for for static NSs in the frame of   $f({\cal R})$.}
\end{itemize}

In conclusion, our investigation implies that $f({\cal R})$  gravity offers a fresh and captivating perspective for comprehending neutron stars. Although this realm of study continues to evolve, our results suggest that the $f({\cal R})$ gravity frameworks can align with both theoretical stability criteria and empirical observations. The future exploration of this domain holds the promise of fine-tuning and scrutinizing these models more rigorously, potentially unveiling new revelations about the fundamental characteristics of gravity and the dynamics of compact objects such as neutron stars.

\section*{Acknowledgements}
SC acknowledges the Istituto Nazionale di Fisica Nucleare (INFN) Sez. di Napoli,  Iniziative Specifiche QGSKY and MOONLIGHT2 for the support.
This paper is based upon work from COST Action CA21136 {\it Addressing observational tensions in cosmology with systematics and fundamental physics} (CosmoVerse) supported by COST (European Cooperation in Science and Technology).


\end{document}



\title{Constraining $f({\cal R})$ gravity by  Pulsar {\textit SAX J1748.9-2021} observations: \\Supplementary material }

\author{Gamal~G.~L.~Nashed}
\email{nashed@bue.edu.eg}
\affiliation {Centre for Theoretical Physics, The British University, P.O. Box
43, El Sherouk City, Cairo 11837, Egypt.}
\author{Salvatore Capozziello}
\email{capozziello@na.infn.it}
\affiliation {Dipartimento di Fisica E. Pancini, Universit`a di Napoli Federico II\\Complesso Universitario di Monte Sant Angelo, Edificio G, Via Cinthia, I-80126, Napoli, Italy,}
\affiliation {Scuola Superiore Meridionale, Largo S. Marcellino 10, I-80138, Napoli, Italy,}
\affiliation {Istituto Nazionale di Fisica Nucleare (INFN), Sez. di Napoli\\Complesso Universitario di Monte Sant Angelo, Edificio G, Via Cinthia, I-80126, Napoli, Italy.}

\date{\today}

\begin{abstract}
 We report  here, for the interested reader,  the various components of the energy-momentum tensor and the form of them after substituting the KB ansatz. Moreover, we report the gradient components of the energy-momentum tensor.
\end{abstract}

\maketitle

\section{The components of the energy momentum tensor}\label{Sec:App_1}
The components  of the energy-momentum tensor  are given as:
\begin{align*}
&{  \rho}  \equiv \dfrac{1}{{r}^{8} c^2\kappa}e^{\frac{1}{2{{r}^{2}}}\bigg\{-\zeta \left( 2 \mu''{r}^{2}+ \mu'^{2}{r}^{2}+ \left( - \nu' {r}^{2}+4r \right) \mu'  -4 \nu'  r+4 \right) {e^{-\nu  }}+4\zeta-8\nu  {r}^{2}\bigg\}} \bigg\{ -\zeta \bigg[ \zeta {r}^{2}{e^{\nu  }}\mu'' -2\left(\zeta+{r}^{2} \right) {e^{2 \nu  }}+\frac{\zeta }2\bigg(  \mu'^{2}{r}^{2}+ \left(4r  - \nu' {r}^{2}\right) \mu''\nonumber\\
&  +4-4 \nu' r \bigg) {e^{\nu  }} \bigg] {r}^{6}\mu''''  + \left( \zeta{r}^{2}\mu''  +\frac{1}2\zeta \mu'^{2}{r}^{2}-\frac{1}2r\zeta \left(  \nu'r-4 \right) \mu'  -2\zeta\nu'r-\left( 2\zeta+3{r}^{2} \right) {e^{\nu  }}+2\zeta \right) {\zeta}^{2 }{r}^{6} \mu'''^ {2}+2 \bigg[ \bigg( 2-\frac{3}2 \nu'r\nonumber\\
& +\mu' r \bigg) {\zeta}^{2}{r}^{4} \mu''^{2}-\frac{\zeta{r}^{3}\mu''}2 \bigg[ r\zeta \left( \mu'r+4 \right)\nu''  -\zeta{r}^{2} \mu'^{3}+ \left( \frac{7}2\zeta{r}^{2}\nu'  -6\zeta r \right)  \mu'^{2}+ \left( -\frac{5}2\zeta{r}^{2} \nu'^{2}+16\zeta \nu'r+ \left( 4\zeta +7{r}^{2} \right) {e^{\nu  }}-8\zeta \right) \mu'\nonumber\\
&  -10\zeta r \nu'^{2}+ \left( 14\zeta-  6\left( \zeta+2{r}^{ 2} \right) {e^{\nu  }}\right) \nu'  +16r{e^{\nu  }} \bigg]   -\frac{1}4 \left(\mu' r+4 \right) \zeta \left( \zeta \mu'^{2}{r}^{2}-r\zeta \left( \nu'r-4 \right) \mu'  -4\zeta\nu'r- \left( 4\zeta+6{r}^{2} \right) {e^{\nu  }}+4\zeta \right) {r}^{2}\nu''\nonumber\\
&  + \left( {r}^{3} \left( {r}^{2}+\zeta \right) \mu'  + \left( -3{r}^{5}-3\zeta{r}^{ 3} \right) \nu'  +4{r}^{4}-8\zeta{r }^{2}-8{\zeta}^{2} \right) {e^{2\nu  }}-\frac{1}4 \bigg[ \zeta \mu'^{4 }{r}^{5}\nu'  + \left(8\zeta{r} ^{4}\nu' -2\zeta{r}^{5} \nu'^{2}  +{r}^{3} \left( {e^{\nu  }}{r}^{2}+4\zeta \right)  \right)  \left( \mu'   \right) ^{3}\nonumber\\
& + \left( \zeta \nu'^{3}{r}^{3}-16\zeta{r}^ {2} \nu'^{2}-10r \left(  \left( {r}^{2}+\frac{2}5\zeta \right) {e^{\nu  }}-\frac{8}5\zeta \right) \nu'  + 8\left({r}^{2}-\zeta \right) {e^{\nu  }}+ 24\zeta \right) {r}^{2} \mu'^{2}+9 \bigg[ {\frac {8}{9}}\zeta \nu'^{3}{r}^{3}+ \left(  \left( {r}^{2}+\frac{4}9 \zeta \right) {e^{\nu  }}-4\zeta \right) {r} ^{2}\nonumber\\
&  \left( \nu'   \right) ^{2}-{\frac {44 }{9}} \left(  \left( 2/11\zeta+{r}^{2} \right) {e^{\nu  }}+2/11\zeta \right) r\nu'  - \left( \frac{16}3\zeta+\frac{4}9{r}^{2} \right) {e^{\nu  }}+\frac{16}3\zeta \bigg] r\mu'  +16\zeta \left( \nu'  \right) ^{3}{r}^{3}+36 \left(  \left( {r}^{2}+\frac{4}9\zeta \right) { e^{\nu  }}-\frac{4}9\zeta \right) {r}^{2} \nu'^{2}\nonumber\\
&-28r \left(  \left( -{ \frac {8}{7}}\zeta+{r}^{2} \right) {e^{\nu  }}+ {\frac {8}{7}}\zeta \right) \nu'  + \left( -32{r}^{2}-64\zeta \right) {e^{\nu  } }+32\zeta \bigg] \zeta \bigg] \zeta{r}^{3}\mu'''  +\frac{1}2 \left(\mu'  r+4 \right) \zeta  {r}^{5}\nu'''\bigg[ \zeta{r}^{2} {e^{\nu  }}\mu''  -2\left( \zeta+{r}^{2} \right) {e^{2\nu  }}\nonumber\\
&+\frac{1}2\zeta \left(\mu'^{2}{r}^{2}+ \left( - \left( \nu'   \right) {r}^{2}+4r \right) \mu'  +4-4 \left( \nu'  \right) r \right) {e^{\nu  }} \bigg]  + \left( \zeta \mu'^{2}{r}^{2}+ \left(4\zeta r  -3 \zeta{r}^{2}\nu' \right) \mu'  +\frac{9}4\zeta{r}^{2}\nu'^{2}+4\zeta-{ e^{\nu  }}{r}^{2}-6\zeta \nu'r \right) {\zeta}^{2}{r}^{6} \mu''^{3}\nonumber\\
&- \bigg[ \left( \zeta\mu'^{2 }{r}^{2}-\frac{3}2r\zeta \left( \nu' r-4 \right) \mu'  -6 \zeta \left( \nu'   \right) r-2{ e^{\nu  }}{r}^{2}+8\zeta \right) \zeta{r}^{2 }\nu'' - \left( 2\zeta{r}^ {2}-\frac{3}2{r}^{4} \right) {e^{2\nu  }}-\frac{1}2 \zeta \bigg[ \zeta\mu'^{4}{r}^{4}+ \bigg( 8\zeta{r}^{3}-6\zeta{r}^{4}\nu'   \bigg)\mu'^{3} \nonumber\\
& + \left( {\frac {41}{4}}\zeta{r}^{ 4} \nu'^{2}-38\zeta \nu'{r}^{3}-7 \left( \left( {r}^{2}+\frac{4}7\zeta \right) {e^{\nu  }}-{ \frac {16}{7}}\zeta \right) {r}^{2} \right) \mu'^{2}-{\frac {21}{4}} \bigg[ \zeta \nu'^{3}{r}^{2}-{\frac {68}{7}}\zeta r\nu'^{2}+ \left(  \left( -{\frac {16}{7}}\zeta-{\frac {32}{7}}{r}^{2} \right) {e^{\nu  }}+{\frac {80}{7}}\zeta \right) \nu' \nonumber\\
& +{\frac {128}{21}}r{ e^{\nu }} \bigg] {r}^{2}\mu'  -21\zeta \nu'^{3}{r}^{3}+ \left(  \left( -9\zeta{r}^{2}-19{r}^{4} \right) {e^{\nu  }}+49\zeta{r}^{2} \right) \nu'^{2}+ \left( -16 \zeta r+56{r}^{3}{e^{\nu  }} \right) \nu' + \left( 16\zeta -28{r}^{2}\right) { e^{\nu  }}-16\zeta \bigg]  \bigg] \zeta{r} ^{4}  \mu''^{2}\nonumber\\
&+ \frac{1}4 \bigg[ {r}^{4}{\zeta}^{2} \left( \mu' r+4 \right) ^{2} \nu''^{2}-2 \bigg[  8\left( \zeta r +{r}^{3} \right) {e^{2\nu  }}+ \bigg( \zeta {r}^{3} \mu'^{4}+ \left( 10\zeta{r}^{2}-\frac{7}2\zeta \nu'{r}^{3} \right)\mu'^{3}+\frac{5 \mu'^{2}r}2 \bigg[\zeta{r}^{2} \nu'^{2}-12\zeta \nu'r+{\frac {64 }{5}}\zeta-e^{\nu  } \bigg( {\frac {18}{5}} {r}^{2} \nonumber\\
&+\frac{8}5\zeta \bigg) \bigg] + \left( 20\zeta{r}^{2} \nu'^{2}+ \left(  \left( 6\zeta r+{\frac { 29}{2}}{r}^{3} \right) {e^{\nu  }}-78\zeta r \right) \nu' - \left( 16\zeta+50{r }^{2} \right) {e^{\nu  }}+32\zeta \right) \mu'  +40\zeta r \nu'^{2}+ \bigg[  \left( 58{r}^{2}+24 \zeta \right) e^{\nu  }-56\zeta \bigg]\nu'\nonumber\\
&  -60r{e^{\nu  }} \bigg) \zeta \bigg] {r}^{3}\nu''  + \bigg[ -2 \mu'^{2}{r}^{6}+ \left(  \left( -18{r}^{6}-20\zeta{r}^{4} \right) \nu'  -64r{\zeta}^{2}+8{r}^{ 5}-80\zeta{r}^{3} \right) \mu'  + \left( 22\zeta{r}^{4}+22{r}^{6} \right)  \nu'^{2}+ \bigg[ 96r{\zeta}^{2}-60{r}^{5 }\nonumber\\
& +88\zeta{r}^{3} \bigg] \nu'  -64{ \zeta}^{2}-176\zeta{r}^{2}-4{r}^{4} \bigg] {e^{2\nu  }}+4{r}^{4}{e^{3\nu }}-2 \zeta \bigg\{\zeta{r}^{6}  \nu'  \mu'^ {5}-4\zeta \left( -1+{r}^{2} \left( \nu' \right) ^{2}-\frac{5}2\nu' r \right) {r}^{4} \mu'^{4}+5 \bigg[ \zeta \nu'^{3}{r}^{3} -{\frac {36}{5}}\zeta{r}^{2}\nu'^{2}\nonumber\\
&-{\frac {17}{10}}r \left(  \left( {\frac {8}{17}}\zeta+{r}^{2} \right) {e^{\nu  }}-{\frac {44}{17}}\zeta \right) \nu'  + \left(\frac{2}5{r}^{2}  -\frac{8}5\zeta \right) {e^ {\nu  }}+{\frac {32}{5}}\zeta \bigg] {r}^{3} \mu'^{3}-2 \bigg[ \zeta{r} ^{4} \nu'^{4}-21\zeta\nu'^{3}{r}^{3}+ \bigg[48\zeta{r}^{2} -\left( 5\zeta{r}^{2}+{\frac {87}{8}}{r}^{4} \right) { e^{\nu  }}\bigg] \nu'^{2}\nonumber\\
&+ \left(  \left( 29{r} ^{3}-2\zeta r \right) {e^{\nu  }}+18\zeta r \right) \nu'  + \left( 32\zeta+8{r} ^{2} \right) {e^{\nu  }}-44\zeta \bigg]{r}^{2 } \left( \mu'   \right) ^{2}-{\frac {53}{4 }}r \bigg[{\frac {64}{53}}\zeta{r}^{4} \nu'^{4}+ \left(  \left( {\frac {24}{53}} \zeta+{r}^{2} \right) {e^{\nu  }}-{\frac {408}{53 }}\zeta \right) {r}^{3} \nu'^{3}\nonumber\\
&-{\frac {436}{53}} \left(  \left( {\frac {28}{109}} \zeta+{r}^{2} \right) {e^{\nu  }}-{\frac {52}{109 }}\zeta \right) {r}^{2} \nu'^{2}+{\frac {240}{53}} \left(  \left( -2\zeta+{r}^{2} \right) {e^{\nu  }}+{\frac {38}{15}}\zeta \right) r\nu'  + \left( {\frac {352}{53}} {r}^{2}+{\frac {512}{53}}\zeta \right) {e^{\nu  }}-{\frac {384}{53}}\zeta \bigg] \mu'  -32\zeta{r}^{4} \nu'^{4}\nonumber\\
&+ \left(56\zeta{r}^{3}- \left( 24\zeta{r}^{3}+53{r}^{5 } \right) {e^{\nu  }} \right) \nu'^{3}+ \left( \left( 95{r}^{4}-48\zeta{r}^{2} \right) {e^{\nu  }}+48\zeta{r}^{2} \right) \nu'^{2}+52 \left(  \left( {\frac {40}{13}} \zeta+{r}^{2} \right) {e^{\nu }}-{\frac {28}{13} }\zeta \right) r\nu'  - \left( 88{r} ^{2}+64\zeta \right) {e^{\nu  }}\nonumber\\
&+32\zeta \bigg\}  \bigg] \zeta{r}^{2}\mu'' +\frac{1}8 \left(  \mu' r+4 \right) ^{2}{\zeta}^{2} \left( \zeta \mu'^{2}{r}^{2}-r\zeta \left( \nu'  r-4 \right) \mu'  -4\zeta \nu'r-\left( 4\zeta+6{r}^{2} \right) { e^{\nu  }}+4\zeta \right) {r}^{4} \nu''^{2}+\frac{1}4 \bigg[ \bigg(  \left( -4\zeta{r}^{4}-4{r}^{6} \right)  \mu'^{2} \nonumber\\
&+ \left(  \left( 14\zeta{ r}^{4}+14{r}^{6} \right) \nu'  +32r{ \zeta}^{2}-24{r}^{5}+24\zeta{r}^{3} \right) \mu'  + \left( 56{r}^{5}+56\zeta{r}^{3} \right) \nu' +176\zeta{r}^{2}-16{r}^{4}+128 {\zeta}^{2} \bigg) {e^{2\nu  }}+\zeta \bigg( \zeta{r}^{6}\nu' \mu'^{5}\nonumber\\
&-2 \left( -6\zeta\nu' r-\frac{1}2{e^{\nu }}{r}^{2}-2\zeta+ \zeta{r}^{2} \nu'^{2} \right) {r}^{4} \mu'^{4 }+ \bigg[ \zeta  \nu'^{3}{r}^{3}-24\zeta{r}^{2} \nu'^{2}-2\frac{1}2r \left(  \left( {\frac {8}{21}}\zeta+ {r}^{2} \right) {e^{\nu }}-{\frac {32}{7}} \zeta \right) \nu'  + \left( -8\zeta+ 10{r}^{2} \right) {e^{\nu  }}\nonumber\\
\end{align*}
\begin{align}\label{rho}
&+40\zeta \bigg] {r}^{3} \mu'^{3 }+19/2 \bigg( {\frac {24}{19}}\zeta\nu'^{3}{r}^{3}+{r}^{2} \left(  \left( {r}^{2}+ {\frac {8}{19}}\zeta \right) {e^{\nu  }}-{ \frac {200}{19}}\zeta \right) \nu'^{2}-{\frac {172}{19}} \left(  \left( {r}^{2}+{ \frac {12}{43}}\zeta \right) {e^{\nu  }}-{ \frac {28}{43}}\zeta \right) r\nu'  + \bigg( {\frac {16}{19}}{r}^{2}\nonumber\\
&-{\frac {160}{19}}\zeta \bigg) { e^{\nu  }}+{\frac {288}{19}}\zeta \bigg) {r}^{ 2} \mu'^{2}+76 \bigg[ {\frac {12}{19}}\zeta \nu'^{3}{r}^{3}+ \left(  \left( {r}^{2}+{\frac {8}{19}}\zeta \right) {e^{\nu  }}-{\frac {40}{19}}\zeta \right) {r}^{2} \nu'^{2 }-{\frac {97}{38}} \left( {e^{\nu  }}{r}^{2}+{ \frac {32}{97}}\zeta \right) r\nu'  - \bigg[ {\frac {26}{19}}{r}^{2}+{\frac {64}{19}}\zeta \bigg]\nonumber\\
& { e^{\nu }}+{\frac {56}{19}}\zeta \bigg] r\mu'  +64\zeta \left( \nu'   \right) ^{3}{r}^{3}+152 \left(  \left( {r}^{2}+{ \frac {8}{19}}\zeta \right) {e^{\nu  }}-{\frac {8}{19}}\zeta \right) {r}^{2} \left( \nu'   \right) ^{2}-72 \bigg[  \left( -{\frac {16}{9}}\zeta+{r }^{2} \right) {e^{\nu  }} +{\frac {16}{9}}\zeta \bigg] r\nu' \nonumber\\
& + \left( -176{r}^{2}-256 \zeta \right) {e^{\nu  }}+128\zeta \bigg) \bigg] \zeta{r}^{2}\nu''-\frac{1}8 \bigg[  \mu'^{4}{r} ^{8}+ \left( 8{r}^{7}-2{r}^{8}\nu'  \right) \mu'^{3}+ \bigg[  \left( -12{r}^{6}\zeta-11{r}^{8} \right) \nu'^{2}+ \left( 4{r}^{7}-64{ \zeta}^{2}{r}^{3}-88{r}^{5}\zeta \right) \nu'\nonumber\\
&  -16\zeta{r}^{4}-64{\zeta}^{2}{r}^{2}+20{r}^{6} \bigg]  \mu'^{2}+12 \bigg[ {r}^{5} \left( {r}^{2}+\zeta \right)\nu'^{3}+ \left( 10/3\zeta{r}^{4}+16/3{ \zeta}^{2}{r}^{2}-5{r}^{6} \right) \nu'^{2}+ \left( -3{r}^{5}-{\frac {104}{3}}\zeta {r}^{3}-16r{\zeta}^{2} \right) \nu'\nonumber\\
&    -{ \frac {128}{3}}{\zeta}^{2}+4/3{r}^{4}-{\frac {112}{3}}\zeta{ r}^{2} \bigg] r\mu'  + \left( 48{r}^{7} +48{r}^{5}\zeta \right)\nu'^{3}+ \left( -32{r}^{6}+256{\zeta}^{2}{r}^{2}+384 \zeta{r}^{4} \right)\nu'^{2}+ \left( -96{r}^{5}-64\zeta{r}^{3}+256r{\zeta}^ {2} \right) \nu'\nonumber\\
&    -64{r}^{4}-896\zeta {r}^{2}-768{\zeta}^{2} \bigg] \zeta{e^{2\nu  }}+ \left( \frac{1}2\zeta  \mu'^{2}{r}^{6}-\frac{1}2{r}^{5}\zeta \left(  \nu'r-4 \right) \mu'  +{r}^{3} \left(r^4 -4\zeta{r}^{2}-4{\zeta}^{2} \right) \nu'  -56{\zeta}^{2}{r} ^{2}-8\zeta{r}^{4}-{r}^{6}-32{\zeta}^{3} \right) {e^{3 \nu  }}\nonumber\\
&  +{e^{4\nu  }}{r}^{6}+1/8{ \zeta}^{2} \bigg[\zeta \nu'^{2}\mu'^{6}{r }^{8}-3\zeta \nu' {r }^{6} \left( -\frac{8}3-4 \nu'  r+{r}^{2} \nu'^{2} \right)  \left( \mu' \right) ^{5}+3  \bigg[ \zeta{r}^{4} \nu'^{4}-12\zeta\nu'^{3}{r}^{3}-3 \bigg[  \left( {r}^{2}+\frac{4}9\zeta \right) { e^{\nu  }}-4\zeta \bigg] {r}^{2} \nu'^{2}\nonumber\\
&  +4/3 \left(  \left( {r }^{2}-4\zeta \right) {e^{\nu  }}+20\zeta \right) r\nu'  +\frac{16}3\zeta \bigg]{r}^ {4} \mu'^{4}-r^3 \bigg[ \zeta \nu'^{5}{r}^{5} -36\zeta{r}^{4}\nu'^{4}-18 \left(  \left( {r}^{2}+\frac{4}9\zeta \right) {e ^{\nu  }}-8\zeta \right) {r}^{3} \nu'^{3}+ \left( 80{r}^{4}{e^{\nu  }}+64\zeta{r}^{2} \right)  \nu'^{2}\nonumber\\
&+36r \left(  \left( {\frac {40}{9} }\zeta+{r}^{2} \right)e^{\nu}-{\frac {68}{ 9}}\zeta \right) \nu'  +64\zeta \left( {e^{\nu  }} -2\right)  \bigg] \mu'^{3}-9 \bigg[ \frac{4}3 \zeta \nu'^{5}{r}^{ 5}+ \left(  \left( {r}^{2}+\frac{4}9\zeta \right) e^{\nu}-{\frac {148}{9}}\zeta \right) {r}^{4} \nu'^{4}-{\frac {148}{9}} \bigg[ \left( {\frac {12}{37}}\zeta+{r}^{2} \right) e^{\nu}\nonumber\\
&-{\frac {44}{37}}\zeta \bigg] {r}^{3} \nu'^{3}+16 \left(  \left( {r}^{2}-{ \frac {14}{9}}\zeta \right) {e^{\nu  }}+\frac{10}3 \zeta \right) {r}^{2} \nu'^{2}+{\frac {104}{3}} \left(  \left( {r}^{2}+{\frac {56}{39 }}\zeta \right) {e^{\nu }}-{\frac {40}{39}} \zeta \right) r\nu'  + \left( {\frac {448 }{9}}\zeta+{\frac {112}{9}}{r}^{2} \right) {e^{\nu  }}-{\frac {128}{3}}\zeta \bigg] {r}^{2}\mu'^{2}\nonumber\\
&-72 \bigg[ \frac{2}3\zeta\nu'^{5}{r}^{5}+ \left(  \left( {r}^{2}+4/9\zeta \right) {e^{\nu  }}-{\frac {28}{9}}\zeta \right) {r}^{4} \nu'^{4}+ \left( -9/2{r}^{5}{e^{\nu  }}-{\frac {16}{9}}\zeta{r}^{3} \right)  \nu'^{3}-{\frac {29}{9}} \left(  \left( {\frac {96}{29}}\zeta+{r}^{2} \right) {e^{\nu  }}-{\frac {88}{29}}\zeta \right) {r}^{2} \nu'^{2}\nonumber\\
&+{\frac {20}{3}} \left(  \left( {\frac {4}{15}}\zeta+{r}^{2} \right) {e^{\nu  }}+2/15\zeta \right) r\nu'  + \left( {\frac {56}{9}}{r}^{2}+{\frac {128}{9}}\zeta \right) {e^{\nu  }}-{\frac {64}{9}}\zeta \bigg] r\mu'  -64\zeta \nu'^{5}{r}^{5}-144 \left( \left( {r}^{2}+4/9\zeta \right) {e^{\nu  }}-4/ 9\zeta \right) {r}^{4} \nu'^{4}\nonumber\\
&+80 \left(  \left( {r}^{2}-{\frac {16}{5}}\zeta \right) {e^{\nu  }}+{\frac {16}{5}}\zeta \right) {r}^{3} \left( \nu'   \right) ^{3 }+448 \left(  \left( {r}^{2}+{\frac {8}{7}}\zeta \right) {e ^{\nu  }}-4/7\zeta \right) {r}^{2}\nu'^{2}-32r \left(  \left( -16\zeta+ {r}^{2} \right) {e^{\nu  }}+8\zeta \right) \nu'\nonumber\\
&  + \left( -768\zeta-448{r}^{2} \right) {e^{\nu  }}+256\zeta \bigg] \bigg\}\,,
\end{align}
\begin{align}\label{pr}
&p_r  \equiv \dfrac{-1}{8 {r}^{6}\kappa}{e^{{\frac {- \left( 2{r}^{2}\mu'' + \mu'^{2}{r}^{2}+ \left(4r -{r}^{2}\nu' \right) \mu' +4-4 \nu' r \right) \zeta{ e^{-\nu  }}+4\zeta-6\nu  {r}^{2 }}{2{r}^{2}}}}} \bigg\{ -2{r}^{3} \left(  \mu' r+4 \right) \zeta\bigg[2\zeta{r}^ {2}\mu''  +\zeta\mu'^{2}{r}^{2}-r\zeta\left( -4+ \left( \nu'   \right) r \right) \mu' -4\zeta \nu' r+4\zeta\nonumber\\
&-4 \left({r}^{2}+\zeta\right) {e^{\nu  }}\bigg] \mu'''  -4{r}^{4}\zeta\left( \zeta\mu'^{2}{r}^{2}-3/2r \zeta\left(  \nu'r-4 \right) \mu' +{e^{\nu  }}{r}^{2}-6\zeta\nu' r+8\zeta \right) \mu''^{2}-2{r}^{3}\zeta\bigg[ -\zeta r \left(  \mu' r+4 \right) ^{2}\nu''  -4r{e ^{2\nu  }}+\zeta {r}^{3} \mu'^{4}\nonumber\\
&+ \left( 10\zeta{r}^{2}-7/2\nu'{r}^{3}\zeta\right)  \mu'^{3}+5/2 r \left( \zeta{r}^{2} \left( \nu'  \right) ^{2}-12\zeta\left( \nu'  \right) r+ \left( -4/5{r}^{2}-8/5\zeta \right) {e^{\nu  }}+{\frac {64}{5}}\zeta\right)  \mu'^{2}+ \bigg\{ 20\zeta{r}^{2}\nu'^{2}+4r \bigg[ \left( {r}^{2}+3/2\zeta\right) {e^{\nu  }}\nonumber\\
&-{ \frac {39}{2}}\zeta\bigg] \nu' + \left( -16\zeta-20{r}^{2} \right) {e^{\nu  } }+32\zeta\bigg\}\mu' +40\zeta r \left( \nu'   \right) ^{2}+ \left( \left( 16{r}^{2}+24\zeta\right) {e^{\nu  }} -56\zeta\right) \nu'  -28{e^{\nu  }}r \bigg]\mu''  +{r}^{2} \left(  \mu'r+4 \right) ^{2}\zeta\nonumber\\
&\left( \zeta\mu'^{2}{r}^{2}-r\zeta\left( -4+ \nu'r \right) \mu' -4\zeta\nu'r+ \left( -4{r}^{2}-4\zeta\right) {e^{\nu  }}+4\zeta \right) \nu''  + \bigg[ 4\zeta\mu'^{2}{r}^{4}+ \left( 32{r}^{3}\zeta+32r{\zeta} ^{2}-4{r}^{4}\zeta\nu' -8{r}^{5} \right) \mu' -8{r}^{4}+128{\zeta}^ {2}\nonumber\\
&-16 \nu'{r}^{3}\zeta+128\zeta{r}^{2} \bigg] {e^{2\nu  }}+8 {e^{3\nu  }}{r}^{4}+ \bigg[ \zeta\mu'^{5}\nu'{r}^{6}-2{r}^{4} \left( -6\zeta\left( \nu'   \right) r-2\zeta+\zeta{r}^{2}\nu'^{2}+1/2 {e^{\nu  }}{r}^{2} \right)  \mu'^{4}+{r}^{3} \bigg( \zeta \nu'^{3}{r}^{3}-24\zeta{r}^ {2} \nu'^{2}\nonumber\\
&-2r \left(  \left( {r}^{2}+2\zeta\right) {e^{\nu  }}-24\zeta\right) \nu'  -4\left(2\zeta+{r}^{2} \right) {e^{\nu  }}+ 40\zeta\bigg)  \mu'^{3}+3{r}^{2} \bigg( 4\zeta\nu'^{3}{r}^{3}+{r}^{2} \left(  \left( 4/3\zeta+{r}^{ 2} \right) {e^{\nu }}-{\frac {100}{3}}\zeta\right)\nu'^{2}\nonumber\\
&-{ \frac {20}{3}}r \left(  \left( \frac{6}5\zeta+{r}^{2} \right) {e^ {\nu  }}-{\frac {14}{5}}\zeta\right) \nu'  + \left( -{\frac {20}{3}}{r}^{2}-{\frac {80}{3}} \zeta\right) {e^{\nu  }}+48\zeta\bigg) \mu'^{2}+24r \bigg[ 2 \zeta \nu'^{3}{r}^{ 3}+{r}^{2} \left(  \left( 4/3\zeta+{r}^{2} \right) {e^{\nu  }}-{\frac {20}{3}}\zeta\right)  \nu'^{2}-\bigg( \frac{8}3r\zeta\nonumber\\
&+{\frac { 11}{6}}{r}^{3}{e^{\nu  }} \bigg) \nu'-\left( {\frac {32}{3}}\zeta+4{r}^{2} \right) {e^{\nu  }}+{\frac {28}{3}}\zeta\bigg]\mu' +64\zeta \nu'^{3}{r}^{3}+48{r}^{2} \left(  \left( \frac{4}3\zeta+{r}^{2} \right) {e^{\nu  }}-\frac{4}3\zeta\right)\nu'^{2}+16r \left(  \left( 8\zeta+{r}^{2} \right) {e^{\nu  }}-8\zeta\right) \nu'\nonumber\\
& - 128\left( 2\zeta+{r}^{2} \right) {e^ {\nu  }}+128\zeta\bigg] \zeta\bigg\}\,,
\end{align}
\begin{align*}
&p_t  \equiv -\dfrac{1}{8 {r}^{8}\kappa}e^{\frac {- \left( 2{r}^{2}\mu'' + \mu'^{2}{r}^{2}+ \left( -{r}^{2}\nu' +4r \right) \mu'+4-4 \nu' r \right) \zeta { e^{-\nu }}+4\zeta -8\nu {r}^{2 }}{2{r}^{2}}} \bigg\{ -4{r}^{6} \bigg[ 2\zeta {r}^{2}{e^{\nu }}\mu'' -4 \left({r}^{2}+\zeta  \right) {e^{2\nu }}+{e^{\nu }}\zeta  \bigg( \mu'^{2}{r}^{2}+ \left( 4r-{r}^{2}\nu'  \right) \mu'+4\nonumber\\
&-4 \nu' r \bigg)  \bigg]\zeta \mu'''' +4{r}^{6}{\zeta }^{2} \left( 2\zeta {r}^{2}\mu'' +\zeta  \mu'^{2}{r}^{2}-r\zeta  \left(  \nu' r -4\right) \mu'-4\zeta  \nu' r- \left(6{r}^{2}+4\zeta  \right) {e^{\nu }}+4\zeta  \right)  \mu'''^{2}-4{r}^{3} \bigg[ -4{r}^{4}{\zeta }^ {2} \bigg(2-\frac{3}2\nu'r\nonumber\\
&+  \mu'r \bigg)\mu''^{2}-2{ r}^{3} \bigg[ -r\zeta  \left(  \left( \mu' \right) r+4 \right) \nu'' +\zeta {r}^{2} \mu'^{3}+ \left( 6r\zeta -\frac{7}2\zeta  \nu' {r}^{2} \right)  \mu'^{2}+ \left( \frac{5}2\zeta {r}^{2} \nu'^{2}-16\zeta \nu' r- \left(4\zeta +\frac{15}2{ r}^{2} \right) {e^{\nu }}+8\zeta  \right) \mu'\nonumber\\
& +10\zeta r \nu'^{2}+ \left(  \left( 12{r}^{2}+6\zeta  \right) {e^{\nu }}-14\zeta  \right) \nu' -15{e^{\nu }}r \bigg] \zeta \mu' +{r}^{2} \left(\mu' r+4 \right) \zeta   \left( \zeta \mu'^{2}{r}^{2}-r\zeta  \left( \nu' r-4 \right) \mu'-4 \zeta  \nu' r- \left(6{r}^{2}+4\zeta  \right) {e^{\nu }}+4\zeta  \right)\nu''\nonumber\\
& + \left(  \left( -6{r}^{5}-6{r}^{3}\zeta  \right) \mu'+ \left( 12{r}^{3}\zeta +12{r}^{5} \right) \nu' -12{r}^{4}+32{\zeta }^{2}+36\zeta {r}^{2} \right) {e^{2\nu }}+ \bigg[ \zeta  \mu'^{4}{r}^{5}\nu' -2{r}^{3} \left( \zeta {r}^{2}\nu'^{2}-2\zeta -3/4{e^{\nu }}{r}^{2}-4\zeta  \nu' r \right)\mu'^{3}\nonumber\\
&+{r}^{2} \left( \zeta  \nu'^{3}{r}^{3}-16\zeta {r}^{2}\nu'^{2}-\frac{21}2r \left(  \left( {\frac {8}{21}}\zeta +{r}^{2} \right) {e^{\nu } }-{\frac {32}{21}}\zeta  \right) \nu' + \left( 9{r}^{2}-8\zeta  \right) {e^{\nu }}+ 24\zeta  \right) \mu'^{2}+9r \bigg( {\frac {8}{9}}\zeta  \nu'^{3}{r}^{3}+{r}^{2} \bigg[  \left( {r}^{2}+ \frac{4}9\zeta  \right) {e^{\nu }}\nonumber\\
&-4\zeta  \bigg] \nu'^{2}-5r \left( \left( {\frac {8}{45}}\zeta +{r}^{2} \right) {e^{\nu }}+{\frac {8}{45}}\zeta  \right) \nu' + \left( -16/3\zeta -2/3{r}^{2} \right) {e^{\nu }}+16/3\zeta  \bigg) \mu'+16\zeta \nu'^{3}{r}^{3}+36{r}^{2} \left(  \left( {r}^{2}+\frac{4}9\zeta  \right) {e^{\nu }}-\frac{4}9\zeta  \right) \nu'^{2}\nonumber\\
&-24r \left(  \left( {r }^{2}-\frac{4}3\zeta  \right) e^{\nu }+\frac{4}3\zeta  \right) \nu'-\left(64\zeta +36{r }^{2} \right) {e^{\nu }}+32\zeta  \bigg] \zeta  \bigg] \zeta \mu'''+2 {r}^{5} \bigg[ 2\zeta {r}^{2}e^{\nu }\mu''- 4\left({r}^{2}+ \zeta  \right) {e^{2\nu }}+e^{\nu }\zeta  \bigg\{  \mu'^{2}{r}^{2}+ 4-4 \nu' r\nonumber\\
&+\left( 4r-{r}^{2}\nu'\right) \mu' \bigg\}  \bigg] \left( \mu' r+4 \right) \zeta \nu''' +8{r}^{ 6}{\zeta }^{2} \left( \zeta  \left( \mu' \right) ^{2}{r}^{2}+ \left( 4r\zeta -3\zeta \nu' {r}^{2} \right)\mu' +\frac{9}4\zeta {r}^{2} \left( \nu' \right) ^{2}+4\zeta -{e^{\nu }}{r}^{2}-6\zeta  \nu' r \right)  \mu'^{3}\nonumber\\
&+4{r}^{4} \bigg[ -2{r}^{2} \left( \zeta  \left( \mu' \right) ^{2}{r}^{2}-3/2r \zeta  \left( -4+ \nu' r \right) \mu'-2{e^{\nu }}{r}^{2}+8\zeta -6\zeta  \nu' r \right) \zeta \nu'' + \left( 4\zeta {r}^{2}+4{r}^{4} \right) { e^{2\nu }}+ \bigg[\zeta  \mu'^{4}{r}^{4}+ \left( 8{r}^{3}\zeta -6 {r}^{4}\zeta \nu'  \right)  \mu'^{3}\nonumber\\
&+ \left( {\frac {41}{4}} \zeta  \nu'^{2}{r}^{ 4}-38 \nu' {r}^{3} \zeta -8{r}^{2} \left(  \left( {r}^{2}+1/2\zeta  \right) {e^ {\nu }}-2\zeta  \right)  \right)  \mu'^{2}-{\frac {21}{4}}{r}^{2} \bigg( \zeta  \nu'^{3}{r}^{2} -{\frac {68}{7}}\zeta r \nu'^{2}+ \left(  \left( -{\frac {16}{7}}\zeta -{\frac {34}{7}} {r}^{2} \right) {e^{\nu }}+{\frac {80}{7}} \zeta  \right) \nu' \nonumber\\
&+{\frac {128}{21}}{ e^{\nu }}r \bigg) \mu'-21\zeta  \left( \nu' \right) ^{3}{r}^{3}+ \left(  \left( -19{r}^{4}-9\zeta {r}^{2} \right) {e^{\nu }}+49\zeta {r}^{2} \right) \nu'^{2}+ \left( -16r \zeta +53{r}^{3}{e^{\nu }} \right) \nu' + \left( 16\zeta -24{r}^{2} \right) {e ^{\nu }}-16\zeta  \bigg] \zeta  \bigg] \zeta  \mu''^{2}\nonumber\\
&-4{ r}^{2} \bigg\{ -\frac{1}2{\zeta }^{3}{r}^{4} \left(  \mu' r+4 \right) ^{2} \nu'^{2}+{r}^{3} \bigg[  \left( 8r \zeta +8{r}^{3} \right) {e^{2\nu }}+ \bigg( \zeta {r}^{3} \left( \mu' \right) ^{4} + \left( 10\zeta {r}^{2}-7/2 \nu' {r}^{3}\zeta  \right)  \mu'^{3}+5/2r \bigg( \zeta {r}^{2} \nu'^{2}-12\zeta  \nu' r\nonumber\\
&+ \left( -8/5\zeta -{ \frac {19}{5}}{r}^{2} \right) {e^{\nu }}+{ \frac {64}{5}}\zeta  \bigg) \mu'^{2}+ \left( 20\zeta {r}^{2}  \nu'^{2}+ \left(  \left( {\frac {29}{2}}{ r}^{3}+6r\zeta  \right) {e^{\nu }}-78r\zeta  \right) \nu' + \left( -51{r}^{2}-16 \zeta  \right) {e^{\nu }}+32\zeta  \right) \mu' +40\zeta r \nu'^{2}\nonumber\\
&+ \left(  \left( 24\zeta +58{r}^{2 } \right) {e^{\nu }}-56\zeta  \right) \nu' -56{e^{\nu }}r \bigg) \zeta  \bigg] \zeta \nu'' -2 \bigg[ {r}^{4} \left( {r}^{2}+\zeta  \right)  \mu'^{2}+ \left( -13/2{r}^{4} \left( {r}^{2}+\zeta  \right) \nu' +7/2 {r}^{5}-16r{\zeta }^{2}-22{r}^{3}\zeta  \right) \mu'\nonumber\\
&+11/2{r}^{4} \left( {r}^{2}+\zeta  \right) \nu'^{2}+ \left( 25{r }^{3}\zeta -{\frac {25}{2}}{r}^{5}+24r{\zeta }^{2} \right) \nu' -16{\zeta }^{2}-44\zeta {r}^{2}-5{r }^{4} \bigg] \zeta {e^{2\nu }}+{r}^{4} \left( {r}^{2}-2\zeta  \right) {e^{3\nu }}+ \bigg[ \zeta  \left( \mu' \right) ^{5 } \nu' {r}^{6}\nonumber\\
&-4{r}^{4} \left(\zeta {r}^{2} \nu'^{2} -\frac{5}2\zeta  \nu' r-\zeta -\frac{1}8{e^{\nu }}{r}^{2} \right)\mu'^{4}+5 r^3 \mu'^{3}\bigg( \zeta  \nu'^{3 }{r}^{3}-{\frac {36}{5}}\zeta {r}^{2} \nu'^{2}-{\frac {41}{20}}r \left(  \left( {r} ^{2}+{\frac {16}{41}}\zeta  \right) {e^{\nu }}- {\frac {88}{41}}\zeta  \right) \nu' + \left( \frac{4}5{r}^{2} -\frac{8}5\zeta \right) {e^{\nu }}\nonumber\\
&+{\frac {32}{5}}\zeta  \bigg)-2{r}^{2} \bigg( \zeta  \nu'^{4}{r}^{4}-21 \zeta  \nu'^{3}{r}^{3} + \left(  \left( -5\zeta {r}^{2}-\frac{23}2{r}^{4} \right) {e^{\nu }}+48\zeta {r}^{2} \right)  \nu'^{2}+ \left(  \left( {\frac {125}{4}}{ r}^{3}-2r\zeta  \right) {e^{\nu }}+18r\zeta  \right) \nu' + \left( 9{r}^{2}+32 \zeta  \right) {e^{\nu }}\nonumber\\
&-44\zeta  \bigg) \left( \mu' \right) ^{2}-{\frac {53}{4} }r \bigg( {\frac {64}{53}}\zeta  \nu'^{4}{r}^{4}+{r}^{3} \left(  \left( {r}^{2}+{\frac { 24}{53}}\zeta  \right) {e^{\nu }}-{\frac {408}{ 53}}\zeta  \right)  \left( \nu' \right) ^{3}-{\frac {446}{53}}{r}^{2} \left(  \left( {\frac {56}{ 223}}\zeta +{r}^{2} \right) {e^{\nu }}-{\frac { 104}{223}}\zeta  \right)\nu'^{2}\nonumber\\
&+{\frac {204}{53}}r \left(  \left( {r}^{2}-{\frac {40}{ 17}}\zeta  \right) {e^{\nu }}+{\frac {152}{51}} \zeta  \right) \nu' + \left( {\frac { 384}{53}}{r}^{2}+{\frac {512}{53}}\zeta  \right) {e^{\nu }}-{\frac {384}{53}}\zeta  \bigg) \mu' -32\zeta  \nu'^{4}{r}^{4}+ \left(  \left( -53{r}^{5}-24{r}^{3} \zeta  \right) {e^{\nu }}+56{r}^{3}\zeta  \right)  \nu'^{3}\nonumber\\
&+ \left(  \left( 85{r}^{4}-48\zeta {r}^{2} \right) {e^{\nu }}+48\zeta {r}^{2} \right)  \nu'^{2}+66r \left(  \left( {\frac {80}{33 }}\zeta +{r}^{2} \right) {e^{\nu }}-{\frac {56} {33}}\zeta  \right) \nu' + \left( -64 \zeta -88{r}^{2} \right) {e^{\nu }}+32\zeta  \bigg] {\zeta }^{2} \bigg\}\mu''+{r}^{4} \left(  \mu'r+4 \right) ^{2}{\zeta }^{2}\nonumber\\
&\times \left( \zeta  \mu'^{2}{r}^{2}-r\zeta  \left( -4+ \nu' r \right) \mu' -4\zeta  \nu' r+ \left( -6{r}^{2}-4\zeta  \right) {e^{\nu }}+4\zeta  \right)  \nu''^{2}+2{r}^{2} \bigg\{  \big[  \left( - 6{r}^{6}-6{r}^{4}\zeta  \right)   \mu'^{2}+ \bigg(  \left( 14{r}^{6}+14{r}^{4}\zeta  \right) \nu' -28{r}^{5}\nonumber\\
&+20{r}^{3} \zeta +32r{\zeta }^{2} \bigg) \mu'+ \left( 56{r}^{5}+56{r}^{3}\zeta  \right) \nu' +128{\zeta }^{2}+192\zeta {r}^{2} \bigg] {e^{2 \nu }}+ \bigg[ \zeta  \mu'^{5} \nu' {r}^{6}-2{r}^{4} \left( -6\zeta  \nu' r+\zeta {r}^{2}\nu'^{2}-2\zeta -3/4{e ^{\nu }}{r}^{2} \right)  \mu'^{4}\nonumber\\
&+{r}^{3} \left( \zeta  \nu'^{3}{r}^{3}-24\zeta {r}^{2} \nu'^{2}+ \left( 48r\zeta-\left( 11{r}^{3}+4r\zeta  \right) {e^{\nu } }  \right) \nu' + \left( 13{ r}^{2}-8\zeta  \right) {e^{\nu }}+40\zeta  \right) \mu'^{3}+\frac{19}2 {r}^{2} \bigg[ {\frac {24}{19}}\zeta  \nu'^{3}{r}^{3}+{r}^{2} \bigg(  \left( {\frac { 8}{19}}\zeta +{r}^{2} \right) {e^{\nu }}\nonumber\\
&-{ \frac {200}{19}}\zeta  \bigg)  \nu'^{2}-{\frac {178}{19}}r \left(  \left( {\frac {24} {89}}\zeta +{r}^{2} \right) {e^{\nu }}-{\frac { 56}{89}}\zeta  \right) \nu' + \left( -{ \frac {160}{19}}\zeta +{\frac {20}{19}}{r}^{2} \right) {e^{\nu }}+{\frac {288}{19}}\zeta  \bigg]  \mu'^{2}+76r \bigg[ {\frac {12}{19}} \zeta  \nu'^{3}{r}^{ 3}+{r}^{2} \bigg(  \left( {\frac {8}{19}}\zeta +{r}^{2} \right) { e^{\nu }}\nonumber\\
&-{\frac {40}{19}}\zeta  \bigg) \nu'^{2}-{\frac {97}{38 }}r \left( {e^{\nu }}{r}^{2}+{\frac {32}{97}} \zeta  \right) \nu' + \left( -{\frac {29} {19}}{r}^{2}-{\frac {64}{19}}\zeta  \right) {e^{\nu }}+{\frac {56}{19}}\zeta  \bigg] \mu'+64\zeta  \nu'^{3}{r}^{3}+152{r}^{2} \left(  \left( {\frac {8}{19}} \zeta +{r}^{2} \right) {e^{\nu }}-{\frac {8}{19}} \zeta  \right)  \nu'^{ 2}\nonumber\\
\end{align*}
\begin{align}\label{pt}
&-56r \left(  \left( {r}^{2}-{\frac {16}{7}}\zeta  \right) { e^{\nu }}+{\frac {16}{7}}\zeta  \right) \nu' + \left( -192{r}^{2}-256\zeta  \right) {e^{\nu }}+128\zeta  \bigg] \zeta  \bigg\} \zeta \nu'' -4\zeta  \bigg[ {r}^{5} \left( r \left( {r}^{2}+\zeta  \right) \nu' +1/2{r}^{2}+2\zeta  \right)  \mu'^{3}-4{r}^{2}\nonumber\\
& \left( {r}^{4} \left( {r}^{2}+\zeta  \right)  \nu'^{2}+ \left( 4r{\zeta }^{2}-\frac{5}4{r}^{5}+\frac{11}2{r}^{3} \zeta  \right) \nu' -\frac{7}4{r}^{4}-\zeta  {r}^{2}+4{\zeta }^{2} \right)  \mu'^{2}+3r \bigg( {r}^{5} \left( {r}^{2}+\zeta  \right)  \nu'^{3}+ \left( \frac{16}3{\zeta }^{2}{r}^{2}+\frac{8}3{r}^{4}\zeta -{\frac {35}{6}} {r}^{6} \right)  \nu'^{2 }\nonumber\\
&+ \left( -16r{\zeta }^{2}-{\frac {116}{3}}{r}^{3}\zeta -{\frac { 17}{3}}{r}^{5} \right) \nu' -40\zeta  {r}^{2}-{\frac {128}{3}}{\zeta }^{2}+2{r}^{4} \bigg) \mu' + \left( 12{r}^{7}+12\zeta {r}^{5} \right)  \nu'^{3}+ \left( 104{r}^{4}\zeta +2{r}^{6}+64{\zeta }^{2}{r}^{2} \right) \nu'^{2}\nonumber\\
&+ \left( -30{ r}^{5}+64r{\zeta }^{2} \right) \nu' - 192{\zeta }^{2}-256\zeta {r}^{2}-28{r}^{4} \bigg] {e^{2 \nu }}+ \bigg(  \left( -2{r}^{8}+4{r}^{6}\zeta  \right) \mu'^{2}+2{r }^{3} \left( {r}^{3} \left( {r}^{2}-2\zeta  \right) \nu' +28\zeta {r}^{2}+16{\zeta }^{2}-2{r}^{4} \right) \mu'\nonumber\\
&+ \left( -56\zeta {r}^{ 5}-32{\zeta }^{2}{r}^{3}+4{r}^{7} \right) \nu' -256{\zeta }^{3}-512{\zeta }^{2}{r}^{2}-96{r}^{4} \zeta  \bigg) {e^{3\nu }}+ \bigg[ -16{r}^{4} {e^{4\nu }}+ \bigg\{ \zeta \nu'^{2}\mu'^{6}{r}^{8}-3{r}^{6} \nu'  \bigg( \zeta {r}^{2} \nu'^{2}-4\zeta  \nu'r-\frac{8}3\zeta \nonumber\\
&-1/3{e^{\nu }}{r}^{2} \bigg)  \mu'^{5}+3{r}^{4} \bigg( \zeta  \nu'^{4}{r}^{4}-12\zeta \nu'^{3}{r}^{3}-11/3{r}^{2} \left( \left( {r}^{2}+4/11\zeta  \right) {e^{\nu }}-{ \frac {36}{11}}\zeta  \right)  \nu'^{2}+10/3r \left(  \left( -8/5\zeta +{r}^{2} \right) {e^{\nu }}+8\zeta  \right) \nu'\nonumber\\
& +4/3{e^{\nu }}{r}^{2}+16 /3\zeta  \bigg) \mu'^{4}-{r}^{3} \bigg[\zeta  \nu'^{5}{r}^{5}-36\zeta  \nu'^{4}{r}^{4}+ \left(  \left( -19{r}^{5}-8{r}^{3} \zeta  \right) {e^{\nu }}+144{r}^{3}\zeta  \right)  \nu'^{3}+ \left( 64\zeta {r}^{2}+92{r}^{4}{e^{\nu }} \right)  \nu'^{2}\nonumber\\
&+36r \left(  \left( {\frac {40}{9}}\zeta +{r}^{2} \right) {e^{\nu }}-{\frac {68}{9}}\zeta  \right) \nu' + \left( 64\zeta -16{r}^{2} \right) {e^{\nu }}-128\zeta  \bigg] \mu'^{3}-9{r}^{2} \bigg[4/3\zeta  \nu'^{5}{r}^{5}+{r}^{4} \left(  \left( {r}^{2}+4/9\zeta  \right) {e^{\nu }}-{\frac {148}{9}}\zeta  \right)  \nu'^{4}\nonumber\\
&-{\frac {154}{9}}{r}^{3} \left( \left( {r}^{2}+{\frac {24}{77}}\zeta  \right) {e^{\nu }}-{\frac {8}{7}}\zeta  \right)  \nu'^{3}+{\frac {148}{9}}{r}^{2} \left( \left( {r}^{2}-{\frac {56}{37}}\zeta  \right) {e^{\nu }}+{\frac {120}{37}}\zeta  \right)  \nu'^{2}+{\frac {352}{9}}r \left(  \left( {r} ^{2}+{\frac {14}{11}}\zeta  \right) {e^{\nu }}- {\frac {10}{11}}\zeta  \right) \nu'\nonumber\\
& + \left( {\frac {448}{9}}\zeta +{\frac {112}{9}}{r}^{2} \right) { e^{\nu }}-{\frac {128}{3}}\zeta  \bigg] \mu'^{2}-72r \bigg[ 2 /3\zeta  \nu'^{5}{r} ^{5}+{r}^{4} \left(  \left( {r}^{2}+4/9\zeta  \right) {e^{\nu }}-{\frac {28}{9}}\zeta  \right)   \nu'^{4}+ \left( -9/2{r}^{5}{e^{ \nu }}-{\frac {16}{9}}{r}^{3}\zeta  \right)  \nu'^{3}\nonumber\\
&-4{r}^{2} \left( \left( {r}^{2}+8/3\zeta  \right) {e^{\nu }}-{ \frac {22}{9}}\zeta  \right)  \nu'^{2}+{\frac {62}{9}}r \left(  \left( {r}^{2}+{ \frac {8}{31}}\zeta  \right) {e^{\nu }}+{\frac {4}{31}}\zeta  \right) \nu' + \left( { \frac {128}{9}}\zeta +{\frac {64}{9}}{r}^{2} \right) {e^{\nu }}-{\frac {64}{9}}\zeta  \bigg] \mu'-64\zeta  \nu'^{5}{r}^{5}\nonumber\\
&-144 \left(  \left( {r}^{2}+4/9\zeta  \right) {e^{\nu }}-4/9\zeta  \right) {r}^{4} \nu'^{4}+48{r}^{3} \left(  \left( -16/3\zeta +{r}^{2} \right) {e^{\nu }}+16/3\zeta  \right)  \nu'^{3}+480{r}^{2} \left(  \left( {\frac {16}{15}} \zeta +{r}^{2} \right) {e^{\nu }}-{\frac {8}{15}} \zeta  \right)  \nu'^{ 2}\nonumber\\
&+32r \left(  \left( 16\zeta +{r}^{2} \right) {e^{\nu }}-8\zeta  \right) \nu' + \left( -512{r}^{2}-768\zeta  \right) {e^{\nu }}+256\zeta  \bigg\} \zeta  \bigg] \zeta \bigg\} \,,
\end{align}
\section{The form of the components of $\rho$, $p_r$ and $p_t$ adopting the  KB ansatz}\label{KBC}
\begin{align}\label{soll}
&  \rho=\frac{1}{ c^2{r}^{8}{R}^{16}\kappa} \bigg\{  -56{R}^{4} \bigg[ {\frac {12}{7}} \left( {R}^{4}-{r}^{2} \left( s_0-2s_2 \right) {R}^{2}+{r}^{4}s_0s_2 \right)  \left( {R}^{2}+{r}^{2} \left( s_0-s_2 \right) \right)  \bigg( {R}^{6}+2 \left( s_0-\frac{s_2}2 \right) {r} ^{2}{R}^{4}+\frac{{r}^{4}}3 \left( 2s_0s_2-2{s_2}^{2 }\right.\nonumber\\
&\left.+{s_0}^{2} \right) {R}^{2}+\frac{{r}^{6}s_0s_2}3 \left( s_0-s_2 \right)  \bigg) \zeta+{r}^{2} \left( {R}^{ 12}+\frac{3}7{r}^{2} \left( s_0+4s_2 \right) {R}^{10}-{\frac { 8}{7}} \left( {\frac {13}{8}}{s_2}^{2}+{s_0}^{2}-{\frac {41}{8}}s_0s_2 \right) {r}^{4}{R}^{8}+{\frac {9}{7}}{r }^{6} \left( {s_0}^{3}-\frac{16}3{s_0}^{2}s_2\right.\right.\nonumber\\
&\left.\left.+{\frac {34}{3 }}s_0{s_2}^{2}-\frac{16}3{s_2}^{3} \right) {R}^{6}+{ \frac {9}{7}}{r}^{8} \left( {\frac {223}{9}}{s_0}^{2}{s_2}^{2}+{s_0}^{4}+4{s_2}^{4}-{\frac {29}{3}}{s_0}^{3 }s_2-{\frac {172}{9}}{s_2}^{3}s_0 \right) {R}^{4}-{ \frac {20}{7}}s_0 \left( s_0-s_2 \right) {r}^{10}s_2 \left( \frac{9}5{s_2}^{2}-{\frac {37}{10}}s_0s_2\right.\right.\nonumber\\
&\left.\left.+{s_0}^{2} \right) {R}^{2}+{\frac {9{r}^{12}}{7}}{s_0}^ {2}{s_2}^{2} \left( s_0-s_2 \right) ^{2} \right) \bigg] {\zeta}^{2}X(r)+32 \left( {R}^{4}-{r}^{2} \left( s_0-2s_2 \right) {R}^{2}+{r}^{4}s_0s_2 \right) ^{2} \left( {R}^{2}+{r}^{2} \left( s_0-s_2 \right) \right) ^{2} \left( {R}^{4}+3 \left( s_0{r}^{2}{R}^{2}\right.\right.\nonumber\\
&\left.\left.-\frac{2s_2}3 \right) +{r}^{4}s_0 \left( s_0-s_2 \right)  \right) {\zeta}^{3}X_1(r)+ \left( 8 \left( 12 \left( {R}^{6}+{r}^{2}s_0{R}^{4}-1/3{r}^{4} \left( -7s_0s_2+{s_0}^{2}+4{s_2}^{2} \right) {R}^{2}+2/3{r }^{6}s_0s_2 \left( s_0-s_2 \right)  \right) {R }^{2}{\zeta}^{2}\right.\right.\nonumber\\
&\left.\left.+14  {r}^{2} \zeta\left( {R}^{8}+3/14{r}^{2} \left( s_0+13 /3s_2 \right) {R}^{6}-4/7 \left( s_0-1/2s_2 \right)  \left( s_0-{\frac {17}{4}}s_2 \right) {r}^{4}{R} ^{4}+{\frac {5}{14}}{r}^{6}s_2 \left( 4/5s_0s_2 +{s_0}^{2}-6/5{s_2}^{2} \right) {R}^{2}\right.\right.\right.\nonumber\\
&\left.\left.\left.+\frac{3}{14}{r}^{8}s_0{s_2}^{2} \left( s_0-s_2 \right)  \right)+ \left( {R}^{8}-3/4 \left( s_0-8/3s_2 \right) {r} ^{2}{R}^{6}+1/2 \left( {s_0}^{2}-{\frac {41}{2}}s_0s_2+18{s_2}^{2} \right) {r}^{4}{R}^{4}-3/2{r}^{6} \left( 4{s_0}^{2}s_2-11s_0{s_2}^{2}\right.\right.\right.\right.\nonumber\\
&\left.\left.\left.\left.+4{s_2}^{ 3}+{s_0}^{3} \right) {R}^{2}-1/4{r}^{8}s_0 \left( s_0-s_2 \right)  \left( s_0+3s_2 \right)  \left( -4 s_2+s_0 \right)  \right) {r}^{4} \right) \zeta X_2(r)+ \left(  \left( -32{R}^{4}{\zeta}^{3}+ \left( -8s_2{r}^{4}{R}^{2}-56{R}^{4}{r}^{2} \right) {\zeta}^{2}\right.\right.\right.\nonumber\\
&\left.\left.\left.-8 \left( { R}^{2}-{r}^{2} \left( s_0-s_2 \right)  \right) {r}^{4} \left( 1/4{r}^{2}s_0+{R}^{2} \right) \zeta+2{R}^{2}s_2 {r}^{8}-{R}^{4}{r}^{6} \right)X_3(r)+{R}^{4}X_4(r){r}^{6} \right) {R}^{4} \right) {R}^{8}\bigg\} \,, \nonumber\\
& { p_r}=\frac{1}{ {r}^{6}{R}^ {12}{\kappa}}\bigg\{ 16{R}^{4}\zeta \left(  \left( {r}^{2}+2\zeta \right) { R}^{8}+5/8 \left(  \left( 6/5s_2+s_0 \right) {r}^{2}+{ \frac {32}{5}} \left( s_0-1/4s_2 \right) \zeta \right) {r}^{2}{R}^{6}-\frac{1}2{r}^{4} \left(  \left( 3{s_2}^{2}+{s_0 }^{2}-{\frac {27}{4}}s_0s_2 \right) {r}^{2}\right.\right.\nonumber\\
&\left.\left.-3 \left( 5 /3s_0s_2+{s_0}^{2}-4/3{s_2}^{2} \right) \zeta \right) {R}^{4}+2 \left(  \left( -3/4s_2+s_0 \right) {r}^{2}+5/4 \left( -4/5s_2+s_0 \right) \zeta \right) s_2{r}^{6}s_0{R}^{2}+1/8 \left(  \left( s_0+3s_2 \right) {r}^{2}+4\zeta s_2 \right) {r}^{8}\right.\nonumber\\
& \left.\left( s_0-s_2 \right) {s_0}^{2} \right) X_2(r)-16 \left( {R}^{2}+1/2{r}^{2}s_0 \right)  \left( {R} ^{4}-{r}^{2} \left( s_0-2s_2 \right) {R}^{2}+{r}^{4}s_0s_2 \right)  \left( {R}^{4}+3 \left( -2/3s_2+s_0 \right) {r}^{2}{R}^{2}+{r}^{4}s_0 \left( s_0-s_2 \right)  \right) {\zeta}^{2}\nonumber\\
&\left( {R}^{2}+{r}^{2} \left( s_0- s_2 \right)  \right) X(r)- \left(  \left(  \left( 16 \zeta{r}^{2}+16{\zeta}^{2}-{r}^{4} \right) {R}^{4}-2{r}^{2} \left( {r}^{4}s_0-5 \left( s_0-2/5s_2 \right) \zeta{r}^{2}-4{\zeta}^{2}s_0 \right) {R}^{2}+2{r}^{6}s_0\zeta \left( s_0-s_2 \right)  \right) X_3(r)\right.\nonumber\\
&\left.+{R}^{4}X_4(r){r}^{4} \right) {R}^ {8}\bigg\}
\,,
\nonumber\\
& {  p_t}=\frac{1}{ {r}^{8}{R}^{16}{\kappa} }\bigg\{ 64 \left(  \left( \frac{3}2\zeta+{r}^{2} \right) {R}^{12}+ \left(  \left( \frac{5}8s_0+{\frac {11}{8}}s_2 \right) {r}^{4 }+3\zeta{r}^{2}s_0 \right) {R}^{10}-{\frac {11}{8}}{r}^{4} \left(  \left( {s_0}^{2}+{\frac {17}{11}}{s_2}^{2}-{ \frac {48}{11}}s_0s_2 \right) {r}^{2}+{\frac {8}{11}} \left( {s_0}^{2}\right.\right.\right.\nonumber\\
&\left.\left.\left.+\frac{11{s_2}^{2}}2-10s_0s_2 \right) \zeta \right) {R}^{8}+\frac{3}4{r}^{6} \left(  \left(15s_0{s_2}^{2} -{\frac { 19}{3}}{s_0}^{2}s_2+{s_0 }^{3}-{\frac {22}{3}}{s_2}^{3} \right) {r}^{2}-4 \left( {s_0}^{3}-16/3{s_0}^{2}s_2-\frac{2}3{s_2}^{3}+\frac{13}3s_0{s_2}^{2} \right) \zeta \right) {R}^{6}\right.\nonumber\\
&\left.+{\frac {11}{8}}  \left(  \left( -{\frac {172}{11}}{s_2}^{3}s_0+{\frac { 36}{11}}{s_2}^{4}+{\frac {228}{11}}{s_0}^{2}{s_2}^{ 2}-{\frac {95}{11}}{s_0}^{3}s_2+{s_0}^{4} \right) {r} ^{2}-4/11\zeta \left( {s_0}^{4}-8{s_0}^{3}s_2+10 {s_2}^{3}s_0-4{s_2}^{4} \right)  \right) {r}^{8}{R} ^{4}\right.\nonumber\\
&\left.+\frac{1}8{r}^{10} \left( s_0-s_2 \right) s_0 \left( \left( {s_0}^{3}+77s_0{s_2}^{2}-36{s_2}^{3} -26{s_0}^{2}s_2 \right) {r}^{2}+24 \left( -2/3s_2 +s_0 \right) {s_2}^{2}\zeta \right) {R}^{2}-1/8s_2 \left(  \left( s_0-9s_2 \right) {r}^{2}-4\zeta s_2 \right) {r}^{12}\right.\nonumber\\
&\left. \left( s_0-s_2 \right) ^{2}{s_0}^{ 2} \right) {R}^{4}{\zeta}^{2}X(r)-32 \left( {R}^{4}-{r}^{2} \left( s_0-2s_2 \right) {R}^{2}+{r}^{4}s_0s_2 \right) ^{2} \left( {R}^{2}+{r}^{2} \left( s_0-s_2 \right) \right) ^{2} \left( {R}^{4}+3 \left( -2/3s_2+s_0 \right) {r}^{2}{R}^{2}\right.\nonumber\\
&\left. +{r}^{4}s_0 \left( s_0-s_2 \right)  \right) {\zeta}^{3}X_1(r)+2 \left( -7 \left( \left( {r}^{4}+{\frac {64}{7}}\zeta{r}^{2}+{\frac {48}{7}}{ \zeta}^{2} \right) {R}^{8}-{\frac {8}{7}} \left(  \left( -{\frac { 15}{8}}s_2+s_0 \right) {r}^{4}-2\zeta \left( s_0 +3s_2 \right) {r}^{2}-6{\zeta}^{2}s_0 \right) {r}^{2}{R }^{6}\right.\right.\nonumber\\
&\left. \left.+\frac{4}7{r}^{4} \left(  \left( s_0-13/2s_2 \right) \left( s_0-s_2 \right) {r}^{4}+\zeta \left( -11{s_0}^{2}+42s_0s_2-19{s_2}^{2} \right) {r}^{2}-4 {\zeta}^{2} \left( -7s_0s_2+{s_0}^{2}+4{s_2 }^{2} \right)  \right) {R}^{4}+\frac{3}7 \left(  \left( {s_0}^{3}+{ \frac {71}{3}}s_0{s_2}^{2}\right.\right.\right.\right.\nonumber\\
&\left. \left.\left.\left.-8{s_2}^{3}-14{s_0}^{2}s_2 \right) {r}^{4}+4 \left( -2{s_2}^{2}+{s_0}^{2}+5/3s_0s_2 \right) s_2\zeta{r}^{2}+{ \frac {32}{3}}s_0{\zeta}^{2}s_2 \left( s_0-s_2 \right)  \right) {r}^{6}{R}^{2}-4/7{r}^{10}s_0s_2 \left( s_0-s_2 \right)  \left( s_0-3s_2 \right) \right.\right.\nonumber\\
&\left.  \left.\left( \zeta+{r}^{2} \right)  \right) \zeta X_2(r)+ \left(  \left(  \left( 16{\zeta}^{3}+32{\zeta}^{2}{r} ^{2}+6\zeta{r}^{4} \right) {R}^{4}+ \left(  \left( s_0-1/2 s_2 \right) {r}^{4}-8 \left( s_0-{\frac {7}{8}}s_2 \right) \zeta{r}^{2}-4{\zeta}^{2} \left( s_0-s_2 \right)  \right) {r}^{4}{R}^{2}+\right.\right.\right.\nonumber\\
&\left. \left.\left.1/2{r}^{8}s_0 \left( -2 \zeta+{r}^{2} \right)  \left( s_0-s_2 \right)  \right)X_3(r)+{R}^{4}X_4(r)\zeta{r}^{4} \right) {R}^{4} \right) {R}^{8}
  \bigg\}\,,
\end{align}
where $X_i(r)$ and $i=1 \cdots 4$ are defined as
\begin{align}
&X(r)={e^{ \left[ -2 \left\{ {R}^{4}+3 \left( -2/3s_2+s_0 \right) {r}^{2}{R}^{2}+{r}^{4}s_0 \left( s_0-s_2 \right)  \right\} \zeta{e^{-{\frac { s_2{r}^{2}}{{R}^{2}}}}}+2\zeta{R}^{4}-3s_2{r}^{4} {R}^{2} \right] {R}^{-4}{r}^{-2}}}\,,\nonumber\\
&X_1(r)={e^{ \left[ -2 \left\{ {R}^{4}+3  \left( -2/3s_2+s_0 \right) {r}^{2}{R}^{2}+{r}^{4}s_0 \left( s_0-s_2 \right)  \right\} \zeta{e^{-{ \frac {s_2{r}^{2}}{{R}^{2}}}}}+2\zeta{R}^{4}-4s_2 {r}^{4}{R}^{2} \right] {R}^{-4}{r}^{-2}}}\,,\nonumber\\
&X_2(r)={e^{ \left[ -2 \left\{ {R}^{4}+3 \left( -2/3s_2+s_0 \right) {r}^{2}{R}^{2}+{r}^{4}s_0 \left( s_0-s_2 \right)  \right\} \zeta{e^{-{\frac {s_2{r}^{2}}{{R}^{2 }}}}}+2\zeta{R}^{4}-2s_2{r}^{4}{R}^{2} \right] {R}^{-4}{ r}^{-2}}}\,,\nonumber\\
&X_3(r)= {e^{ \left[ -2 \left\{ {R}^{4} +3 \left( -2/3s_2+s_0 \right) {r}^{2}{R}^{2}+{r}^{4}s_0 \left( s_0-s_2 \right)  \right\} \zeta{e^{ -{\frac {s_2{r}^{2}}{{R}^{2}}}}}+2\zeta{R}^{4}-s_2{ r}^{4}{R}^{2} \right] {R}^{-4}{r}^{-2}}}\,,\nonumber\\
&X_4(r)={e^{-2 \left[ \left\{ {R}^{4}+3 \left( -2/3s_2+s_0 \right) {r}^{2}{R}^ {2}+{r}^{4}s_0 \left( s_0-s_2 \right)  \right\} { e^{-{\frac {s_2{r}^{2}}{{R}^{2}}}}}-{R}^{4} \right] \zeta{R}^{-4}{r}^{-2}}}\,.\nonumber\\
&
\end{align}

It has been shown that the KB ansatz relates the pressures and the density which effectively induces the EoSs as given by Eqs. {\color{blue} (19)} and   {\color{blue} (20)}. The coefficients in those equations are related to the model parameters as listed as below.
\begin{align}
&a_1=- \left\{ \left( -23616{\zeta_1}^{3}{s_2}^{3}{s_0}^{2 }+16704{\zeta_1}^{3}{s_2}^{2}{s_0}^{3}-5184{\zeta_1}^{3}s_2{s_0}^{4}+15120{\zeta_1}^{3}{s_2}^{4}s_0+576{\zeta_1}^{3}{s_0}^{5}-3600{\zeta_1}^{3}{s_2}^{5}-9252{\zeta_1}^{2}{s_0}^{2}{s_2}^{2}\right.\right.\nonumber\\
&\left.\left.+8988{\zeta_1}^{2}s_0{s_2}^{3}+3384
{\zeta_1}^{2}{s_0}^{3}s_2-2832{\zeta_1}^{2}{s_2}^{4}-360{\zeta_1}^{2}{s_0}^{4}+24\zeta_1{s_0}^{3}-362\zeta_1{s_2}^{3}+822\zeta_1s_0{s_2}^{2}-408\zeta_1{s_0}^{2}s_2+12s_0s_2\right.\right.\nonumber\\
&\left.\left.-3{ s_2}^{2} \right)\right\}\left\{-59040{\zeta_1}^{3}{s_2}^{3}{s_0}^{2}+41760{\zeta_1}^{3} {s_2}^{2}{s_0}^{3}-12960{\zeta_1}^{3}s_2{s_0}^{4}+37800{\zeta_1}^{3}{s_2}^{4}s_0+1440{\zeta_1}^{3}{s_0}^{5}-9000{\zeta_1}^{3}{s_2}^{5}-7620 {\zeta_1}^{2}{s_2}^{4}\right.\nonumber\\
&\left.-504{\zeta_1}^{2}{s_0}^{ 4}+22884{\zeta_1}^{2}s_0{s_2}^{3}+6696{\zeta_1}^{2}{s_0}^{3}s_2-21636{\zeta_1}^{2}{s_0} ^{2}{s_2}^{2}-72\zeta_1{s_0}^{3}+1878\zeta_1 s_0{s_2}^{2}-1130\zeta_1{s_2}^{3}-384\zeta_1{s_0}^{2}s_2\right. \nonumber\\
&\left.-15{s_2}^{2}\right\}^{-1}\,,\nonumber\\
&a_2=2\left\{ \left( -19642{\zeta_1}^{2}{s_2}^{4}s_0+ 4080{\zeta_1}^{2}s_2{s_0}^{4}+27112{\zeta_1} ^{2}{s_2}^{3}{s_0}^{2}-16116{\zeta_1}^{2}{s_2}^{ 2}{s_0}^{3}-3414528{\zeta_1}^{5}{s_2}^{3}{s_0}^{ 5}+6113664{\zeta_1}^{5}{s_2}^{4}{s_0}^{4}\right.\right.\nonumber\\
&\left.\left.+4207680{\zeta_1}^{5}{s_2}^{6}{s_0}^{2}-1468800{\zeta_1}^{ 5}{s_2}^{7}s_0+1112832{\zeta_1}^{5}{s_2}^{2}{s_0}^{6}-193536{\zeta_1}^{5}{s_0}^{7}s_2-1232352 {\zeta_1}^{4}s_0{s_2}^{6}-3180672{\zeta_1}^{ 4}{s_0}^{3}{s_2}^{4}\right.\right.\nonumber\\
&\left.\left.-632448{\zeta_1}^{4}{s_0}^{5 }{s_2}^{2}+2768832{\zeta_1}^{4}{s_0}^{2}{s_2}^{5 }+94464{\zeta_1}^{4}{s_0}^{6}s_2+1966464{\zeta_1}^{4}{s_0}^{4}{s_2}^{3}-251040{\zeta_1}^{3}{s_2}^{5}s_0+420024{\zeta_1}^{3}{s_2}^{4}{s_0 }^{2}\right.\right.\nonumber\\
&\left.\left.+110448{\zeta_1}^{3}{s_2}^{2}{s_0}^{4}-320784{ \zeta_1}^{3}{s_2}^{3}{s_0}^{3}-15264{\zeta_1}^{3 }{s_0}^{5}s_2+1056\zeta_1{s_0}^{2}{s_2}^{ 2}-704\zeta_1s_0{s_2}^{3}-276\zeta_1{s_0}^{3}s_2+3{s_2}^{3}+3s_0{s_2}^{2}\right.\right.\nonumber\\
&\left.\left.-72 {\zeta_1}^{2}{s_0}^{5}+5152{\zeta_1}^{2}{s_2}^ {5}+216000{\zeta_1}^{5}{s_2}^{8}+13824{\zeta_1}^{5} {s_0}^{8}-4608{\zeta_1}^{4}{s_0}^{7}+220320{\zeta_1}^{4}{s_2}^{7}+56040{\zeta_1}^{3}{s_2}^{6}+288 {\zeta_1}^{3}{s_0}^{6}-72\zeta_1{s_0}^{4}\right.\right.\nonumber\\
&\left.\left.+ 142\zeta_1{s_2}^{4}-6587136{\zeta_1}^{5}{s_2 }^{5}{s_0}^{3} \right) {e^{-6 \left( -s_2+s_0 \right) \zeta_1}}\right\}\left\{ \left( -59040{\zeta_1}^{3}{s_2}^ {3}{s_0}^{2}+41760{\zeta_1}^{3}{s_2}^{2}{s_0}^{3 }-12960{\zeta_1}^{3}s_2{s_0}^{4}+37800{\zeta_1}^{3}{s_2}^{4}s_0\right.\right.\nonumber\\
&\left.\left.+1440{\zeta_1}^{3}{s_0}^ {5}-9000{\zeta_1}^{3}{s_2}^{5}-7620{\zeta_1}^{2}{s_2}^{4}-504{\zeta_1}^{2}{s_0}^{4}+22884{\zeta_1}^{2}s_0{s_2}^{3}+6696{\zeta_1}^{2}{s_0}^{3}s_2-21636{\zeta_1}^{2}{s_0}^{2}{s_2}^{2}-72\zeta_1{s_0}^{3}\right.\right.\nonumber\\
&\left.\left.+1878\zeta_1s_0{s_2}^{ 2}-1130\zeta_1{s_2}^{3}-384\zeta_1{s_0}^{2 }s_2-15{s_2}^{2} \right) {R}^{2}{\kappa}^{2}\right\}^{-1}\,,\nonumber\\
&a_3=-2\left\{-23616{\zeta_1}^{3}{s_0}^{2}{s_2}^{3}+ 16704{\zeta_1}^{3}{s_0}^{3}{s_2}^{2}+15120{\zeta_1}^{3}s_0{s_2}^{4}-5184{\zeta_1}^{3}{s_0 }^{4}s_2-3600{\zeta_1}^{3}{s_2}^{5}+576{\zeta_1}^{3}{s_0}^{5}-2832{\zeta_1}^{2}{s_2}^{4}\right.\nonumber\\
&\left.+8898{\zeta_1}^{2}{s_2}^{3}s_0-9018{\zeta_1}^{2}{s_2}^{2}{s_0}^{2}+3204{\zeta_1}^{2}{s_0}^{3}s_2- 324{\zeta_1}^{2}{s_0}^{4}+30\zeta_1{s_0}^{3} -362\zeta_1{s_2}^{3}-360\zeta_1s_2{s_0}^{2}+774\zeta_1{s_2}^{2}s_0-3{s_2}^{2}\right.\nonumber\\
&\left.+9 s_0s_2-3{s_0}^{2}\right\}\left\{-72\zeta_1{s_0}^ {3}+6696{\zeta_1}^{2}{s_0}^{3}s_2+1878\zeta_1 {s_2}^{2}s_0-384\zeta_1s_2{s_0}^{2}- 1130\zeta_1{s_2}^{3}-7620{\zeta_1}^{2}{s_2}^ {4}-504{\zeta_1}^{2}{s_0}^{4}\right.\nonumber\\
&\left.+37800{\zeta_1}^{3}s_0{s_2}^{4}-59040{\zeta_1}^{3}{s_0}^{2}{s_2}^{3}+41760{\zeta_1}^{3}{s_0}^{3}{s_2}^{2}-12960 {\zeta_1}^{3}{s_0}^{4}s_2-15{s_2}^{2}+22884{\zeta_1}^{2}{s_2}^{3}s_0-21636{\zeta_1}^{2}{s_2}^{2}{s_0}^{2}\right.\nonumber\\
&\left.+1440{\zeta_1}^{3}{s_0}^{5}-9000{\zeta_1}^{3}{s_2}^{5}\right\}^{-1}\,,\nonumber\\
&a_4=-\left\{ \left( -504{\zeta_1}^{2}{s_0}^{5}-6587136{\zeta_1}^{5}{s_2}^{5}{s_0}^{3}-3414528{\zeta_1}^{ 5}{s_2}^{3}{s_0}^{5}+6113664{\zeta_1}^{5}{s_2}^{ 4}{s_0}^{4}-1468800{\zeta_1}^{5}{s_2}^{7}s_0+ 4207680{\zeta_1}^{5}{s_2}^{6}{s_0}^{2}\right.\right.\nonumber\\
&\left.\left.+1112832{\zeta_1}^{5}{s_2}^{2}{s_0}^{6}-193536{\zeta_1}^{5 }{s_0}^{7}s_2+3{s_2}^{3}-767904{\zeta_1}^{3}{s_2}^{3}{s_0}^{3}-330\zeta_1{s_2}^{2}{s_0} ^{2}+54616{\zeta_1}^{2}{s_2}^{3}{s_0}^{2}-47232{\zeta_1}^{3}{s_0}^{5}s_2\right.\right.\nonumber\\
&\left.\left.+281952{\zeta_1}^{3}{s_2}^{2}{s_0}^{4}-60\zeta_1{s_0}^{3}s_2- 592752{\zeta_1}^{3}{s_2}^{5}s_0+988104{\zeta_1 }^{3}{s_2}^{4}{s_0}^{2}+982\zeta_1{s_2}^{4}- 1364\zeta_1{s_2}^{3}s_0-12876{\zeta_1}^{2}{s_2}^{2}{s_0}^{3}\right.\right.\nonumber\\
&\left.\left.-24{\zeta_1}^{2}s_2{s_0}^ {4}+3086208{\zeta_1}^{4}{s_0}^{4}{s_2}^{3}+208512{\zeta_1}^{4}{s_0}^{6}s_2-1130112{\zeta_1}^{4}{s_0}^{5}{s_2}^{2}+3803040{\zeta_1}^{4}{s_0}^{2}{s_2}^{5}-4603680{\zeta_1}^{4}{s_0}^{3}{s_2}^{4}\right.\right.\nonumber\\
&\left.\left.- 1634112{\zeta_1}^{4}s_0{s_2}^{6}+285120{\zeta_1}^{4}{s_2}^{7}-14976{\zeta_1}^{4}{s_0}^{7}+18 {s_0}^{2}s_2-24s_0{s_2}^{2}+180\zeta_1{s_0}^{4}+20452{\zeta_1}^{2}{s_2}^{5}+133800{\zeta_1}^{3}{s_2}^{6}\right.\right.\nonumber\\
&\left.\left.+3744{\zeta_1}^{3}{s_0}^{6}- 61420{\zeta_1}^{2}{s_2}^{4}s_0+216000{\zeta_1} ^{5}{s_2}^{8}+13824{\zeta_1}^{5}{s_0}^{8} \right) { e^{-6 \left( -s_2+s_0 \right) \zeta_1}}\right\}\left\{{R}^{2 } \left( -72\zeta_1{s_0}^{3}+6696{\zeta_1}^{2}{s_0}^{3}s_2+1878\zeta_1{s_2}^{2}s_0\right.\right.\nonumber\\
&\left.\left.-384\zeta_1s_2{s_0}^{2}-1130\zeta_1{s_2}^{ 3}-7620{\zeta_1}^{2}{s_2}^{4}-504{\zeta_1}^{2}{s_0}^{4}+37800{\zeta_1}^{3}s_0{s_2}^{4}-59040 {\zeta_1}^{3}{s_0}^{2}{s_2}^{3}+41760{\zeta_1}^{ 3}{s_0}^{3}{s_2}^{2}-12960{\zeta_1}^{3}{s_0}^{4} s_2\right.\right.\nonumber\\
&\left.\left.-15{s_2}^{2}+22884{\zeta_1}^{2}{s_2}^{3}s_0-21636{\zeta_1}^{2}{s_2}^{2}{s_0}^{2}+1440{\zeta_1}^{3}{s_0}^{5}-9000{\zeta_1}^{3}{s_2}^{5} \right) {\kappa}^{2}\right\}^{-1}\,.
\end{align}
Using the above set of equations, one can find the physical quantities appear in Eq. {\color{blue} (19)} and   {\color{blue} (20)} in terms of the model parameters, where $v_r^2=c_1$, $\rho_1=\rho_s=-c_2/c_1$, $v_t^2=c_3$ and $\rho_2=-c_4/c_3$.
\section{The density and pressures gradients}\label{Sec:App_2}
Recalling the matter density and the pressures as obtained for the quadratic polynomial $f(R)$ gravity, namely Eqs. \eqref{rho}, \eqref{pr}, and \eqref{pt} we obtain the gradients of these quantities with respect to the radial distance as below
\begin{align}\label{eq:dens_grad}
&\rho'=\frac{1}{{r}^{11}{R}^{14}{c}^{2}{\kappa}^{2}}\bigg\{ -512{R}^{2} \left( {R}^{4}-{r}^{2} \left( s_0-2s_2 \right) {R}^{2}+{r}^{4}s_0s_2 \right)  \bigg[ \left( {R}^{6}+\frac{9}4{r}^{2} \left( s_0-\frac{5}9s_2 \right) {R}^{4}+\frac{1}2 {r}^{4} \left({s_0}^{2} -{s_2}^{2}+\frac{1}2s_0s_2 \right) {R}^{2}\right.\nonumber\\
&\left.+1/4{r}^{6}s_0s_2 \left(s_0  -s_2\right)  \right) {R}^{2} \left( {R}^{4}-{r}^{2} \left( s_0-2s_2 \right) {R}^{2}+{r}^{4}s_0s_2 \right) \left( {R}^{2}+{r}^{2} \left( -s_2+s_0 \right)  \right) \zeta_1+{\frac {15}{16}}{r}^{2} \left( {R}^{12}+7/5{r}^{2} \left( s_0+4/7s_2 \right) {R}^{10}\right.\nonumber\\
&\left.-{\frac {4}{15}}{r}^{ 4} \left( {\frac {25}{4}}{s_2}^{2}+{s_0}^{2}-{\frac {67}{4 }}s_0s_2 \right) {R}^{8}+{r}^{6} \left( {s_0}^{3}- \frac{16}3s_2{s_0}^{2}-{\frac {88}{15}}{s_2}^{3}+{ \frac {182}{15}}s_0{s_2}^{2} \right) {R}^{6}+{\frac {13} {15}}{r}^{8} \left({\frac {68}{13}}{s_2}^ {4} -{\frac {320}{13}}{s_2}^{3}s_0-{ \frac {149}{13}}{s_0}^{3}s_2\right.\right.\nonumber\\
&\left.\left.+{s_0}^{4}+{\frac {405}{13}}{s_2}^{2}{s_0}^{2} \right) {R}^{4}-{\frac {12}{5}} \left( -s_2+s_0 \right) { r}^{10}s_2 \left( {\frac {17}{9}}{s_2}^{2}-{\frac {23}{6 }}s_0s_2+{s_0}^{2} \right) s_0{R}^{2}+{ \frac {17}{15}}{r}^{12}{s_0}^{2}{s_2}^{2} \left( -s_2 +s_0 \right) ^{2} \right)  \bigg]X_5 {\zeta_1}^{3} \left( {R}^ {2}\right.\nonumber\\
&\left.+{r}^{2} \left( s_0-s_2 \right)  \right) +768{R}^{4}{\zeta_1}^{2} \left( {R}^{6} \left( {R}^{4}-{r} ^{2} \left( s_0-2s_2 \right) {R}^{2}+{r}^{4}s_0s_2 \right)  \left( {R}^{6}+3/2{r}^{2} \left( s_0-1/3s_2 \right) {R}^{4}+3/2{r}^{4} \left( -2/3s_2+s_0 \right) s_2{R}^{2}\right.\right.\nonumber\\
&\left.\left.+\frac{{r}^{6}}2s_0s_2 \left( s_0-s_2 \right)  \right)  \left( {R}^{2}+{r}^{2} \left( s_0-s_2 \right)  \right) {\zeta_1}^{2}+{\frac {15}{8}} {r}^{2}{R}^{2} \left( {R}^{14}+{\frac {14}{15}}{r}^{2} \left( s_0+{\frac {19}{14}}s_2 \right) {R}^{12}-{\frac {38}{45}}{r}^ {4} \left( {s_0}^{2}-{\frac {127}{19}}s_0s_2+{ \frac {89}{38}}{s_2}^{2} \right) {R}^{10} \right.\right.\nonumber\\
&\left.\left.-\frac{2{r}^{6}}{15} \left({s_0}^{3}+{\frac {155}{6}}{s_2}^{3}-{\frac {122}{3}}s_0{s_2}^{2}-\frac{7}3s_2{s_0}^{2} \right) {R}^{8}+ {\frac {17}{45}}{r}^{8} \left( {\frac {22}{17}}{s_2}^{2}-{ \frac {88}{17}}s_0s_2+{s_0}^{2} \right)  \left( 4 {s_2}^{2}-6s_0s_2+{s_0}^{2} \right) {R}^{6}-{ \frac {13}{15}}{r}^{10} \right.\right.\nonumber\\
&\left.\left.\times\left({\frac {50}{39}}{s_2}^{ 3}s_0 -{\frac {12}{13}}{s_2}^{4}-{ \frac {64}{13}}{s_0}^{3}s_2+{s_0}^{4}+\frac{10}3{s_2}^{2}{s_0}^{2} \right) s_2{R}^{4}+{\frac {7}{45}} \left(s_0 -s_2 \right) { r}^{12} \left( -{\frac {36}{7}}{s_2}^{2}+{\frac {47}{7}}s_0s_2+{s_0}^{2} \right) {s_2}^{2}s_0{R}^{2}\right.\right.\nonumber\\
&\left.\left.+1 /5{r}^{14}{s_0}^{2}{s_2}^{3} \left( -s_2+s_0 \right) ^{2} \right) \zeta_1+{\frac {23}{48}}{r}^{4} \left( {R }^{14}+{\frac {9}{46}}{r}^{2} \left( {\frac {34}{3}}s_2+s_0 \right) {R}^{12}-{\frac {9}{23}}{r}^{4} \left( {s_0}^{2}-{ \frac {41}{9}}{s_2}^{2}-4s_0s_2 \right) {R}^{10}- {\frac {3}{46}}{r}^{6} \left( {\frac {86}{3}}s_2{s_0}^ {2}\right.\right.\right.\nonumber\\
&\left.\left.\left.+{s_0}^{3}-{\frac {305}{3}}s_0{s_2}^{2}+{\frac { 82}{3}}{s_2}^{3} \right) {R}^{8}-{\frac {21}{46}}{r}^{8} \left( {\frac {152}{7}}{s_2}^{4}-{\frac {90}{7}}{s_0}^{ 3}s_2+{s_0}^{4}-{\frac {1228}{21}}{s_2}^{3}s_0+{ \frac {940}{21}}{s_2}^{2}{s_0}^{2} \right) {R}^{6}+{\frac {3}{23}}{r}^{10} \left( 367{s_0}^{2}{s_2}^{3}\right.\right.\right.\nonumber\\
&\left.\left.\left.+44{s_2}^{5}-{\frac {363}{2}}{s_0}^{3}{s_2}^{2}+{s_0}^{5} -242{s_2}^{4}s_0+{\frac {45}{2}}{s_0}^{4}s_2 \right) {R}^{4}+1/46{r}^{12}s_0 \left( -s_2+s_0 \right)  \left( -264{s_2}^{4}+610{s_2}^{3}s_0-204 {s_2}^{2}{s_0}^{2}+{s_0}^{ 4}\right.\right.\right.\nonumber\\
&\left.\left.\left.-11{s_0}^{3}s_2 \right) {R}^{2}-1/46{r}^{14}{s_0}^{2}s_2 \left( {s_0}^{2}-s_0s_2-66{s_2}^{2} \right)  \left( -s_2+s_0 \right) ^{2} \right)  \right) X_5+128 \left( {R}^{4}-{r}^{2} \left( s_0-2s_2 \right) {R}^{2}+{r }^{4}s_0s_2 \right) ^{3} \left( {R}^{4}+3{r}^{2} \left(s_0 \right.\right.\nonumber\\
&\left.\left.-2/3s_2+ \right) {R}^{2}+{r}^{4}s_0 \left( -s_2+s_0 \right)  \right) {\zeta_1}^{4} \left( {R}^{2}+{r }^{2} \left( -s_2+s_0 \right)  \right) ^{3}X_6-512{R}^{6} \left(  \left( {R}^{10} \left( {R}^{6}+3/4{r}^{2} \left( s_0+1/3s_2 \right) {R}^{4}-1/2{r}^{4} \left( 3 {s_2}^{2}\right.\right.\right.\right.\nonumber\\
&\left.\left.\left.\left.+{s_0}^{2}-\frac{11}2s_0s_2 \right) {R}^{2} +\frac{3}4{r}^{6}s_0s_2 \left(s_0 -s_2 \right) \right) {\zeta_1}^{3}+{\frac {45}{16}}{r}^{2}{R}^{6} \left( {R }^{8}+{\frac {7}{15}}{r}^{2} \left( s_0+{\frac {11}{7}}s_2 \right) {R}^{6}-{\frac {19}{45}}{r}^{4} \left( {s_0}^{2}-{ \frac {106}{19}}s_0s_2+{\frac {46}{19}}{s_2}^{2} \right) {R}^{4}\right.\right.\right.\nonumber\\
&\left.\left.\left.+{\frac {11}{45}}{r}^{6} \left( s_0-{\frac {8}{ 11}}s_2 \right) s_2 \left( 3s_2+s_0 \right) {R}^{2}+{\frac {4}{15}}{r}^{8}s_0{s_2}^{2} \left( -s_2+s_0 \right)  \right) {\zeta_1}^{2}+{\frac {23}{16}}{r} ^{4}{R}^{2} \left( {R}^{10}+{\frac {3}{46}}{r}^{2} \left( s_0+{ \frac {62}{3}}s_2 \right) {R}^{8}-{\frac {19}{92}}{r}^{4} \left({s_0}^{2} \right.\right.\right.\right.\nonumber\\
&\left.\left.\left.\left.-{\frac {61}{19}}s_0s_2-{\frac {50}{19}}{s_2}^{2} \right) {R}^{6}-{\frac {3}{92}}{r}^{6} \left( 23s_2{s_0}^{2}+{\frac {100}{3}}{s_2}^{3}+ {s_0}^{3}-79s_0{s_2}^{2} \right) {R}^{4}-{\frac {3} {46}}{r}^{8}s_2 \left( -{\frac {19}{3}}s_2{s_0} ^{2}+4{s_2}^{3}-2/3s_0{s_2}^{2}+{s_0}^{3} \right) {R}^{2}\right.\right.\right.\nonumber\\
&\left.\left.\left.-{\frac {1}{92}}{r}^{10}s_0s_2 \left( -s_2+s_0 \right)  \left( 3s_2+s_0 \right) \left( -4s_2+s_0 \right)  \right) \zeta_1+{\frac {9} {128}}{r}^{6} \left( {R}^{10}-1/3{r}^{2} \left( s_0-5s_2 \right) {R}^{8}-1/9{r}^{4} \left( 2s_0s_2-12{s_2}^{2}+{s_0}^{2} \right) {R}^{6}\right.\right.\right.\nonumber\\
&\left.\left.\left.+2/3{r}^{6} \left( { \frac {31}{6}}s_2{s_0}^{2}+{\frac {50}{3}}{s_2}^{ 3}+{s_0}^{3}-{\frac {157}{6}}s_0{s_2}^{2} \right) { R}^{4}+2/9{r}^{8} \left( {s_0}^{4}-24{s_2}^{4}-8{s_0}^{3}s_2+79{s_2}^{3}s_0-36{s_2}^{2}{s_0}^{2} \right) {R}^{2}-2/9{r}^{10}s_0s_2 \left( s_0\right.\right.\right.\right.\nonumber\\
&\left.\left.\left.\left.-s_2 \right)  \left( 3s_2+s_0 \right)  \left( -4 s_2+s_0 \right)  \right)  \right) \zeta_1X_7-\frac{1}4{R}^{4} \left(  \left( {R}^{12}{\zeta_1}^{4}+ \left(\frac{3}4{R}^{8}{r}^{4}s_2+{\frac {15}{4}}{r}^{2}{R}^{10} \right) {\zeta_1}^{3}+{\frac {23}{8}}{r}^{4}{R}^{4} \left( {R}^{4}-{ \frac {3}{46}}{r}^{2} \left( s_0-{\frac {22}{3}}s_2 \right) {R}^{2}\right.\right.\right.\right.\nonumber\\
&\left.\left.\left.\left.-1/46{r}^{4} \left( s_0+s_2 \right) \left( s_0-2s_2 \right)  \right) {\zeta_1}^{2}+{ \frac {5}{16}}{r}^{6}{R}^{2} \left( {R}^{4}-3/10{r}^{2} \left( s_0-7/3s_2 \right) {R}^{2}-1/10{r}^{4} \left( s_0+s_2 \right)  \left( s_0-2s_2 \right)  \right) \zeta_1+{\frac {1}{64}}{r}^{8} \left( {R}^{2}\right.\right.\right.\right.\nonumber\\
&\left.\left.\left.\left.+2{r}^{2}s_2 \right)  \left( {R}^{2}-{r}^{2}s_2 \right)  \right) X_8-1/32{r}^{6} \left( \zeta_1{R}^{2}+1/2{r}^{2} \right) {R} ^{4}X_9 \right)  \right) \bigg\}\,,
\end{align}
\begin{align}\label{eq:pr_grad}
  & p'_r=\frac{1}{ {R}^{12}{r}^{9}{\kappa}^{2}}\left( 192{R}^{2} \left( \zeta_1{R}^{16}+\frac{5}2{r}^{2} \left( \zeta_1s_0+\frac{1}3 \right) {R}^{14}+\frac{{r}^{4}}3 \left(  \left( 20s_0s_2-11{s_2}^{2}+{s_0}^{ 2} \right) \zeta_1+\frac{9}4s_2+{\frac {33}{8}}s_0 \right) {R}^{12}-\frac{7}3{r}^{6} \left(  \left(\frac{9}2s_0{s_2}^{2}\right.\right.\right.\right.\nonumber\\
&\left.\left.\left.\left. -6s_2{s_0} ^{2} +{s_0}^{3}-\frac{4}7{s_2}^{3} \right) \zeta_1+{\frac {3}{28}}{s_0}^{2}-{\frac {16}{7}} s_0s_2+{\frac {25}{28}}{s_2}^{2} \right) {R}^{10} -\frac{4}3{r}^{8} \left(  \left( {\frac {13}{4}}{s_2}^{2}{s_0} ^{2}-6{s_0}^{3}s_2+2{s_2}^{3}s_0+{s_0}^{4 }-{s_2}^{4} \right) \zeta_1+9/4{s_2}^{3}\right.\right.\right.\nonumber\\
&\left.\left.\left. -{\frac {13}{8}}s_2{s_0}^{2}-{\frac {87}{32}}s_0{s_2}^{2}+{\frac {5}{32} }{s_0}^{3}\right) {R}^{8}-1/6{r}^{10} \left( s_0 \left( {s_0}^{4}+30{s_2}^{3}s_0-8 {s_0}^{3}s_2-12{s_2}^{2}{s_0}^{2}-12{s_2}^{4} \right) \zeta_1-{\frac {15}{4}}{s_0}^{4}-{\frac { 223}{2}}{s_2}^{2}{s_0}^{2}\right.\right.\right.\nonumber\\
&\left.\left.\left.-18{s_2}^{4}+90{s_2}^{3}s_0+34{s_0}^{3}s_2 \right) {R}^{6}+1/6s_0{r}^{12} \left( 8s_0 \left( s_0-3/4s_2 \right) {s_2}^{2} \left( -s_2+s_0 \right) \zeta_1 +{s_0}^{4}-{\frac {89}{4}}{s_0}^{3}s_2+{\frac {361}{4 }}{s_2}^{2}{s_0}^{2}+27{s_2}^{4}\right.\right.\right.\nonumber\\
&\left.\left.\left.-{\frac {187}{2}} {s_2}^{3}s_0 \right) {R}^{4}-1/24{s_0}^{2} \left( -4 s_0{s_2}^{2} \left( -s_2+s_0 \right) \zeta_1+{s_0}^{3}+54{s_2}^{3}+11s_2{s_0}^{2} -86s_0{s_2}^{2} \right) {r}^{14} \left( -s_2+s_0 \right) {R}^{2}+1/24{r}^{16}{s_0}^{3}s_2 \left( s_0\right.\right.\right.\nonumber\\
&\left.\left.\left.+9s_2 \right)  \left( -s_2+s_0 \right) ^{2} \right) {\zeta_1}^{2}X_{10}-192{R}^{4}\zeta_1 \left( {\zeta_1}^{2}{R}^{14}+3/2 \left( \zeta_1 s_0+{\frac {10}{9}} \right) {r}^{2}\zeta_1{R}^{12}+1/6 {r}^{4} \left( {\frac {15}{8}}+ \left( 14s_0s_2+{s_0}^{2}-8{s_2}^{2} \right) {\zeta_1}^{2}+ \left( 5s_2\left( s_0\right.\right.\right.\right.\right.\nonumber\\
&\left.\left.\left.\left.+21/2s_0 \right) \zeta_1 \right) {R}^{10}-\frac{1}6 \left( s_0 \left( 8{s_2}^{2}-11s_0s_2+{s_0}^{ 2} \right) {\zeta_1}^{2}+ \left( \frac{23}2{s_2}^{2}-{\frac {105 }{4}}s_0s_2+5/4{s_0}^{2} \right) \zeta_1-{ \frac {21}{8}}s_2-\frac{3}8s_0 \right) {r}^{6}{R}^{8}+\frac{1}3{r} ^{8}\left( {s_0}^{2}s_2 \right.\right.\right.\nonumber\\
&\left.\left. \left.\left( -s_2+s_0 \right) {\zeta_1}^{2}+ \left( -5/8{s_0}^{3}-{\frac {19}{8 }}s_0{s_2}^{2}+{\frac {59}{8}}s_2{s_0}^{2} -3/2{s_2}^{3} \right) \zeta_1+1/16{s_0}^{2}+{\frac {7}{8}}{s_2}^{2}+1/4s_0s_2 \right) {R}^{6}+{ \frac {5}{12}}{r}^{10} \left( s_0s_2 \left( 3/5s_0s_2\right.\right.\right.\right.\nonumber\\
&\left.\left.\left. \left.-6/5{s_2}^{2}+{s_0}^{2} \right) \zeta_1- {\frac {33}{20}}s_2{s_0}^{2}-6/5{s_2}^{3}+{\frac {71}{20}}s_0{s_2}^{2}+1/10{s_0}^{3} \right) {R}^{ 4}+1/24s_0 \left( s_0s_2 \left( 3s_2+s_0 \right)  \left( -s_2+s_0 \right) \zeta_1-{s_0}^{3}-12{s_2}^{3}\right.\right.\right.\nonumber\\
&\left.\left. \left.-3s_2{s_0}^{2}+20s_0 {s_2}^{2} \right) {r}^{12}{R}^{2}+1/24{r}^{14}{s_0}^{2}s_2 \left( 3s_2+s_0 \right)  \left( -s_2+s_0 \right)  \right)X_7 -64{\zeta_1}^{3} \left( {R}^{4}+3{r}^{2} \left( -2/3s_2+s_0 \right) {R}^{2}+{r}^ {4}s_0 \left( -s_2+s_0 \right)  \right) \right.\nonumber\\
& \left. \left( {R}^{ 4}-{r}^{2} \left( s_0-2s_2 \right) {R}^{2}+{r}^{4}s_0 s_2 \right) ^{2} \left( {R}^{2}+1/2s_0{r}^{2} \right) \left( {R}^{2}+{r}^{2} \left( -s_2+s_0 \right)  \right) ^{2 }X_5+64{R}^{8} \left(  \left( {\zeta_1}^ {3}{R}^{10}+1/2{r}^{2}{\zeta_1}^{2} \left( \zeta_1s_0+5 \right) {R}^{8}\right.\right.\right.\nonumber\\
& \left.\left.\left.+{\frac {9}{8}}{r}^{4}\zeta_1 \left( { \frac {7}{9}}+ \left( \frac{2}9s_2+s_0 \right) \zeta_1 \right) {R}^{6}+\frac{1}8 \left( s_0 \left( s_0+s_2 \right) {\zeta_1}^{2}-\frac{1}4++ \left( \frac{3}2s_0+5/2s_2 \right) \zeta_1 \right) {r}^{6}{R}^{4}+\frac{{r}^{8}}{16} \left( s_0 \left( s_0+s_2 \right) \zeta_1-\frac{1}2s_2 \right)R^2 \right.\right.\right.\nonumber\\
& \left.\left.\left. -1/16{r}^{10}s_0s_2 \right) X_8+1/16{R}^{4}X_9{r }^{4} \left( \zeta_1{R}^{2}+1/2{r}^{2} \right)  \right) \right) \,,
\end{align}
\begin{align}\label{eq:pt_grad}
&p'_t=\frac{1}{ {R}^{14}{r}^{11}{\kappa}^{2}}\left( 512{R}^{2} \left( {R}^{14}\zeta_1+\frac{9}4{r}^{2} \left(\frac{4}9+ \left( s_0-\frac{1}9s_2 \right) \zeta_1 \right) {R}^{12 }-\frac{1}2 \left(  \left( {s_0}^{2}+15/2{s_2}^{2}-13s_0s_2\right) \zeta_1-{\frac {23}{8}}s_0-{\frac {11} {8}}s_2 \right) {r}^{4}{R}^{10}\right.\right.\nonumber\\
& \left.\left. -9/4{r}^{6} \left(  \left( -{ \frac {8}{9}}{s_2}^{3}-16/3s_2{s_0}^{2}+{s_0 }^{3}+14/3s_0{s_2}^{2} \right) \zeta_1-{\frac {37}{ 18}}s_0s_2+{\frac {7}{36}}{s_0}^{2}+{\frac {29}{ 36}}{s_2}^{2} \right) {R}^{8}-1/2 \left(  \left( -2{s_2}^{4}+{s_0}^{4}+11/2{s_2}^{2}{s_0}^{2}\right.\right.\right.\right.\nonumber\\
& \left.\left.\left.\left.-8{s_0} ^{3}s_2+3{s_2}^{3}s_0 \right) \zeta_1-{\frac {85 }{4}}s_0{s_2}^{2}-3/2{s_0}^{3}+{\frac {35}{4}}s_2{s_0}^{2}+21/2{s_2}^{3} \right) {r}^{8}{R}^{6}+1/ 4{r}^{10} \left( s_0s_2 \left( -s_2+s_0 \right)  \left( -4{s_2}^{2}+5s_0s_2+{s_0}^{ 2} \right) \zeta_1\right.\right.\right.\nonumber\\
& \left.\left.\left.+17{s_2}^{4}+{\frac {205}{2}}{s_2 }^{2}{s_0}^{2}+{\frac {15}{4}}{s_0}^{4}-{\frac {157}{4}} {s_0}^{3}s_2-80{s_2}^{3}s_0 \right) {R}^{4}+\frac{ s_0{r}^{12}}{16} \left( 4s_0{s_2}^{2} \left(s_0 -s_2 \right) \zeta_1+{s_0}^{3}-68{s_2}^{3}-42 s_2{s_0}^{2}+141s_0{s_2}^{2} \right) \right.\right.\nonumber\\
& \left.\left.\left( -s_2+s_0 \right) {R}^{2}-1/16{s_0}^{2}{r}^{14 }s_2 \left( s_0-17s_2 \right)  \left( -s_2+s_0 \right) ^{2} \right)  \left( {R}^{4}-{r}^{2} \left( s_0-2 s_2 \right) {R}^{2}+{r}^{4}s_0s_2 \right) {\zeta_1}^{3} \left( {R}^{2}+{r}^{2} \left( -s_2+s_0 \right) \right) X_5\right.\nonumber\\
& \left.-768{R}^{4}{\zeta_1}^{2} \left( {R}^ {18}{\zeta_1}^{2}+\frac{3}2{r}^{2}\zeta_1 \left( \frac{4}3+ \left( \frac{1}3s_2+s_0 \right) \zeta_1 \right) {R}^{16}-{r}^{4} \left( -{\frac {55}{96}}+ \left( {s_0}^{2 }-7s_0s_2+\frac{7}2{s_2}^{2} \right) {\zeta_1}^{2}-\left({\frac {19}{ 8}}s_2+{\frac {15}{8}}s_0 \right) \zeta_1 \right) { R}^{14}\right.\right.\nonumber\\
& \left.\left.-3/2 \left( s_0 \left( 10/3{s_2}^{2}-16/3s_0s_2+{s_0}^{2} \right) {\zeta_1}^{2}+ \left( { \frac {25}{9}}{s_2}^{2}+{\frac {23}{18}}{s_0}^{2}-{ \frac {137}{18}}s_0s_2 \right) \zeta_1-1/12s_0-{\frac {19}{24}}s_2 \right) {r}^{6}{R}^{12}-{\frac {5}{12}}  \left(  \left( {\frac {84}{5}}{s_2}^{3}s_0\right.\right.\right.\right.\nonumber\\
&\left.\left. \left.\left.-{\frac {66}{ 5}}{s_2}^{2}{s_0}^{2}-{\frac {24}{5}}{s_2}^{4} \right) {\zeta_1}^{2}+ \left( -{\frac {13}{5}}s_2{s_0}^{2}-{\frac {229}{10}}s_0{s_2}^{2}+{s_0}^{3}+{ \frac {151}{10}}{s_2}^{3} \right) \zeta_1-{\frac {59}{40}} {s_2}^{2}+\frac{5}8{s_0}^{2}-{\frac {29}{10}}s_0s_2 \right) {r}^{8}{R}^{10}-\frac{1}2{r}^{10} \left( s_0s_2 \right.\right.\right.\nonumber\\
&\left. \left.\left.\left( s_0-s_2 \right)  \left( 4{s_2}^{2}-7s_0s_2+{s_0}^{2} \right) {\zeta_1}^{2}+ \left( {\frac {83}{2}}{s_2}^{3}s_0-{ \frac {161}{3}}{s_2}^{2}{s_0}^{2}+{\frac {107}{6}}{s_0}^{3}s_2-{\frac {11}{ 6}}{s_0}^{4}-{\frac {23}{3}}{s_2}^{4} \right) \zeta_1-{\frac {199}{24}}s_0{s_2}^{2}+{\frac {65}{24}} {s_2}^{3}+3s_2{s_0}^{2}\right.\right.\right.\nonumber\\
&\left. \left.\left.-{\frac {11}{48}}{s_0 }^{3} \right) {R}^{8}+\frac{1}{24} \left( 12{s_0}^{2}{s_2}^{2} \left( -s_2+s_0 \right) ^{2}{\zeta_1}^{2}+ \left( -47 {s_0}^{4}s_2+36{s_2}^{5}-50{s_2}^{4}s_0 +{s_0}^{5}-130{s_0}^{2}{s_2}^{3}+198{s_0}^{3}{ s_2}^{2} \right) \zeta_1-\frac{13}2{s_0}^{4}\right.\right.\right.\nonumber\\
&\left. \left.\left.-{\frac {841}{4 }}{s_2}^{2}{s_0}^{2}-103{s_2}^{4}+{\frac {263}{4}} {s_0}^{3}s_2+273{s_2}^{3}s_0 \right) {r}^{12}{ R}^{6}-{\frac {7}{96}}{r}^{14} \left( {\frac {8}{7}}{s_2}^{2 }s_0 \left(s_2+s_0 \right)  \left( 25s_0s_2-18{s_2}^ {2}+{s_0}^{2} \right) \zeta_1+{\frac {1452}{7}}{s_2}^{4}s_0\right.\right.\right.\nonumber\\
&\left. \left.\left.+{s_0}^{5}-{\frac {264}{7}}{s_2}^{5}-{\frac {232}{7}}{s_0}^{4}s_2+{\frac {1265}{7} }{s_0}^{3}{s_2}^{2}-{\frac {2284}{7}}{s_0}^{2}{s_2}^{3} \right) {R}^{4}-1/24s_0{r}^{16} \left( s_0 s_2 \left(s_0 -s_2 \right)  \left( s_0-9s_2 \right) \zeta_1-{\frac {643}{4}}s_0{s_2}^{2 }-{\frac {19}{4}}{s_0}^{3}\right.\right.\right.\nonumber\\
&\left. \left.\left.+{\frac {143}{2}}s_2{s_0}^{2}+66{s_2}^{3} \right) s_2 \left(s_0 -s_2 \right) {R}^{2}-{\frac {5}{48}}{s_0}^{2} \left( s_0-{ \frac {33}{5}}s_2 \right) {r}^{18}{s_2}^{2} \left( s_0-s_2 \right) ^{2} \right)X_{10}-128 \left( {R}^{4} -{r}^{2} \left( s_0-2s_2 \right) {R}^{2}+{r}^{4}s_0 s_2 \right) ^{3}\right.\nonumber\\
&\left. {\zeta_1}^{4} \left( {R}^{2}+{r}^{2} \left(s_0 -s_2 \right)  \right) ^{3} \left( {R}^{4}+3{r}^{2} \left(s_0 -\frac{2}3s_2 \right) {R}^{2}+{r}^{4}s_0 \left( s_0-s_2 \right)  \right) X_6-128 {R}^{6} \left( -4\zeta_1 \left( {R}^{16}{\zeta_1}^{3}+\frac{3}4 \left( 4+ \left( \frac{1}3s_2+s_0 \right)\right.\right.\right.\right.\nonumber\\
&\left.\left. \left.\left. \zeta_1 \right) {r}^{2}{\zeta_1}^{2}{R}^{14}-1/2 \left( -{\frac {109}{ 32}}+ \left( 3{s_2}^{2}-11/2s_0s_2+{s_0}^{2} \right) {\zeta_1}^{2}+ \left( -{\frac {21}{8}}s_0-{\frac {33}{8}}s_2 \right) \zeta_1 \right) {r}^{4}\zeta_1{ R}^{12}+3/4{r}^{6} \left( {\frac {7}{48}}+s_0s_2 \left(s_0 -s_2 \right) {\zeta_1}^{3}\right.\right.\right.\right.\nonumber\\
&\left.\left. \left.\left. + \left({\frac {113}{12}}s_0s_2 -{\frac {25}{6}}{s_2}^{2}-{\frac {23}{12}}{s_0}^{2} \right) {\zeta_1}^{2}+{\frac {35}{12}} \zeta_1s_2 \right) {R}^{10}+1/16 \left(  \left( 6s_2{s_0}^{2}+{s_0}^{3}-24{s_2}^{3}+27s_0 {s_2}^{2} \right) {\zeta_1}^{2}+  \zeta_1\left({\frac {31}{4}}{s_2}^{2} -{\frac {25}{4}} {s_0}^{2}\right.\right.\right.\right.\right.\nonumber\\
&\left.\left.\left. \left.\left. +{\frac {87}{4}}s_0s_2 \right)-3/4s_0+7/2s_2 \right) {r}^{8}{R}^{8}-1/8{r}^{10} \left( s_0s_2 \left( -s_2+s_0 \right)  \left( s_0-6s_2 \right) {\zeta_1}^{2}+ \left( 15s_2{s_0}^{2}+{ \frac {57}{4}}{s_2}^{3}-{\frac {139}{4}}s_0{s_2}^ {2}-{\frac {11}{8}}{s_0}^{3} \right) \zeta_1\right.\right.\right.\right.\nonumber\\
&\left.\left. \left.\left.-{\frac {29}{ 16}}{s_2}^{2}-1/16{s_0}^{2}+{\frac {15}{16}}s_0 s_2 \right) {R}^{6}+{\frac {1}{64}}{r}^{12} \left(  \left( 5{ s_2}^{3}s_0-15{s_0}^{3}s_2-24{s_2}^{4}+41 {s_2}^{2}{s_0}^{2}+{s_0}^{4} \right) \zeta_1-5/2 {s_0}^{3}-70s_0{s_2}^{2}+30s_2{s_0} ^{2}\right.\right.\right.\right.\nonumber\\
&\left.\left. \left.\left.+39{s_2}^{3} \right) {R}^{4}-{\frac {1}{64}}{r}^{14} \left( s_0s_2 \left(s_0 -s_2 \right) \left( 3s_0s_2-12{s_2}^{2}+{s_0}^{2} \right) \zeta_1+24{s_2}^{4}-\frac{21}2{s_0}^{3}s_2+ {\frac {125}{2}}{s_2}^{2}{s_0}^{2}+\frac{1}2{s_0}^{4}-{ \frac {169}{2}}{s_2}^{3}s_0 \right) {R}^{2}\right.\right.\right.\nonumber\\
&\left. \left.\left.+{\frac {1}{128 }}{r}^{16}s_0s_2 \left( -s_2+s_0 \right) \left( 24{s_2}^{2}-9s_0s_2+{s_0}^{2} \right)  \right)X_7+{R}^{2} \left(  \left( {\zeta_1}^{4 }{R}^{14}+4{R}^{12}{\zeta_1}^{3}{r}^{2}-1/4 \left( -{\frac { 53}{4}}+ \left( s_0-3s_2 \right) \zeta_1 \right) {r}^ {4}{\zeta_1}^{2}{R}^{10}\right.\right.\right.\right.\nonumber\\
&\left.\left. \left.\left.+\frac{3}4{r}^{6}\zeta_1 \left( \frac{1}2+ \left( s_0-{\frac {13}{6}}s_2 \right) \zeta_1 \right) {R}^{8}-{\frac {5}{16}}{r}^{8}\zeta_1 \left( \left( s_0-4/5s_2 \right) s_2\zeta_1+\frac{3}5s_0-\frac{6}5s_2 \right) {R}^{6}-\frac{1}{16} \left( s_0s_2  \left(s_0 -s_2 \right) \zeta_1-\frac{7}2{s_2}^{2}- \frac{1}2{s_0}^{2}\right.\right.\right.\right.\right.\nonumber\\
&\left.\left.\left. \left.\left.+9/2s_0s_2 \right) {r}^{10}\zeta_1{R}^{4}-1/32{r}^{12} \left( s_0s_2 \left( -s_2+s_0 \right) \zeta_1+1/2{s_0}^{2}+1/2{s_2}^{2}-3/2s_0s_2 \right) {R}^{2}+{\frac {1}{64}}{r}^ {14}s_0s_2 \left( -s_2+s_0 \right)  \right) X_8\right.\right.\right. \nonumber\\
& \left. \left.\left.+1/16{R}^{8}{r}^{4}\zeta_1X_9\left( {r}^{2}+\zeta_1{R}^{2} \right)  \right)  \right)  \right) \,.
\end{align}
\begin{align}
&X_5={e^{ \left[ -2\zeta_1 \left\{ {R}^{4}+3{r}^{2} \left( -2/3s_2+s_0 \right) {R}^{2}+{r}^{4}s_0 \left( -s_2+s_0 \right)  \right\} {e^{-{\frac {{r}^{2}s_2}{{R}^{2}}} }}+2\zeta_1{R}^{4}-4s_2{r}^{4} \right] {R}^{-2}{r}^{ -2}}}\,,\nonumber\\
&X_6={e^{ \left[ -2\zeta_1 \left\{ {R}^{4}+3{r}^{2} \left( -2/3s_2+s_0 \right) {R}^{2}+{r}^{4}s_0 \left( -s_2+s_0 \right)  \right\} {e^{-{\frac {{r}^{2}s_2}{{R}^{2}}} }}+2\zeta_1{R}^{4}-5s_2{r}^{4} \right] {R}^{-2}{r}^{ -2}}}\,,\nonumber\\
&X_7={e^{ \left[ -2\zeta_1 \left\{ {R}^{4}+3{r}^{2} \left( -2/3s_2+s_0 \right) {R}^{2}+{r}^{4}s_0 \left( -s_2+s_0 \right)  \right\} {e^{-{\frac {{r}^{2}s_2}{{R}^{2}}} }}+2\zeta_1{R}^{4}-2s_2{r}^{4} \right] {R}^{-2}{r}^{ -2}}}\,,\nonumber\\
&X_8={e^{ \left[ -2\zeta_1 \left( {R}^{4}+3{r}^{2} \left( -2/3s_2+s_0 \right) {R}^{2}+{r}^{4}s_0 \left( -s_2+s_0 \right)  \right) {e^{-{\frac {{r}^{2}s_2}{{R}^{2}}} }}+2\zeta_1{R}^{4}-s_2{r}^{4} \right] {R}^{-2}{r}^{-2} }}\,,\nonumber\\
&X_9={e^{-2 \left[  \left( {R}^{4}+3{r}^{2} \left( -2/3s_2+s_0 \right) {R}^{2}+{r}^{4}s_0 \left( -s_2+s_0 \right)  \right) {e^{-{\frac {{r}^{2}s_2}{{R}^{2}}} }}-{R}^{4} \right] \zeta_1{R}^{-2}{r}^{-2}}}\,,\nonumber\\
&X_{10}={e^{ \left[ -2\zeta_1 \left\{ {R}^{4}+3{r}^{2} \left( -2/3s_2+s_0 \right) {R}^ {2}+{r}^{4}s_0 \left( -s_2+s_0 \right)  \right\} { e^{-{\frac {{r}^{2}s_2}{{R}^{2}}}}}+2\zeta_1{R}^{4 }-3s_2{r}^{4} \right] {R}^{-2}{r}^{-2}}}\nonumber\\
&
\end{align}